\documentclass[sn-mathphys-num]{sn-jnl}
\usepackage{graphicx}%
\usepackage{multirow}%
\usepackage{mathrsfs}%
\usepackage[title]{appendix}%
\usepackage{textcomp}%
\usepackage{manyfoot}%
\usepackage{booktabs}%
\usepackage{algorithmicx}%
\usepackage{algpseudocode}%
\usepackage{listings}%
\usepackage{siunitx}
\usepackage{bm}
\usepackage{rotating}
\usepackage{tikz}
\usetikzlibrary{shapes.geometric, arrows}
\usepackage{pgfplots}
\usepackage{float}
\usepackage{amssymb,amsfonts,bbm,amsmath,amsthm,amscd}
\usepackage{natbib}

\usepackage{xcolor}  % Add this line to include the xcolor package
\usepackage{caption}
\usepackage{subcaption}
\usepackage{minibox}
\usepackage{adjustbox} % Include adjustbox for the adjustbox command

\usepackage[linesnumbered,ruled,vlined]{algorithm2e}

\usepackage{colortbl}
\usepackage{xcolor}
\usepackage{booktabs}
\usepackage[textwidth=1.7cm, textsize=footnotesize, obeyFinal]{todonotes}

\tikzstyle{startstop} = [rectangle, rounded corners, minimum width=3cm, minimum height=1cm,text centered, draw=black, fill=red!30]
\tikzstyle{process} = [rectangle, minimum width=3cm, minimum height=1cm, text centered, draw=black, fill=orange!30]
\tikzstyle{decision} = [diamond, minimum width=3cm, minimum height=1cm, text centered, draw=black, fill=green!30]
\tikzstyle{arrow} = [thick,->,>=stealth]
\newcommand{\DPS}{\displaystyle}
\newcommand{\derp}[2]{\frac{\partial #1}{\partial #2}}

\newcommand{\expcomp}{\eta} % Une composante de l'espace des variables explicatives
\newcommand{\St}{\mathbf{S}} % Tenseur des taux de deformation
\newcommand{\Ot}{\boldsymbol{\Omega}} % Tenseur des taux de rotation
\newcommand{\vel}{U} % Vitesses moyennes
\newcommand{\pres}{P} % Pression moyenne
\newcommand{\ReviewerOne}[1]{{\color{black} #1}}
\newcommand{\ReviewerTwo}[1]{{\color{black} #1}}
\newcommand{\NewModif}[1]{{\color{black} #1}}

\newcommand{\tauij}{\tau_{ij}}
\newcommand{\aij}{a_{ij}}
\newcommand{\bij}{b_{ij}}
\newcommand{\Sij}{S_{ij}}

\newcommand{\komegasst}{$k$-$\omega$ SST}

\newcommand{\bdelta}{b^\Delta_{ij}}
\newcommand{\bR}{b^R_{ij}}
\newcommand{\bdeltak}{b^{\Delta,M}_{ij}}
\newcommand{\bRk}{b^{R,M}_{ij}}

\newcommand{\MSST}{M_\text{SST}}
\newcommand{\MSEP}{M_\text{SEP}}
\newcommand{\MANSJ}{M_\text{ANSJ}}
\newcommand{\Mblend}{M_\text{blend}}

\tikzstyle{decision} = [diamond, aspect=2,draw, node distance=1.5cm, inner sep=0pt, fill=blue!20,]
\tikzstyle{block} = [rectangle, draw, node distance=1.cm, rounded corners, fill=green!20,]
\tikzstyle{line} = [draw, -latex']
\tikzstyle{cloud} = [draw, ellipse,fill=red!20, node distance=3cm, minimum height=2em]
\tikzstyle{blk} = [rectangle, draw, node distance=1.cm, rounded corners, fill=green!20, text width=13em]

\definecolor{redHESAM}{RGB}{210,0,37}
\definecolor{EmeraldGreen}{RGB}{0, 128, 0}

\colorlet{MyColorOne}{blue!50} % ???
\newcommand{\darkercolor}[3]{\colorlet{#3}{#1!#2!black}} % Reference Color, Percentage, New Color Name
\darkercolor{MyColorOne}{50}{MyColorOneDark}

\tikzset{
  basetraits/.style ={
    baseline=-0.6ex,
    very thick
  }
}

\tikzset{
  pics/solid/.style ={
    code = {
      \useasboundingbox (0,0) -- (0.54,0) ;
      \draw [pic actions] (0,0) -- (0.54, 0) ;
    }
  }
}

\DeclareRobustCommand{\dotteddef}[1]{\tikz[basetraits] \draw[color=#1, dotted] pic{solid} ;}

\DeclareRobustCommand{\dasheddef}[1]{\tikz[basetraits] \draw[color=#1, dashed] pic{solid} ;}

\DeclareRobustCommand{\dashdotteddef}[1]{\tikz[basetraits] \draw[color=#1, dash dot] pic{solid} ;}

\definecolor{theoryColor}{RGB}{165,190,20}  %{165,190,20}   % Yellow, similar to Python's yellow
\definecolor{HFcolor}{RGB}{0, 0, 0}           % Black for HF
\definecolor{HF1color}{RGB}{128, 128, 128} % Gray for HF
\definecolor{HF2color}{RGB}{255, 0, 255} % Magenta for HF
\definecolor{ANSJcolor}{RGB}{214, 39, 40}     % Red for ANSJ
\definecolor{CHANcolor}{RGB}{34, 139, 34}%{44, 160, 44}     % Green for CHAN
\definecolor{SEPcolor}{RGB}{0,0,255}%{0, 68, 136}  % Darker blue, slightly darker 
\definecolor{intXMAcolor}{RGB}{255, 165, 0}%{255, 127, 14}  % Orange for intXMA

\DeclareRobustCommand{\linetheory}{\dasheddef{theoryColor}}
\DeclareRobustCommand{\lineHF}{\dotteddef{HFcolor}}
\DeclareRobustCommand{\lineHFgray}{\dotteddef{HF1color}}
\DeclareRobustCommand{\lineHFm}{\dotteddef{HF2color}}
\DeclareRobustCommand{\lineANSJ}{\dasheddef{ANSJcolor}}
\DeclareRobustCommand{\lineCHAN}{\dasheddef{CHANcolor}}
\DeclareRobustCommand{\lineSEP}{\dasheddef{SEPcolor}}
\DeclareRobustCommand{\lineintXMA}{\dashdotteddef{intXMAcolor}}

\DeclareRobustCommand{\MarkerSquare}[1]{%
  $\tikz[baseline=-0.6ex] \node[color=#1, scale=1.2] at (0,0) {\pgfuseplotmark{square}} ;$%
}

\DeclareRobustCommand{\MarkerCircle}[1]{%
  $\tikz[baseline=-0.6ex] \node[color=#1, scale=1.2] at (0,0) {\pgfuseplotmark{o}} ;$%
}

\DeclareRobustCommand{\legendtheory}{\linetheory\linetheory}
\DeclareRobustCommand{\legendHF}{\lineHF \MarkerSquare{HFcolor}\lineHF}
\DeclareRobustCommand{\legendHFgray}{\lineHFgray \MarkerCircle{HF1color}\lineHFgray}

\DeclareRobustCommand{\legendHFm}{\lineHFm \MarkerCircle{HF2color}\lineHFm}

\DeclareRobustCommand{\legendANSJ}{\lineANSJ}
\DeclareRobustCommand{\legendCHAN}{\lineCHAN}
\DeclareRobustCommand{\legendSEP}{\lineSEP}
\DeclareRobustCommand{\legendintXMA}{\lineintXMA}

\DeclareRobustCommand{\legendCHANMarker}{\MarkerCircle{CHANcolor}}
\DeclareRobustCommand{\legendSEPMarker}{\MarkerCircle{SEPcolor}}
\DeclareRobustCommand{\legendintXMAMarker}{\MarkerCircle{intXMAcolor}}

\tikzset{
  basehach/.style ={
    baseline=0.4ex
  }
}

\tikzset{
  pics/rectangle/.style ={
    code = {
      \draw[pic actions] (0,0) rectangle (0.35,0.35) ;
    }
  }
}

\pgfdeclarepatternformonly{south east lines}{\pgfqpoint{-0pt}{-0pt}}{\pgfqpoint{3pt}{3pt}}{\pgfqpoint{3pt}{3pt}}{
  \pgfsetlinewidth{0.4pt}
  \pgfpathmoveto{\pgfqpoint{0pt}{3pt}}
  \pgfpathlineto{\pgfqpoint{3pt}{0pt}}
  \pgfpathmoveto{\pgfqpoint{.2pt}{-.2pt}}
  \pgfpathlineto{\pgfqpoint{-.2pt}{.2pt}}
  \pgfpathmoveto{\pgfqpoint{3.2pt}{2.8pt}}
  \pgfpathlineto{\pgfqpoint{2.8pt}{3.2pt}}
  \pgfusepath{stroke}
}

\tikzset{
  pics/pic_Intrusive/.style ={
    code = {
      \useasboundingbox (-0.25,0) -- (0.11,0) ;
      \draw[line width=0.8pt] (0,0.1) -- (0,-0.1);
      \draw[line width=0.4pt] (-0.1,0.1) -- (0.1,0.1);
      \draw[line width=0.4pt] (-0.1,-0.1) -- (0.1,-0.1);
      \draw[line width=0.8pt] (-0.25,0.) -- (0.25,0.);
      \fill[black] (0,0.08) -- (-0.07,-0.06) -- (0.07,-0.06) -- cycle;
    }
  }
}

\tikzset{
  pics/pic_nonIntrusive/.style ={
    code = {%
      \useasboundingbox (-0.25,0) -- (0.11,0) ;
      \draw[line width=0.8pt, color=orange] (0,0.1) -- (0,-0.1);
      \draw[line width=0.4pt, color=orange] (-0.1,0.1) -- (0.1,0.1);
      \draw[line width=0.4pt, color=orange] (-0.1,-0.1) -- (0.1,-0.1);
      \draw[line width=0.8pt, color=orange,dashed] (-0.25,0.) -- (0.25,0.);
      \draw[line width=0.4pt, color=orange] (0,0) circle (0.05);
    }
  }
}

\tikzset{
  pics/pic_nonIntrusiveRFR/.style ={
    code = {%
    \useasboundingbox (0,0) -- (0.4,0) ;
    \draw[color={rgb, 255:red,131;green,7;blue,131}, dashed] (0,0) -- (0.5,0) ;
    \node[rotate=180, color={rgb, 255:red,131;green,7;blue,131}, scale=1.5, thin] at (0.25,0) {\pgfuseplotmark{triangle}} ;
    }
  }
}

\definecolor{yellowRFR}{RGB}{255,255,0}
\tikzset{
  pics/pic_IntrusiveRFR/.style ={
    code = {%
    \useasboundingbox (0,0) -- (0.4,0) ;
    \draw[color={rgb,255:red,255;green,255;blue,0}] (0,0) -- (0.5,0) ;
    \node[color={rgb,255:red,255;green,255;blue,0}, scale=1.5, thin] at (0.25,0) {\pgfuseplotmark{*}} ;
    }
  }
}

\tikzset{
  pics/pic_triangle/.style ={
    code = {%
      \draw[color=#1] pic{solid} node[scale=1.2, rotate=-90] at (0.27,0) {\pgfuseplotmark{triangle*}} ;
    }
  }
}

\tikzset{
  pics/pic_multimapdot/.style ={
    code={%
      \useasboundingbox (0,0) -- (0.4,0) ;
      \draw[color=#1] (0,0) -- (0.5,0) ;
      \filldraw [color=#1] (0.25,0) circle [radius=0.5ex];
    }
  }
}

\tikzset{
  invisible/.style={opacity=0,text opacity=0},
  visible on/.style={alt={#1{}{invisible}}},
  alt/.code args={<#1>#2#3}{%
    \alt<#1>{\pgfkeysalso{#2}}{\pgfkeysalso{#3}} % \pgfkeysalso doesn't change the path
  },
}

\begin{document}

\title[Machine-learning-assisted Blending of Data-Driven Turbulence Models]{Machine-learning-assisted Blending of Data-Driven Turbulence Models}

%\title{{\vspace{-0.0cm}\sc{
%Machine-learning-assisted Blending of Data-Driven Turbulence Models
%}}}
\author*[1,2]{\fnm{Mourad} \sur{Oulghelou}}\email{mourad.oulghelou@sorbonne-universite.fr}
\equalcont{These authors contributed equally to this work.}
\author[3]{\fnm{Soufiane} \sur{Cherroud}}\email{soufiane.cherroud@ensam.eu}
\equalcont{These authors contributed equally to this work.}

\author[3]{\fnm{Xavier} \sur{Merle}}\email{xavier.merle@ensam.eu}

\author[1]{\fnm{Paola} \sur{Cinnella}}\email{paola.cinnella@sorbonne-universite.fr}

\affil*[1]{\orgdiv{Institut Jean Le Rond D'Alembert}, \orgname{Sorbonne Universit\'e}, \\\orgaddress{\street{4 Place Jussieu}, \city{Paris}, \postcode{75005}, \country{France}}}

\affil[2]{\orgdiv{Computing and Data Science Institute}, \orgname{Sorbonne Universit\'e}, \\\orgaddress{\street{4 Place Jussieu}, \city{Paris}, \postcode{75005}, \country{France}}}

\affil[3]{\orgdiv{DynFluid Laboratory}, \orgname{Arts et Métiers Institute of Technology}, \\\orgaddress{\street{155 bd. de l'H\^opital}, \city{Paris}, \postcode{75013}
, \country{France}}}

\abstract{
\NewModif{
We present a machine learning–based framework for blending data-driven turbulent closures in the Reynolds-Averaged Navier–Stokes (RANS) equations, aimed at improving their generalizability across diverse flow regimes. Specialized models (hereafter referred to as ``\textit{experts}'') are trained via sparse Bayesian learning and symbolic regression for distinct flow classes, including turbulent channel flows, separated flows, and a near sonic axisymmetric jet. These experts are then combined \emph{intrusively} within the RANS equations using weighting functions, initially derived via a Gaussian kernel on a dataset spanning equilibrium shear conditions to separated flows. Finally, a Random Forest Regressor is trained to map local physical features to these weighting functions, enabling deployment in previously unseen scenarios.
We evaluate the resulting blended model on three representative test cases: a turbulent zero-pressure-gradient flat plate, a wall-mounted hump, and a NACA0012 airfoil at various angles of attack, ranging from fully attached to near-stall conditions. Results for these 2D flows show that the proposed strategy adapts to local flow characteristics, effectively leveraging the strengths of individual models and consistently selecting the most suitable expert in each region. Notably, the blended model also demonstrates robustness for flow configurations not included in the training set, underscoring its potential as a practical and generalizable framework for RANS turbulence modeling.
}
}


\maketitle

\section{Introduction}
Turbulence models are a crucial component of Computational Fluid Dynamics (CFD) solvers for engineering applications, which mostly rely on the Reynolds-Averaged Navier-Stokes (RANS) equations.  Most modeling efforts target unclosed terms, known as Reynolds Stresses, which represent the contribution of the unresolved turbulent scales to the transport of mean flow momentum. Although a considerable number of models have been developed with varying degrees of success in predicting flows (see \cite{spalart2000,pope2000turbulent,wilcox2006turbulence,spalart2015,durbin2018some} for overviews), none has been shown to be systematically superior or hopeless \cite{spalart2000}, and turbulence modeling continues to be an active area of research \cite{rumsey2022nasa}. Most RANS turbulence models are known to face challenges with complex flows involving turbulence nonequilibrium, strong gradients, separations, shocks, and 3D effects. The shortcomings are particularly significant for the so-called linear eddy viscosity models (LEVM),  which rely on Boussinesq's hypothesis of a linear relationship between the turbulent stresses and the mean flow deformation  \cite{wilcox2006turbulence,schmitt2007boussinesq}. Despite their limitations, LEVM remain the most widely used turbulence models in industrial flow solvers because of their good tradeoff between accuracy, computational cost, and robustness. 
Efforts to enhance LEVM include rotation corrections \cite{spalart_shur1997}, non-linear models \cite{speziale1987}, elliptic relaxation models \cite{durbin1991near}, algebraic and explicit algebraic Reynolds Stress models \cite{rodi1976algebraic,pope1975more,gatski1993explicit,wallin2000explicit}, and full Reynolds Stress Models \cite{speziale1995}. However, most of these more sophisticated models have had limited success in CFD applications due to their higher cost and lower robustness;  they also involve numerous adjustable coefficients, often calibrated against a limited set of so-called ``canonical flows'', which correspond to simplified turbulence states that, while occurring in subregions of more general flows, are not fully representative of the flow behavior in complex configurations.
Overall, the variety of model formulations and associated closure coefficients represent a significant source of uncertainty in flow simulations \cite{xiao2019quantification}. 
%Scale-resolving approaches (LES, wall-modeled LES, hybrid RANS/LES) resolve larger turbulent structures but are computationally expensive for extensive simulations.

Over the past decade, Machine Learning (ML) has emerged as an attractive option for enhancing turbulence models by relaxing some of the simplifying modeling assumptions and by leveraging the information contained in high-fidelity data sets  \cite{duraisamy2019turbulence,xiao2019quantification,duraisamy2020perspectives,sandberg2022review}. Various flavors of ML-supported turbulence models have been proposed, most of which consist of learning some form of correction terms from the data to improve the representation of the Reynolds Stresses, the auxiliary equations used to describe the evolution of turbulent scales, or both. The corrections are generally learned for specific flow classes, e.g. separated flows \cite{weatheritt2016novel,schmelzer2020discovery,volpiani2021machinePRF,cherroud2022sparse,volpiani2022neuralIJHFF}, airfoils \cite{singh2017machine,holland2019towards}, wind turbine wakes \cite{Steiner2022}, secondary flows \cite{amarloo2023data} and, more rarely, on several flows simultaneously \cite{wang2017physics,wu2018physics}. While data-driven models learned on narrow flow classes can significantly improve accuracy over baseline models, they do not generalize to radically different flows, and often deteriorate the representation of canonical flows for which the baseline models where calibrated. On the other hand, multi-flow learning may lead to a soft compromise solution,  thus limiting potential improvements over standard models \cite{fang2023toward}.  
Enriching the set of features used to represent the data-driven corrections helps improving generalizability but, even so,  distinct models are needed to accurately represent flows involving different physical processes, i.e. free-shear or wall-bounded flows. 
In addition to multi-flow training, so-called progressive augmentation techniques \cite{bin2022progressive}, consisting in learning corrective terms for specific flow classes (e.g. separated flows \cite{amarloo2023progressive} or secondary flows \cite{rincon2023progressive}) while preserving the attached boundary layer representation, have been shown able to capture a wider set of flows. However, complex flows involving various processes simultaneously (e.g. separation and secondary flows) may still require the introduction of thresholding functions to activate or deactivate each correction as required.\\

An attractive alternative is represented by so-called mixture models \cite{matai2019zonal,de2023space,cherroud2025space,lozano2023machine,ho2024probabilistic}, which consist in blending together the predictions of a set of competing models. Initially introduced for quantifying model-form uncertainties (i.e. uncertainties associated with the model mathematical structure) \cite{edeling2014bayesian,edeling2014predictive,edeling2018bayesian}), they also offer promise for improving the model predictive performance by promoting (downgrading) the component models in the mixture in the regions where they perform the worst. 

\ReviewerOne{In a previous work, some of the present authors showed the potential of adopting mixture-of-expert approaches \cite{yuksel_twenty_2012,JordanJacobs_1994expertregions} in turbulence modeling \NewModif{\cite{de2023space,cherroud2025space}}. Inspired to Bayesian model averaging techniques \cite{hoeting1999bayesian}, a convex linear combination of the component model outputs is used to estimate the posterior probability distribution of some quantity of interest (QoI), with combination weights trained on high-fidelity data. The component models are assigned high weights in regions where they perform best, and low ones in regions of high inaccuracy. Such regions are identified through a set of local flow features, and weighting functions are learned under the form of input/output relationships between the input features and the local model weights. The latter can be interpreted as local model probabilities, and an estimate of the turbulence modeling uncertainty can be obtained as the variance of the individual component model solutions.
While such ``external'' or ``non-intrusive'' mixtures of turbulence model solutions have proven effective in predicting a variety of flows not seen at training and in estimating modeling uncertainties \cite{de2023space,cherroud2025space}, their aggregated flow fields are in general not a solution of the governing equations, i.e. they do not satisfy mass and energy conservation principles. Furthermore, similar to other uncertainty quantification approaches, external model mixtures require a new flow to be simulated multiple times using each component model. If $N_M$ models are used, the computational cost of the aggregated prediction is equivalent to that of $N_M$ simulations.}

On the other hand, some authors have proposed \ReviewerOne{ internal model blending strategies, which consist of first combining a set of alternative turbulence models and then propagating the mixture through the mean flow equations.} For instance, \cite{matai2019zonal} and \cite{lozano2023machine} have trained classifiers alongside data-driven models, to identify regions where the flow is similar to the conditions seen at training. Ref. \cite{ho2024probabilistic} has proposed an internal model mixture using the posterior variances of stochastic data-driven models (represented as Gaussian processes) to build the weights. \ReviewerOne{ It is worth noting that several well-established turbulence models in the literature also use heuristic blending functions to activate certain terms in specific flow regions. One the best known examples is the $k-\omega$ SST model, which was originally developed as a mixture of the $k-\epsilon$ and $k-\omega$ models \cite{menter1992improved}}.

\begin{figure}
\begin{center}
\includegraphics[width=9.2cm]{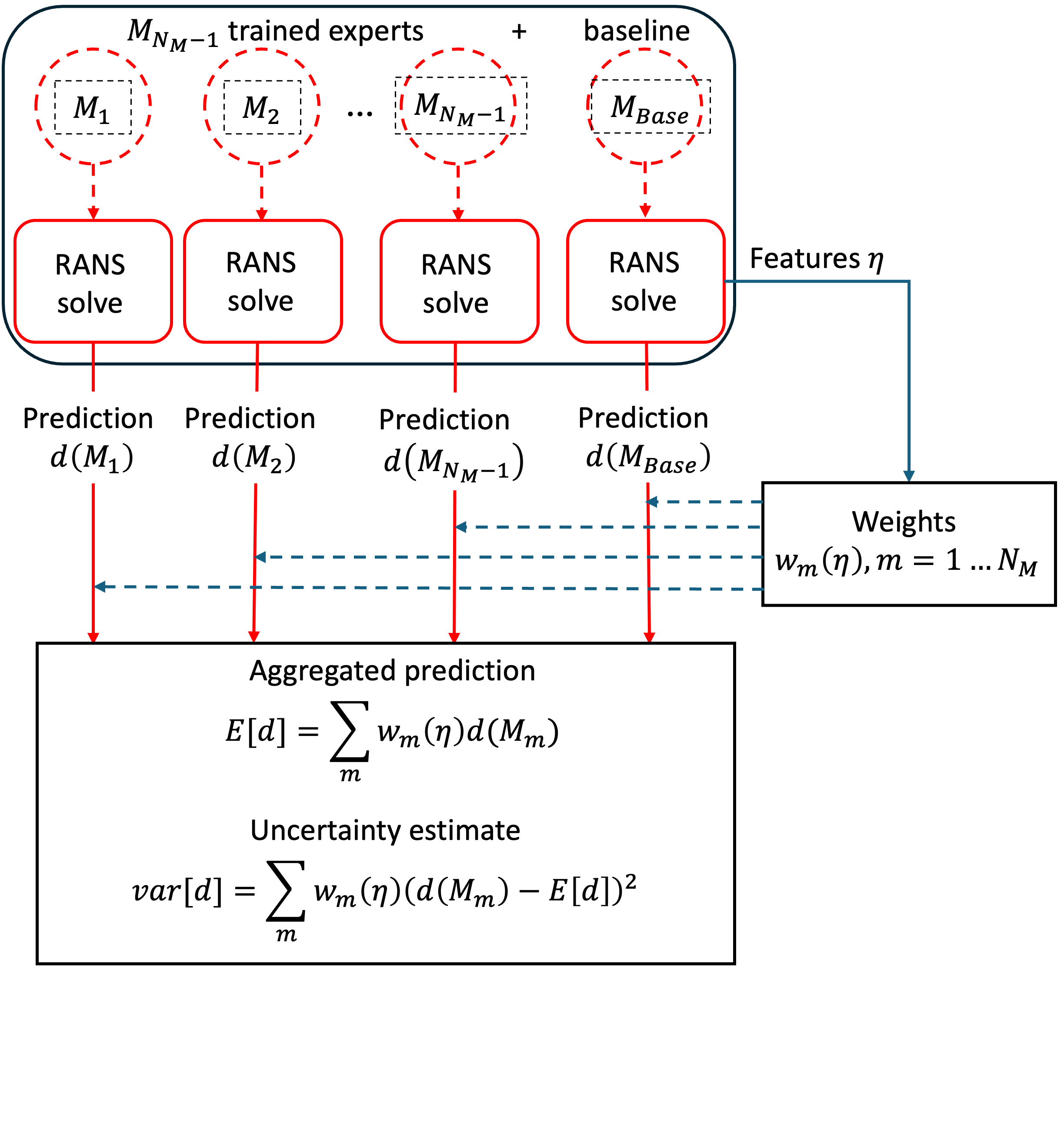}
\vspace{-1cm}
\caption{Workflow of the external model aggregation of \cite{de2023space,cherroud2025space}.}
\label{fig:ext}
\end{center}
\end{figure}

\begin{figure}
\begin{center}
\includegraphics[width=9.2cm]{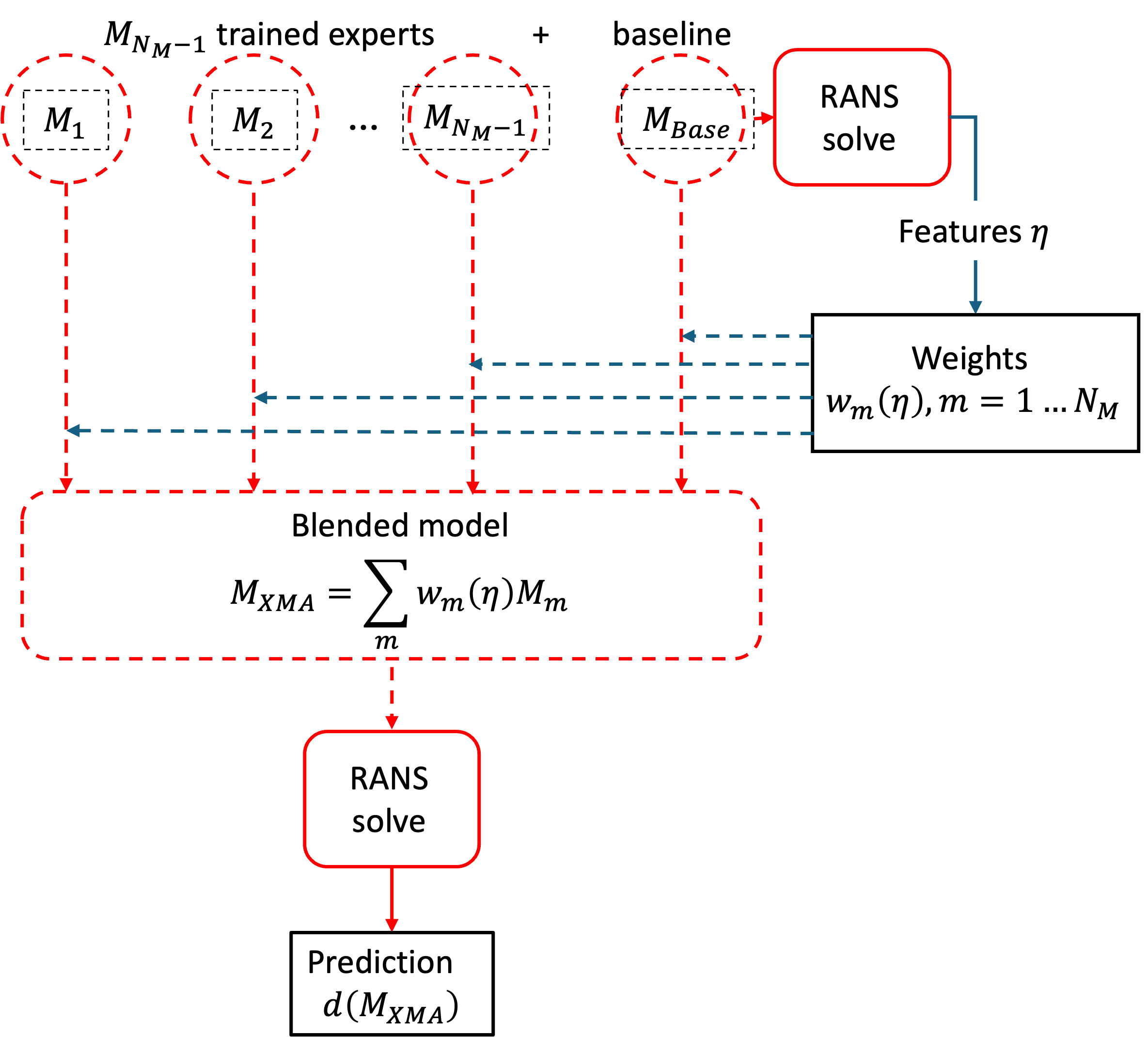}
\vspace{0.2cm}
\caption{Workflow of internal model blending (present work).}
\label{fig:int}
\end{center}
\end{figure}

In this work, \ReviewerOne{we build on} the space-dependent model aggregation (XMA) approach initially proposed in \cite{de2023space} \ReviewerOne{ for mixing the predictions of a set of competing turbulence models, selected among some popular models from the literature. The methodology has been further developed in \cite{cherroud2025space}, where the XMA has been used to aggregate the solutions of a set of data-driven turbulence models trained on selected flow classes. Here, we propose to use XMA for internally blending a set of flow-specific data-driven turbulence model corrections, or "experts". Like in external XMA, we use weighting functions trained across several flow cases. The blended data-driven corrections are then applied to a baseline model and propagated altogether through the RANS equations to predict a new flow. Because the resulting flow fields are obtained as solutions of the RANS equations supplemented with the blended data-driven model, they satisfy the conservation  principles. Furthermore, since a single simulation is now necessary to formulate predictions, the computational cost of the internal blending method is reduced compared to external XMA. As a downside, the proposed approach does not deliver estimates of the predictive uncertainty as naturally as the external model aggregation. The workflows in Figure \ref{fig:ext} and Figure \ref{fig:int} illustrate the difference between the external model aggregation and the present internal model blending}. 


The remainder of this paper is organized as follows. In Section 2, we recall the data-driven framework used to train expert models for selected flow classes. Section 3 introduces the proposed blending strategy and the methodology used to learn the model weighting functions. In Section 4, we present and discuss the results for training and test \ReviewerOne{ two-dimensional} turbulent flow scenarios, highlighting the improvements in predictive accuracy and physical fidelity. Finally, in Section 5, we draw conclusions and outline potential avenues for future research.

\section{Data-driven augmentation of RANS models}

We investigate data-driven turbulence models supplementing the incompressible RANS equations for a constant-property fluid, written as:
\begin{align}
\displaystyle
{
\begin{cases}
	\partial_i U_i &= 0, \nonumber \\
	U_j \partial_j U_i &= \partial_j \left[ -\frac{1}{\rho} P + \nu \partial_j U_i - \tauij \right],
\end{cases}
}
\end{align}
\noindent where $U_i$ is the mean velocity, $\rho$ is the constant density, $P$ is the mean pressure and $\nu$ is the constant kinematic viscosity. The Reynolds-stress $\tauij$ is the target of modelling. This symmetric, positive semi-definite, second-order tensor can be decomposed into an anisotropic part $\aij = 2k\bij$ and an isotropic part $\frac{2}{3} k \delta_{ij}$
\begin{align}
	\tauij &= 2k \left( \bij  + \frac{1}{3} \delta_{ij} \right), \label{eq::reynoldsstress}
\end{align}
\noindent where $k$ is the turbulent kinetic energy and $\delta_{ij}$ the Kroenecker symbol. In the baseline model, the non-dimensional anisotropy tensor $\bij$ is represented as a linear function of the mean-strain rate tensor $\DPS \Sij= \frac{1}{2} \left( \frac{\partial U_i}{\partial x_j} + \frac{\partial U_j}{\partial x_i} \right)$ via the scalar eddy viscosity $\nu_t$: 
$$\bij\approx \bij^0 = -\frac{\nu_t}{k} S_{ij}$$ 
which is known to be a rather simplistic representation of the physical reality (see e.g. \cite{wilcox2006turbulence}).

Furthermore, additional modeling assumptions are used to relate $\nu_t$ to the mean flow variables and turbulent properties. In the \komegasst{} model \cite{menter1992improved}, $\nu_t$ is computed using two auxiliary transport equations for $k$ and for the specific dissipation rate $\omega$. Both equations contain several unclosed terms that are generally modeled on the ground of dimensional considerations and empirical closure coefficients. As a consequence, the modeled turbulent kinetic energy budget differs from the exact one, e.g. obtained from a well-resolved DNS, and this leads to significant discrepancies between the turbulent scales derived from the modeled $k$ and $\omega$ and the exact ones.

Hereafter, we follow the approach initially proposed in \cite{schmelzer2020discovery}, and we augment the baseline  constitutive relation by adding a corrective term $\bdelta$: 
\begin{align}
	\bij   &= \bij^0 + \bdelta.\label{eq::nonlinearconstrel}
\end{align}
This additive correction, leaving the linear term unaltered, has the benefit of enhanced numerical stability \cite{weatheritt2017machine}. Furthermore, we also augment the $k-\omega$ transport equations to account for modeling discrepancies with the exact turbulent kinetic energy equation. Finally, the augmented transport equations are:
\begin{align}
	U_j \partial_j k &= P_k + R_k - \beta^* \omega k + \partial_j \left[ D_k^{\text{Eff}} \partial_j k \right],\label{eq::augmentedkeq} \\
	 U_j \partial_j \omega &= \gamma (\widetilde{P}_k + R_k) - \beta \omega^2 + \partial_j \left[ D_{\omega}^{\text{Eff}} \partial_j \omega \right] + CD_{k\omega},
	 \label{eq::augmentedkOmegaSST}
\end{align}
where 
\begin{align}
	&P_k = \min \left\{ 2 ( \nu_t S_{ij} - k \bdelta) \partial_j U_i ;  c_1 \beta^* k \omega \right\},\nonumber
 \\
   &\widetilde{P}_k = \min \left\{ 2 (S_{ij} - \frac{k}{\nu_t} \bdelta) \partial_j U_i ;  \frac{c_1}{a_1} \beta^* \omega \max\{a_1 \omega; b_1 F_2 \sqrt{S}\} \right\}. 
	\nonumber \\ %\label{eq::productionlimiter}\\
	&CD_{k\omega} = 2 \alpha_{\omega 2} \frac{1}{\omega}(\partial_i k)(\partial_i \omega), \nonumber\\
\end{align}
and the remaining auxiliary equations and parameters of \komegasst{} model are given in Appendix \ref{Appendix_kwSST_csts}. Given $k$ and $\omega$, the eddy viscosity $\nu_t$ is updated as follows 
\begin{equation}\label{nut_equ}
\nu_t = \frac{a_1 k}{\text{max}(a_1 \omega, \sqrt{S} F_{23})}, \quad \textrm{with } S = {2S_{ij}S_{ij}}
\end{equation}
\ReviewerOne{ It is important to note that the transport equations \eqref{eq::augmentedkeq} and \eqref{eq::augmentedkOmegaSST} are augmented with the correction term $R_k$, which corrects the inadequacies in the modeled turbulence kinetic energy equation. The target corrective terms  $\bdelta$ and $R_k$ are extracted from high-fidelity data using the $k$-corrective frozen RANS approach of  \cite{schmelzer2020discovery}, which consists in passively solving the transport equations for the auxiliary turbulent quantities using the high-fidelity mean flow data and turbulent kinetic energy. This allows to estimate a modeled turbulence time scale $\omega^{-1}$ as well as the residual term $R_k$. %detailed in Appendix \ref{Frozen}. 
The reader is referred to  \cite{schmelzer2020discovery,cherroud2022sparse} for more details.}

\ReviewerTwo{
Note that the present data-driven augmentation procedure is not restricted to the $k\text{-}\omega$ SST model and can be extended to other two-equation models, such as the
$ k\text{-}\epsilon$ model and its variants (see e.g. \cite{stocker2024dns}). Of course, in that case the experts models must be retrained using the SBL-SpaRTA algorithm adapted to the new model. In the following, we focus on the $k\text{-}\omega$ SST model, for which data-driven corrections for several flow classes have been provided in previous work \cite{cherroud2025space}. .
}
%%%%% Representation
\subsection{Data-driven models : specialized experts}
\ReviewerOne{ Modeling ansatzs for the corrective terms $\bdelta$ and $R_k$ are obtained by projecting them onto an invariant tensor basis.
Specifically, the extra anisotropy $\bdelta$} is projected onto the minimal integrity bases initially proposed in \cite{pope1975more} and reformulated in \cite{gatski1993explicit}:
\begin{align}
    b_{ij}^{\Delta} &= \sum_{n=1}^{10} \alpha_n^\Delta(I_1, \cdots, I_5) \, T_{ij}^{(n)}(S_{ij}^*, \Omega_{ij}^*),
\end{align}
\ReviewerOne{ where the $T_{ij}^{(n)}$ tensor basis is constructed from} the non-dimensional strain and rotation rates ($S_{ij}^*$ and $\Omega_{ij}^*$, respectively), and the functions $\alpha_n$ depend on a set of associated invariants $I_1, \cdots, I_5$. 
For the two-dimensional flows of interest here, only three of these tensor basis functions and two invariants are needed:
\begin{align}
    T_{ij}^{(1)} &= S_{ij}^*, &I_1 = S_{mn}^* S_{nm}^*, \nonumber\\
    T_{ij}^{(2)} &= S_{ik}^* \Omega_{kj}^* - \Omega_{ik}^* S_{kj}^*, &I_2 = \Omega_{mn}^* \Omega_{nm}^*, \nonumber\\
    T_{ij}^{(3)} &= S_{ik}^* S_{kj}^* - \frac{1}{3} \delta_{ij} S_{mn}^* S_{nm}^*, &\nonumber
\end{align}
with $S_{ij}^* = {S_{ij}}/{\omega}, \quad \Omega_{ij}^* = {\Omega_{ij}}/{\omega}$
and 
$\DPS  \Omega_{ij} = \frac{1}{2} \left( \partial_j U_i - \partial_i U_j \right)$.
Then, the extra anisotropy correction tensor reduces to:
\begin{align}
    \bdelta &= \sum_{n=1}^{3} \alpha_n^\Delta(I_1, I_2) \, T_{ij}^{(n)}(S_{ij}^*, \Omega_{ij}^*).
\end{align}
Following \cite{schmelzer2020discovery} and \cite{cherroud2022sparse}, the scalar correction term $R_k$ is expressed as a "production-like" term, modeled as the inner product of a second-order symmetric tensor $\bij^{R}$ and the mean velocity gradient (see \cite{stocker2024dns} for an alternative, "dissipation-like" formulation):
\begin{equation}
    R_k = 2 k \, \bij^{R} \, \partial_j U_i,
\end{equation}
where $\bij^{R}$ is represented with the same tensor basis as $\bdelta$
\begin{equation}\label{bR}
    \bij^{R} = \sum_{n=1}^{3} \alpha_{n}^{R} (I_{1}, I_{2}) \, T_{ij}^{(n)}(S_{ij}^*, \Omega_{ij}^*).
\end{equation}
Using the preceding representations, the modeling target is now represented by the functions $\alpha_n^{\Delta,R}$.\\

\ReviewerOne{ The $\alpha_n^\Delta$ and $\alpha_n^R$ are learned from data by means of a symbolic regression algorithm. Specifically, the Symbolic Bayesian Learning Sparse Regression of Turbulent stress Anisotropy (SBL-SpaRTA) of  \cite{cherroud2022sparse} is used to regress the $\alpha$ functions from a large dictionary of monomials of the invariants $I_1$ and $I_2$.  The reader is referred to \cite{cherroud2022sparse} for details about the SBL-SpaRTA algorithm. }

\section{Machine-learning-based model blending}

\ReviewerOne{ In \cite{cherroud2025space}, SBL-SpaRTA has been used to train data-driven corrections of the \komegasst{} for various flow cases, including turbulent channel flows, zero-pressure-gradient and adverse-pressure-gradient boundary layers, an axisymmetric near-sonic jet, and separated flows. As an outcome, sparse symbolic models have been identified whose parameters are endowed with Gaussian posterior probability distributions.  Such models have been shown to outperform the other models for flows similar to those used for their training, but to be inaccurate when applied to flows significantly different from those in their training sets. For instance, the separated flow model and jet model, do not predict correctly the law of the wall for attached wall-bounded flows; conversely, the attached-flow models perform poorly for separated flows or jets. For that reason, we call them "expert", "specialized" or "customized" models.
Notably, models for channel flows and boundary layers are found to be quite similar and relatively close to the baseline. For this reason,  \cite{cherroud2025space} proposes a model aggregation method (XMA) to account for the uncertainty introduced by data-driven RANS models for complex flows involving multiple physical processes simultaneously (in particular, attached boundary layers, separation, and free shear).
While the XMA approach shows potential for capturing a diverse set of two-dimensional flows not seen at training, it does not provide a single model, but rather delivers an aggregate of individual model predictions along with an uncertainty estimate. The averaged prediction is obtained by simulating the new flow several times with different expert models, and is not expected to satisfy the governing equations, as also shown by the workflow in Fig. \ref{fig:ext}.

\vspace*{0.5cm}
\noindent
To overcome the limitations of individual expert models and promote generality, we follow an idea similar to XMA, but we aggregate the data-driven corrections (internal model aggregation or blending) instead than their solutions.  This approach is no longer an uncertainty quantification method, but provides instead a blended data-driven correction that is expected to be more universal than the component expert models.}
For simplicity, we consider only three "expert" models in the following of the study, namely, the baseline model (denoted $\MSST$), which is selected as the "expert" for attached wall-bounded flows, and the expert models for separated flows and jets, noted  $\MSEP$ and $\MANSJ$. 
The expression of these three models, alongside the sources of the data used for their training, are given in Table  \ref{tab:training_cases_ML}. More details about model training can be found in  \cite{cherroud2025space}. In the table, the data-driven model parameters are reported with the corresponding standard deviations, since SBL-SpaRTA delivers stochastic models endowed with posterior probability distributions. Given that the posterior model variances are small compared to the mean, the model parameters will be treated as deterministic in the following to simplify the presentation. However, it is also possible to apply the proposed aggregation methodology to models with parametric uncertainties, see \cite{cherroudthesis} for more details.

\begin{table*}                  %[H]
 \centering
 \begin{tabular}{ lcc }
  \toprule
  Models & Data-driven corrections & Training cases \\
  \midrule
  $\MSST$ & $ \begin{aligned}
        & {b^{\Delta}} = [0] \\ 
        & b_{R}^{\Delta} = [0]
    \end{aligned} $ & 
    %DNS of turbulent channel flows at \\
  %& & $ 180 \leq Re_{\tau} \leq 5000 $ \cite{moser1999directLee_Moser_2015} 
  \ReviewerOne{N/A}\\
  \midrule
  $\MANSJ$ & $ \begin{aligned}
  & {b^{\Delta}} = [(0.25 \pm 7.26 \cdot 10^{-3})] \, {T^{(1)}}  \\ 
  & b_{R}^{\Delta} = [0] \pm 6.55 \cdot 10^{-5}
    \end{aligned} $ & PIV of near sonic axisymmetric jet \cite{bridges2010establishing} \\
  \midrule
  $\MSEP$ & $ \begin{aligned}
  & {b^{\Delta}} = [(5.21 \pm 0.0173)] \, {T^{(2)}}  \\ 
  & b_{R}^{\Delta} = [(0.681 \pm 0.02)] \, {T^{(1)}}
    \end{aligned} $ & 
    \begin{minipage}{6cm}
    \vspace{0.2cm}
    \centering
  LES of periodic-hills (PH) at $Re=10595$ \cite{breuer2009flow, temmerman_sgs_models_2001}  \\
  DNS of converging-diverging \\
  \ReviewerOne{channel} (CD) at $Re=13600$ \cite{laval2011direct} \\
  LES of curved backward-facing step (CBFS) at $Re = 13700$ \cite{bentaleb2012large}   \\
  \end{minipage}
  \\
 \bottomrule
 \end{tabular}
 \vspace{0.2cm}
 \caption{Experts selected for model aggregation. \label{tab:training_cases_ML}}
\end{table*}

\vspace*{0.5cm}
\noindent

\ReviewerOne{
The proposed internal blending strategy is based on the space-dependent aggregation of a set of expert data-driven model corrections, namely  
$\mathcal{M} = \left\{ \MSST, \MSEP, \MANSJ \right\}$, as reported in Table \ref{tab:training_cases_ML}.  
The resulting blended model is denoted as $\Mblend$. It is worth noting that the $\MANSJ$ model used in this study differs slightly from the version reported in \cite{cherroud2025space}.  
However, this variation does not significantly impact the results, as both models yield nearly identical outcomes.  

Unlike the external, non-intrusive model aggregation approach of \cite{cherroud2025space}, which consists in blending the output quantities of interest (QoI) (such as velocity, pressure, etc.) obtained by solving the RANS equations under different models,
the internal blending strategy mixes a set of alternative corrections of the anisotropy tensor and of the turbulent kinetic energy equation. Specifically, the corrective tensors $\bdelta$ and $\bR$ are formulated as a convex combination of the experts corrections, using a data-driven set of weighting functions $w_M$ that satisfy the convexity constraint $\sum_{M\in\mathcal{M}} w_M = 1$:
\begin{equation}\label{eq:blending}
\bdelta = \sum_{M\in\mathcal{M}} w_M \bdeltak, \quad  
\bR = \sum_{M\in\mathcal{M}} w_M \bRk.
\end{equation}  
}

The weighting functions are designed to dynamically adjust in space, based on a set of local flow features, so to assign high weights to the experts that are most likely to perform accurately based on the local flow conditions, and low weights to the unsuitable ones. 
Following \cite{cherroud2025space}, we adopt a set of eleven features derived from \cite{ling2015evaluation}, noted $\boldsymbol{\eta}(\mathbf{x}) = \bigl(\eta_1(\mathbf{x}),...,\eta_{11}(\mathbf{x})\bigr)$ and described in the Appendix \ref{ext:methodo}, which have been widely used in the literature to represent data-driven corrections. 
Second, we define the exact model weights as:  
\begin{equation*}
\omega_L = \frac{\mathcal{G}(\delta^L, \bar{\delta})}{\DPS \sum_{M\in\mathcal{M}} \mathcal{G}(\delta^M, \bar{\delta})}
\end{equation*}
where $\mathcal{G}$ is a function that depends on the model outputs for some observed QoI $\{ \delta^M \}_{M \in \mathcal{M}}$, and on the corresponding ground truth values $\bar{\delta}$, extracted from a high-fidelity database. Note that by construction the weights $\omega_L$ sum up to $1$.We chose $\mathcal{G}$ to be the Gaussian kernel, given for $\sigma>0$ by
\begin{equation}
\mathcal{G}(\delta^L, \bar{\delta}) =  \exp  \left( - \, \frac{||\delta^L - \bar{\delta} ||^2_2}{2\sigma^2} \right)
\end{equation}
The function $\mathcal{G}$ assigns higher weights to models that are closer to the ground truth data $\bar{\delta}$ and lower weights to those that are farther away. The bandwidth parameter $\sigma$  controls how fast the weight decreases with increasing errors, and it is sought from a set of user-defined values
$\mathcal{E}=\left\{1,10^{-1},10^{-2},10^{-3},10^{-4}  \right\}$ using  a grid  search method. The optimal choice of $\sigma$ is  based on the minimal distance between high-fidelity QoI and the average expert model output based on the Gaussian weights $w_M$, i.e,
$$ \sigma_{opt} = \underset{\sigma\in \mathcal{E}}{\textrm{\textbf{argmin}}} \, \left|\left| \left(\sum_{M\in\mathcal{M}} w_M(\sigma) \, \delta^M\right) - \bar{\delta} \right|\right|_2$$

In the current paper, the \ReviewerOne{ velocity along the $x$-axis} is selected as the quantity for determining the optimal weights. Throughout the document, we use the notation \(U_{x,M}\) to refer to the \ReviewerOne{modelled $x$-velocity for} model \(M\), \(\overline{U}_x\) for the high-fidelity counterpart, and \(U^{\text{ref}}\) (with the superscript "ref") to denote the reference velocity used for normalization (e.g., inlet velocity, velocity at the jet nozzle, etc).
\\
The next step in the blending strategy consists in selecting and training a machine learning model to express the input/output relation between the features and the Gaussian model weights. 
To accomplish this, we first collect a diverse dataset of high-fidelity data, including experiments, DNS and LES data. The selected flows for training are described in Table \ref{Tab_training_info}.
The training flows are then cross-simulated with each of the experts in the set ${\cal M}$. The modeled \ReviewerOne{$x$-velocities} and their high-fidelity counterparts are then used to calculate the exact Gaussian weights for each model at the observation points locations. The flow physical features can then be estimated at these same points. Obviously, the expert models in ${\cal M}$ do not predict the same flow features. In the non-intrusive space-dependent aggregation approach from \cite{de2023space}, the features are assembled as a uniformly weighted average of those of the expert models, which mitigates  the risk of overlooking flow patterns that are not captured by the baseline model. In the present work, we adopt the simpler choice of  \cite{cherroud2025space} and use the features predicted by the baseline model. The argument for this choice is that the expert data-driven models have been learned to correct the baseline in the training environments. Further studies of feature selection and estimation in prediction environments are warranted as future work.\\
Finally, given the baseline features and the exact weights at each observation location in the training set, we solve a machine learning regression problem of the form
\begin{equation}
\underbrace{\boldsymbol{\eta}(\mathbf{x}) = \bigl(\eta_1(\mathbf{x}),...,\eta_{11}(\mathbf{x})\bigr)}_{\text{baseline model local features}}
\xrightarrow[\mathcal{W}]{\textrm{ML}} 
\underbrace{
\{w_M(\mathbf{x})\}_{M\in \mathcal{M}}}_{\text{local model weights}} \label{eq:MLweights}
\end{equation}
where $\mathcal{W}:\boldsymbol{\eta}\longrightarrow \{w_M\}_{M\in \mathcal{M}}$ is a highly nonlinear mapping to be determined. 
In the present study we approximate the features/weights mapping by means of a Random Forest Regressor (RFR), as previously done in \cite{de2023space,cherroud2025space}. \ReviewerTwo{ RFRs are chosen due to their ability to capture complex, non-linear relationships, providing a robust and flexible predictive model. However, they can exhibit non-smooth behavior. } Alternative choices, such as Gaussian process regressors or neural networks could be also used (see \cite{cherroudthesis} for a comparison of RFR and Gaussian process regressors).\\
Once the mapping $\mathcal{W}$ has been learned, the RFR weighting functions are used to build the blended model corrections (\ref{eq:blending}). 
The latter are finally propagated through the RANS solver to make predictions of the training flow cases or of new, unseen flows.
The training and prediction workflows of the blending strategy are summarized through the Algorithms \ref{alg:training} and \ref{alg:prediction}.

%%%%%% ALGOS
\begin{algorithm}[hbt!]
\SetAlgoLined
\KwIn{Set of expert models $\mathcal{M}$, training HF flow data}
\KwOut{Trained RFR model $\mathcal{W}$}
\textbf{Step 1: Simulations on training cases} \\
\For{each training flow case (e.g., turbulent channel, CD, CBFS, PH, ANSJ)}{
    \For{each model $M \in \mathcal{M}$}{
        Run RANS to convergence using model $M$\;
        Extract streamwise velocity $U_{x,M}$\;
    }
    \textbf{Step 2: Compute Gaussian weights}\\
    Using the modelled \ReviewerOne{$x$-velocity} $\{U_{x,M}\}_{M\in\mathcal{M}}$ and its high-fidelity counterpart, evaluate the corresponding Gaussian weights $\{\omega_M\}_{M\in \mathcal{M}}$ at all observation points for the current training case\;
    \textbf{Step 3: Extract the features}\\
     Evaluate local flow features $\boldsymbol{\eta}$ using the baseline model solution at all observation points\;
}

\textbf{Step 4: Learn the mapping $\mathcal{W}$} \\
Train the Random Forest to learn the mapping
\[
{\mathcal{W}} : \boldsymbol{\eta} \rightarrow \{w_M\}_{M\in \mathcal{M}}
\]\\
Save the learned model $\mathcal{W}$\;

\caption{Training of the features/weights mapping ${\cal W}$.}
\label{alg:training}
\end{algorithm}

\begin{algorithm}[hbt!]
\SetAlgoLined
\KwIn{Flow scenario information; Trained RFR  mapping $\mathcal{W}$.}
\KwOut{RANS blended-model solution}
\textbf{Step 1:} Run RANS simulation using the baseline $k-\omega$ SST model.\\
\textbf{Step 2:} Compute the set of local flow features $\boldsymbol{\eta}$ from the converged baseline solution. \\

\textbf{Step 3:} Use the trained RFR $\mathcal{W}(\boldsymbol{\eta})$ to predict local model weights at each prediction location, $\{w_M\}_{M\in \mathcal{M}}$.
\\

\textbf{Step 4:} Assemble the blended RANS corrections $\bdelta$ and $\bR$ and use them to augment the baseline model \komegasst\\
\[
\bdelta = \sum_{M\in\mathcal{M}} \, w_M \, \bdeltak,  \quad \bR = \sum_{M\in\mathcal{M}} \, w_M \,\bRk;
\]
\\

\textbf{Step 5:} Solve the augmented RANS equations for the flow scenario until convergence
\caption{Prediction of a flow scenario}
\label{alg:prediction}
\end{algorithm}
%%%%%%%

\ReviewerOne{
\subsection{Training of weighting functions}

The features/weights mapping $\mathcal{W}$ is trained using the RFR function from the \texttt{scikit-learn} Python library \cite{scikit-learn}. The RFR model has been configured with 100 estimators (trees) and a minimum of one sample per leaf. The quality of the splits is assessed using the squared error criterion. At each split, seven features were randomly selected, and a fixed random state has been employed to ensure reproducibility. The model has been trained on 80\% of the dataset, while the remaining 20\% is reserved for testing. 
The dataset includes all available observations of the \ReviewerOne{ velocity along the $x$-axis} and the corresponding eleven physical features from \cite{ling2015evaluation}, for all training cases. These cases include RANS data for a turbulent channel flow at $Re_\tau = 1000$, PIV data for the Axisymmetric Near Sonic Jet (ANSJ) case \cite{bridges2010establishing}, and LES or DNS data for the Curved Backward-Facing Step (CBFS), Converging-Diverging (CD) channel, and Periodic Hills (PH) cases (refer to Table \ref{tab:training_cases_ML} for reference). The specific mesh points from which the data were retrieved for each case are given in Table \ref{Tab_training_info}.
%
%\begin{table}[h!]
%\centering
%\begin{tabular}{lcc}
%\toprule
%\textbf{Case} & \textbf{Number of Data Points} & \textbf{Location of Points} \\ 
%\midrule
%ANSJ    & $1443$   & Subdomain: $(x/D_{jet}, y/D_{jet}) \in [1,22] \times [0,1.4]$ \\
%CD      & $14000$  & Whole domain \\ 
%Channel & $2400$   & Whole domain \\ 
%PH      & $15600$  & Whole domain \\ 
%CBFS    & $21000$  & Whole domain \\ 
%\bottomrule
%\end{tabular}
%\caption{Information on the training data used to learn the features/weights mapping by the RFR.}
%\label{Tab_training_info}
%\end{table}
%

\begin{table}[h!]
\centering
\begin{tabular}{lc}
\toprule
\textbf{Case}  & \textbf{Location of Points} \\ 
\midrule
ANSJ       & DOFs within the subdomain: $(x/D_{jet}, y/D_{jet}) \in [1,22] \times [0,1.4]$ \\
CD        & All DOFs \\ 
Channel    & All DOFs \\ 
PH        & All DOFs \\ 
CBFS      & All DOFs \\ 
\bottomrule
\end{tabular}
\caption{Information on the training data used to learn the features/weights mapping by the RFR.}
\label{Tab_training_info}
\end{table}

The performance of the RFR model was evaluated using the $R^2$ criterion, which is commonly used to assess the predictive power of regression models.
The $R^2$ score, also known as the coefficient of determination, is calculated as:
\[
R^2 = 1 - \frac{\sum_{i=1}^{N} (y_i - \hat{y}_i)^2}{\sum_{i=1}^{N} (y_i - \bar{y})^2}
\]
where $y_i$ are the true values, $\hat{y}_i$ are the predicted values, and $\bar{y}$ is the mean of the true values.
It measures how well the model fits the data by comparing the variance explained by the model to the total variance in the true values.
For our application, the \(R^2\) score for the training data was 0.991, indicating an excellent fit to the data the model was trained on. For the test data, the \(R^2\) score was 0.99, demonstrating as well a strong predictive accuracy on unseen data by the RFR.

For a more detailed evaluation of the learned RFR performance, we investigate the distribution of the predicted weights versus the exact Gaussian weights, as shown in Fig. \ref{fig:scatter_train}. The scattered weights correspond to all models $\MANSJ$, $\MSST$, and $\MSEP$. It can be seen that most of the training samples, representing 80\% of the data, lie within a narrow band along the diagonal. This indicates that the RFR has successfully captured the relationship between the flow physical features and the corresponding model weights for the training set.
Furthermore, in Fig. \ref{fig:hist_train}, we report the histogram depicting the discrepancies between the predicted and exact weights. All three weights are represented here as well. The histogram is strongly peaked at zero, suggesting that the deviations between the predicted and actual weights for the training set are small throughout.
To further evaluate the generalization performance of the trained RFR, we examine the test set, representing the remaining 20\% of the total data. In Fig. \ref{fig:scatter_test}, the scatter plot for the predicted versus exact weights for the test samples is shown. Similar to the training set, the test samples also follow the diagonal closely, and the distribution of discrepancies shown in Fig. \ref{fig:hist_test} is strongly peaked at zero, indicating that the RFR has generalized well to the unseen test data.
These results provide strong confidence in deploying the trained RFR across both the training and test cases.

\begin{figure}[hbtp!]
    \centering
    % First row: Training set
    \begin{subfigure}[b]{0.48\textwidth}
        \centering
        \includegraphics[scale=0.5]{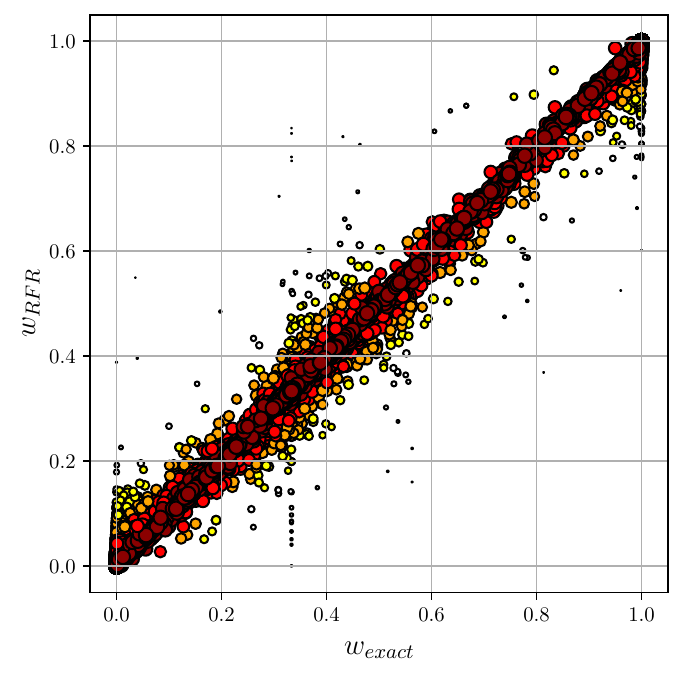}
        \caption{Scatter plot of predicted vs. exact weights for the training set, representing 80\% of the total data set.}
        \label{fig:scatter_train}
    \end{subfigure}
    \hspace{0.02\textwidth}
    \begin{subfigure}[b]{0.48\textwidth}
        \centering
        \includegraphics[scale=0.5]{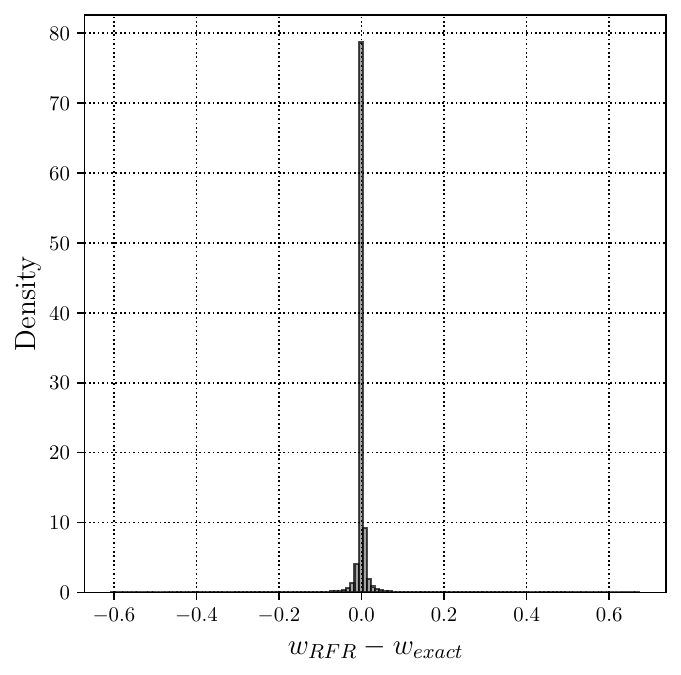}
        \caption{Error histogram for the discrepancies between predicted and exact weights for the training set.}
        \label{fig:hist_train}
    \end{subfigure}
    
    % Second row: Test set
    \begin{subfigure}[b]{0.48\textwidth}
        \centering
        \includegraphics[scale=0.5]{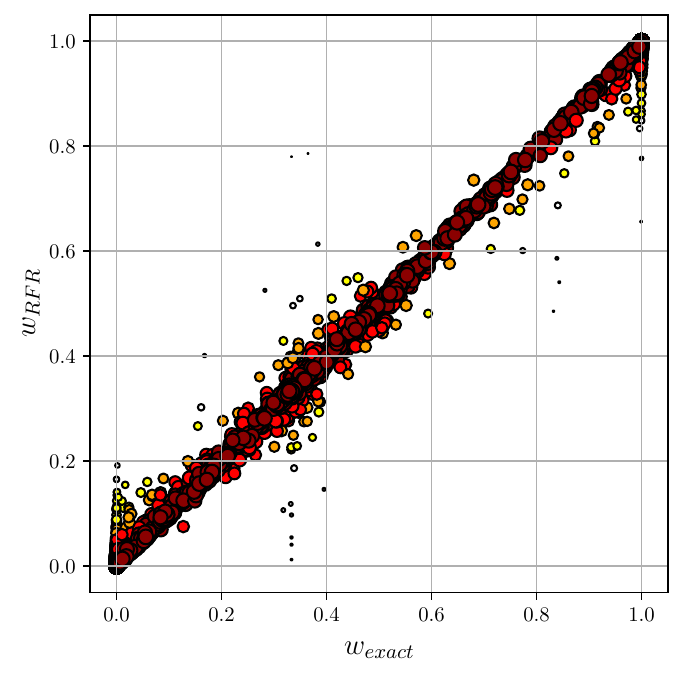}
        \caption{Scatter plot of predicted vs. exact weights for the test set, representing 20\% of the total data set.}
        \label{fig:scatter_test}
    \end{subfigure}
    \hspace{0.02\textwidth}
    \begin{subfigure}[b]{0.48\textwidth}
        \centering
        \includegraphics[scale=0.5]{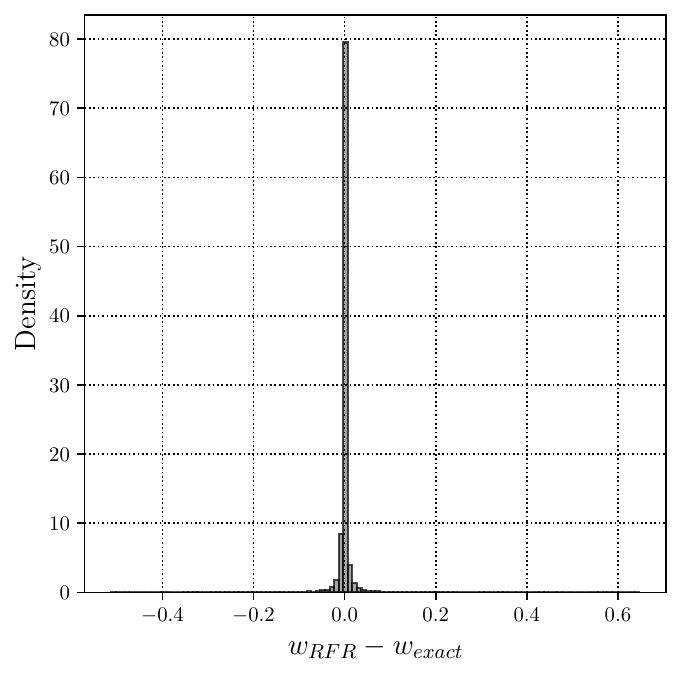}
        \caption{Error histogram for the discrepancies between predicted and exact weights for the test set.}
        \label{fig:hist_test}
    \end{subfigure}
    
    \caption{Comparison of predicted weights versus exact weights using the RFR, and the distribution of prediction errors. The scatter plots highlight the accuracy of the predictions, and the histograms show the density of discrepancies between exact and predicted weights.}
    \label{RFR_weights_and_hist}
\end{figure}

}

\section{Numerical Results}
\ReviewerOne{In this section, we evaluate the performance of the proposed blending model across a series of 2D turbulent flow cases, including both training and test scenarios.  
We begin by applying the blended model to the flow cases used for training both the weighting functions and the expert models. We verify that the blended model consistently assigns higher weights to the expert model specifically trained for the given flow configuration, demonstrating its ability to adapt to known cases.  
Then, we assess the generalization capability of the blending model by testing it on unseen flow cases, highlighting its robustness and predictive accuracy beyond the training dataset.
}

\subsection{Flow solver}
The data-driven models resulting from the SBL-SpaRTA algorithm are implemented within a modified version of the open-source finite-volume solver OpenFoam v23-06 \cite{OpenFOAM}. 
In the numerical tests presented below, the governing equations are solved using SIMPLE algorithm. The linear upwinding is applied to the convective terms, while viscous terms are approximated with a second-order central difference scheme. The solution is advanced to the steady state by using a Gauss-Seidel smoother. For each case, we used the same computational settings and the same grids as in \cite{schmelzer2020discovery}, to which we refer for more details. The grids are fine enough to ensure that the discretization error is negligible compared to the effect of the turbulence model. 

\subsection{Validation of the $\Mblend$ on the training flows}

\subsubsection{Turbulent Channel Flow \( Re_{\tau} = 1000 \)}
The first case considered in the training of the blending model is the turbulent channel flow at \( Re_{\tau} = 1000 \), where the baseline $\MSST$ model is expected to perform optimally. \ReviewerOne{The grid used in RANS simulations is  composed of $40\times 60$ cells and it is refined near the walls to resolve the viscous sublayer.}

\begin{figure}[hbtp!]
 \centering
 \begin{subfigure}[t]{0.48\textwidth}
  \centering
  \includegraphics[scale=0.35]{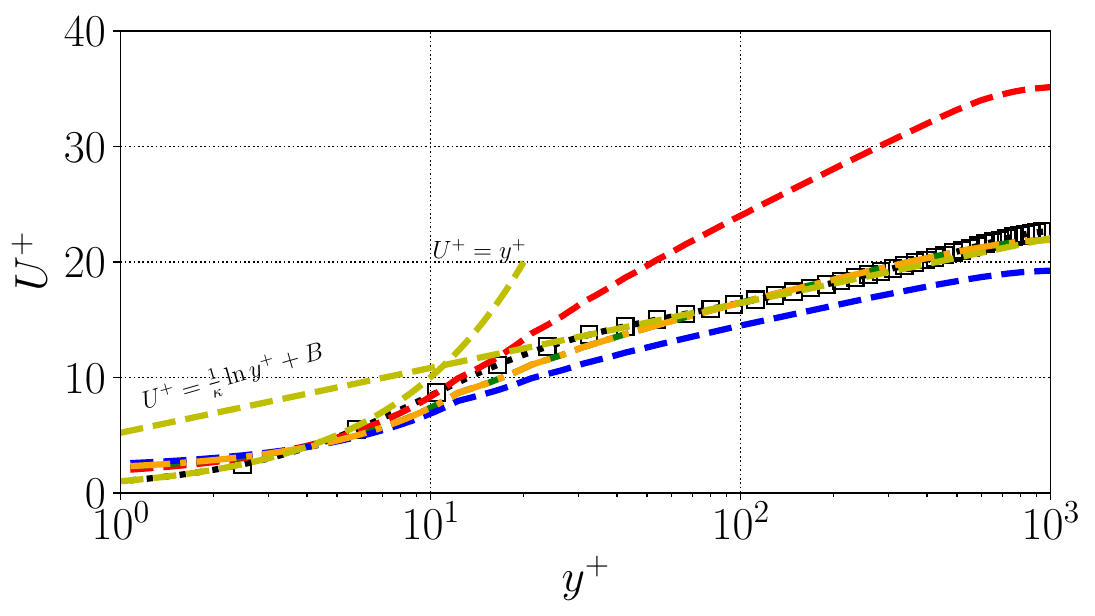}
  \caption{Wall-normal profiles of the velocity}
  \label{Fig.UPlus_vs_yPlus_Channel}
 \end{subfigure}
 \hfill
 \begin{subfigure}[t]{0.48\textwidth}
  \includegraphics[scale=0.35]{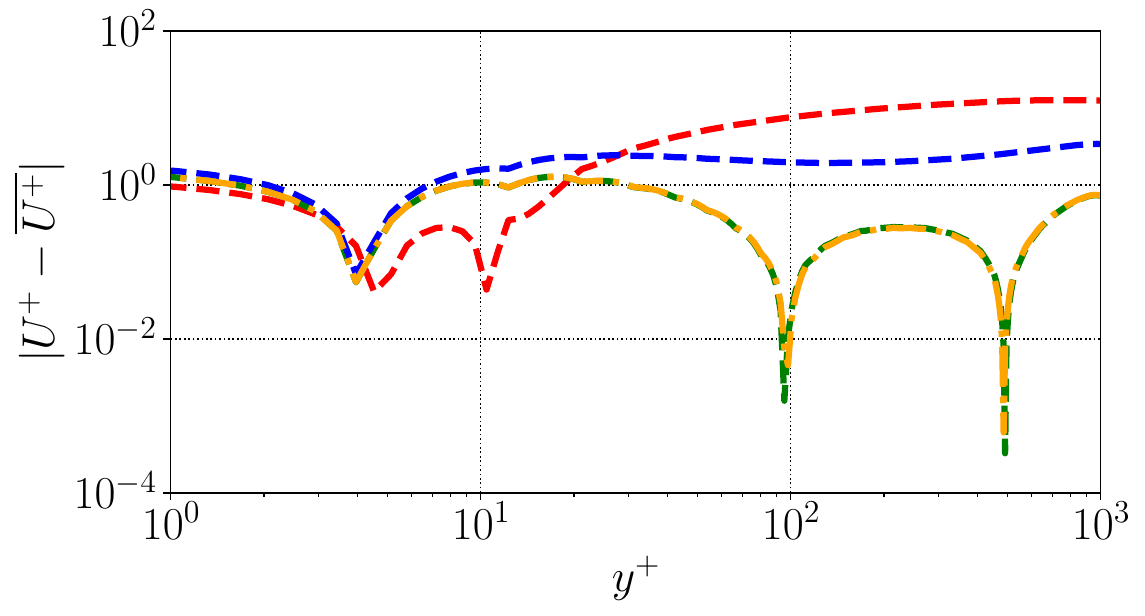}
  \caption{Absolute errors}
  \label{Fig.AbsErrors_UPlus_vs_yPlus_Channel}
 \end{subfigure}
 \caption{Turbulent channel flow at \( Re_{\tau} = 1000 \). The legend denotes the following: (\legendHF) DNS \cite{Lee_Moser_2015}, (\legendANSJ) \(\MANSJ\), (\legendCHAN) \(\MSST\), (\legendSEP) \(\MSEP\), (\legendintXMA) \(\Mblend\), and (\legendtheory) theoretical \NewModif{law of the wall} with the Kármán constants \(\kappa = 0.41\) and \(B = 5.2\).}

 \label{Fig.U_vs_yPlus_Channel}
\end{figure}

The non-dimensionalized velocity profiles at $x=0.97$ in Figure \ref{Fig.UPlus_vs_yPlus_Channel} show that, as expected, the $\MSST$ model aligns almost perfectly with the \NewModif{ theoretical profiles and reference DNS data}, requiring no significant adjustments. The accuracy of this model is further illustrated by its low mean absolute error (\textit{mae}) value, as shown in Table \ref{tab:MAE_Channel}, and the absolute errors of the velocity profiles in Figure \ref{Fig.AbsErrors_UPlus_vs_yPlus_Channel}, which reveal consistently small deviations across \(y^+\), especially in the log-law region. 
In contrast, while the $\MSEP$ model provides a reasonable prediction, it exhibits a deviation from ground truth results, particularly noticeable in log-law region. The deviation is \NewModif{due to the increased} amount of eddy viscosity generated by the corrected model, which has been trained to compensate the under-dissipative nature of the baseline model for strongly separated flows (see \cite{cherroud2022sparse,cherroud2025space}).
The discrepancy is mirrored in the \textit{mae} values and error profiles, which, though higher than those of the $\MSST$ model, still indicate acceptable performance. The $\MANSJ$ model, on the other hand, significantly overpredicts the flow velocity, and its error profiles reflect large discrepancies, particularly in the log-law region, showcasing its inadequacy for this flow scenario. In fact, the $\MANSJ$ correction tends to strongly reduce the eddy viscosity of the baseline model to compensate for the so-called jet anomaly \cite{wilcox2006turbulence}. However, the correction destroys the accuracy for attached wall-bounded flows.

Of note, for this training flow case the blended model predominantly selects the baseline model, which is assigned a weight very close to one throughout the channel height. The contributions from the inaccurate $\MSEP$ and $\MANSJ$  are effectively ruled out, according to their low performance scores.

\begin{table}[h!]
\centering \renewcommand{\arraystretch}{1.2}
\begin{tabular}{lcccc}
\toprule
QoI & $\MANSJ$ & $\MSST$ & $\MSEP$ & $\Mblend$ \\
\midrule
$U^+$ & \cellcolor[gray]{0.55} $8.97$ & \cellcolor[gray]{0.95} $0.44$ & \cellcolor[gray]{0.86} $2.41$ & \cellcolor[gray]{0.95} $0.44$ \\
\bottomrule
\end{tabular}
\caption{\textit{mae} for $U^+$ for the turbulent channel flow case \( Re_{\tau} = 1000 \). Shading represents error magnitude, with darker shades indicating higher error.}
\label{tab:MAE_Channel}
\end{table}

\subsubsection{Axis-symmetric near sonic jet flow}
\ReviewerOne{  The axisymmetric near-sonic jet flow case from \cite{bridges2010establishing}, also used to train the expert model correction $\MANSJ$, features a jet from the ARN2 nozzle with a radius of 1 inch and a jet exit Mach number of $M_{jet} = 0.985$. 
%This case involves complex flow phenomena near the jet exit, including turbulence interactions and potential shock-related effects, which are challenging to capture accurately with the baseline model. DES SHOCKS AVEC SIMPLE??? MIEUX VAUT NE RIEN DIRE!!!
The numerical setup follows the description given in the NASA turbulence modeling resource \url{https://turbmodels.larc.nasa.gov/jetnearsonic_val.html}, and we use a structured grid constituted of three blocks with $25 \times 25$, $16 \times 25$, and $65 \times 57$ points, which was found sufficient for the present purpose.
}

\begin{figure}[hbtp!]
  \centering
  \begin{subfigure}[T]{0.48\textwidth}
    \includegraphics[width=\textwidth]{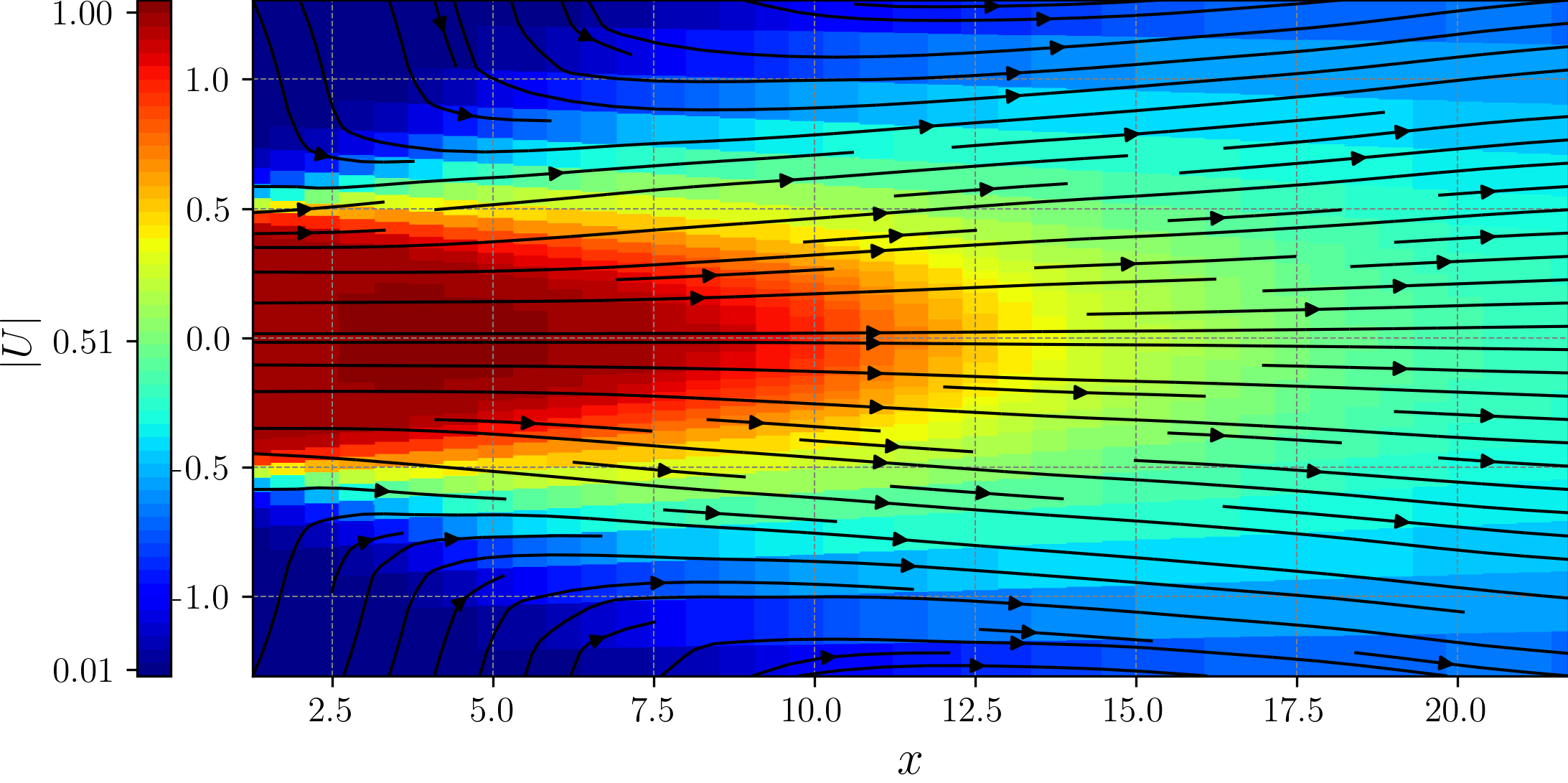}
    \caption{PIV time-average velocity \cite{bridges2010establishing}.}
    \label{Fig. ANSJ_VelocityField}
  \end{subfigure}\hfill
  \begin{subfigure}[T]{0.48\textwidth}
    \includegraphics[scale=0.35]{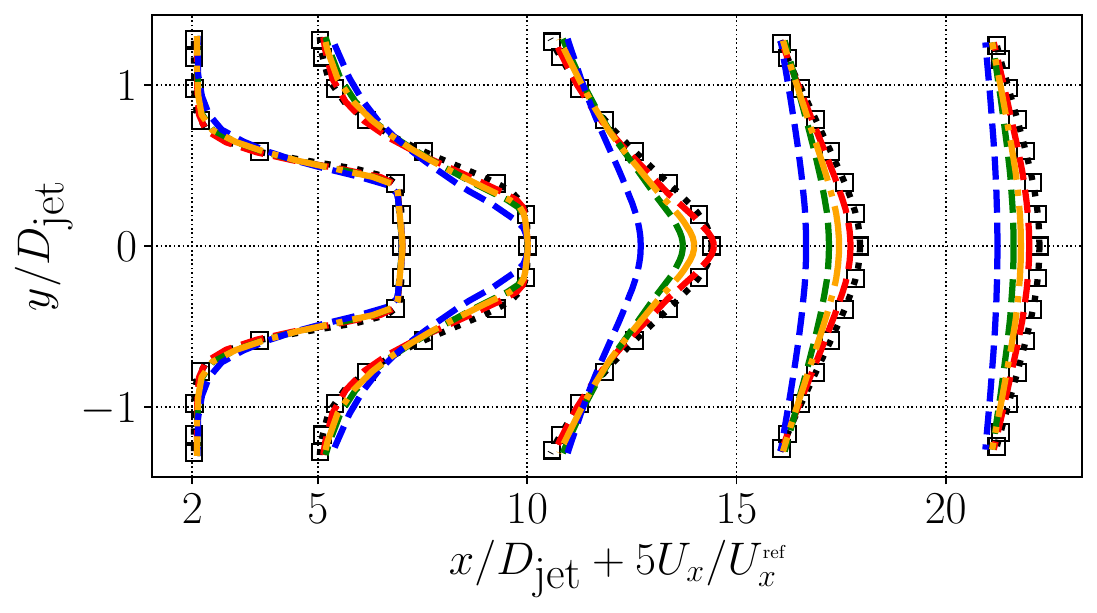}
    \caption{\ReviewerOne{Streamwise} velocity profiles across $x/D_{jet}$}
    \label{Fig.ANSJ_VelocityProfiles}
  \end{subfigure} \\
  \begin{subfigure}[T]{0.48\textwidth}
    \includegraphics[scale=0.35]{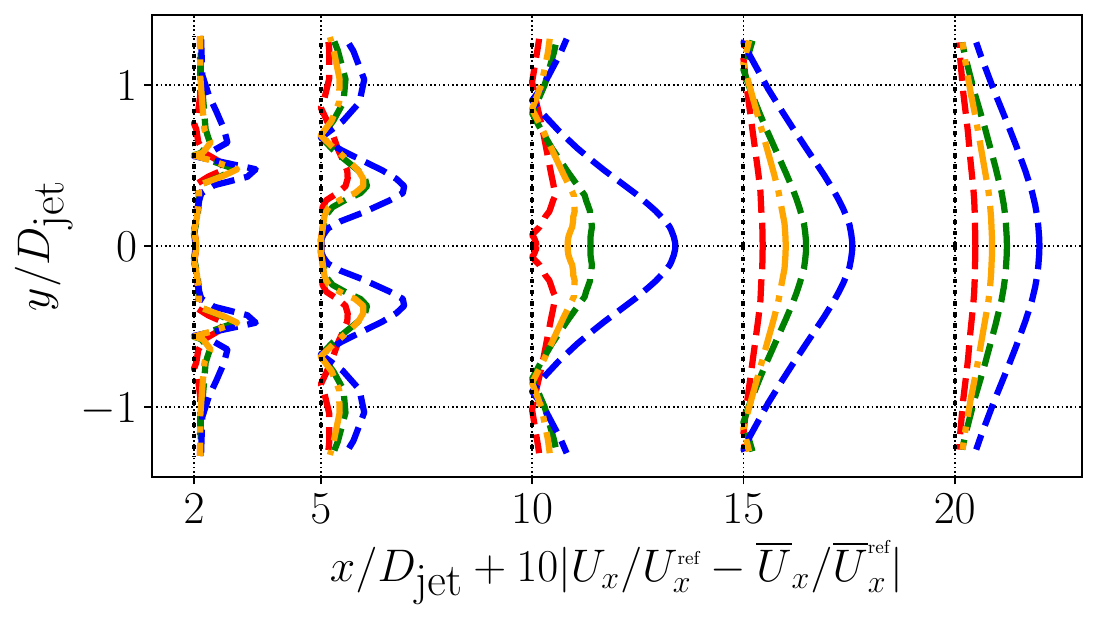}
    \caption{Velocity absolute errors across $x/D_{jet}$}
    \label{Fig.ANSJ_VelocityErrors}
  \end{subfigure}\hfill
  \begin{subfigure}[T]{0.48\textwidth}
    \includegraphics[scale=0.35]{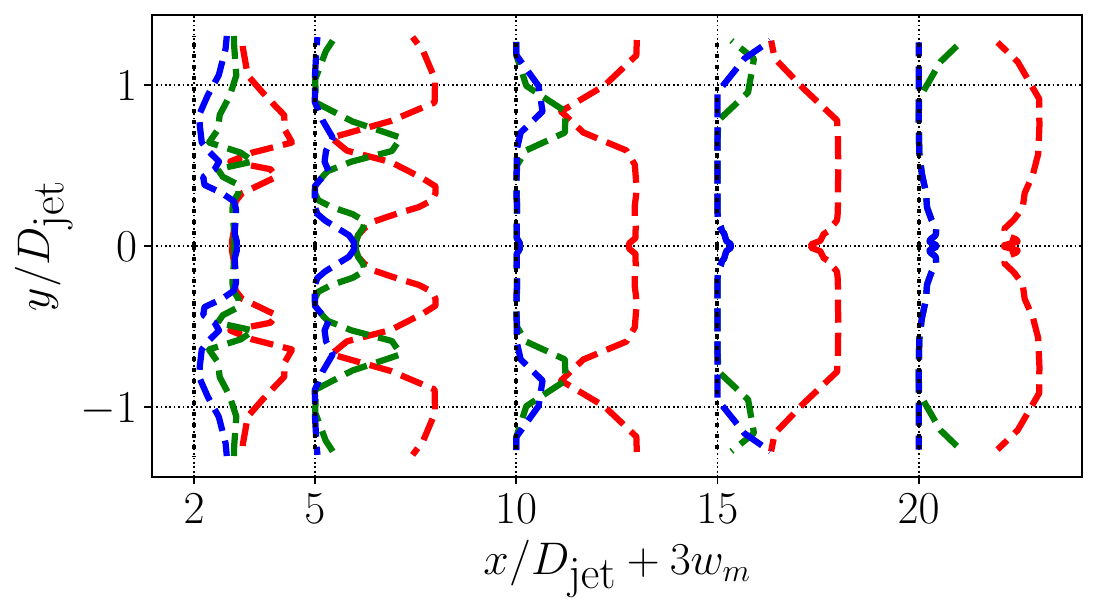}
    \caption{Weight distributions across $x/D_{jet}$}
    \label{Fig.ANSJ_WeightDistributions}
  \end{subfigure}	
  \caption{\ReviewerOne{ Streamwise} velocity, absolute errors and blending weights distributions for the axisymmetric near-sonic jet flow at stations $x/D_{jet} \in \{ 2, 5, 10, 15, 20\}$ with $D_{jet}$ is the nozzle radius. The legend denotes the following: (\legendHF) PIV \cite{bridges2010establishing}, (\legendANSJ) \(\MANSJ\), (\legendCHAN) \(\MSST\), (\legendSEP) \(\MSEP\), (\legendintXMA) \(\Mblend\).}
  \label{Fig.Velocity_and_weights_profiles}
\end{figure}

As shown in Figure \ref{Fig.ANSJ_VelocityProfiles}, the expert $\MANSJ$ model provides as expected the most accurate velocity profiles at least in the far jet region, with a low \textit{mae} (Table \ref{tab:mae_jet}). The $\MSST$ model, underpredicts the velocity, resulting in a higher \textit{mae} (a known anomaly of most two-equation models), while the $\MSEP$ model significantly underpredicts the velocity leading to the largest \textit{mae}. These trends are confirmed by the absolute error profiles (Figure \ref{Fig.ANSJ_VelocityErrors}) where  $\MANSJ$  maintains the smallest errors, while  $\MSEP$ model exhibits the largest deviations throughout.

The blended model achieves an intermediate \textit{mae} which is close to that of the best performing model. As depicted in Figure \ref{Fig.ANSJ_WeightDistributions}, the blending weights are initially distributed fairly equally among the models at the jet entrance, reflecting an effort to combine their contributions. As the flow develops through the shear layers, the relevance of  $\MSEP$ and $\MSST$  decreases, leading to a reduction in their respective weights. Downstream of the jet, the $\MANSJ$ model becomes the dominant. 

\begin{figure}[hbtp!]
\centering
\begin{subfigure}{0.5\textwidth}
\includegraphics[scale=0.35]{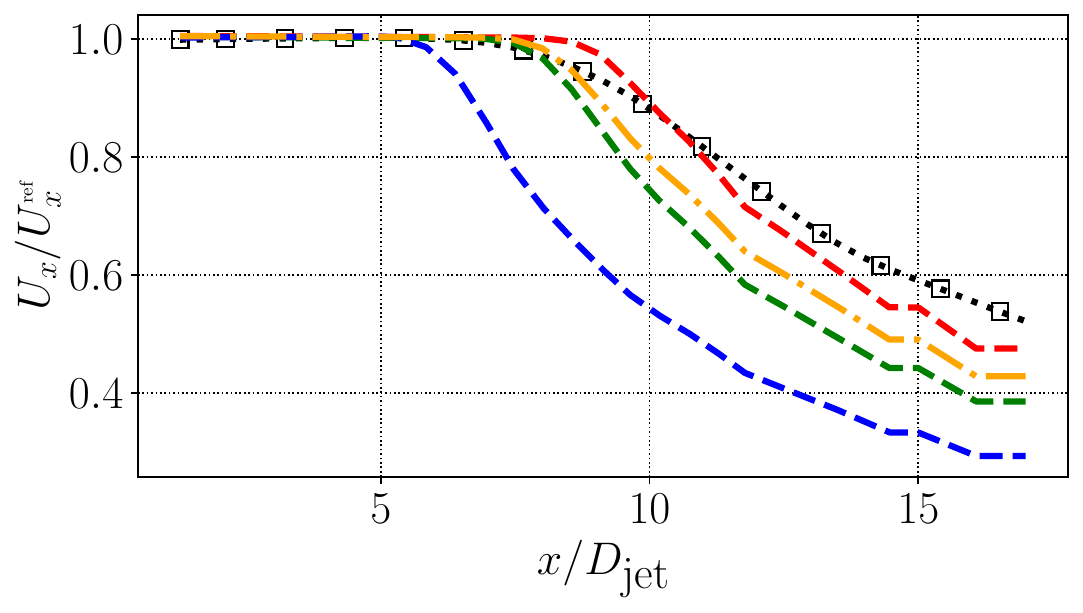}
\caption{\ReviewerOne{ Streamwise} velocity along jet-axis}
\label{Fig.ANSJ_VelocityAxis}
\end{subfigure}\hfill
\begin{subfigure}{0.5\textwidth}
\includegraphics[scale=0.35]{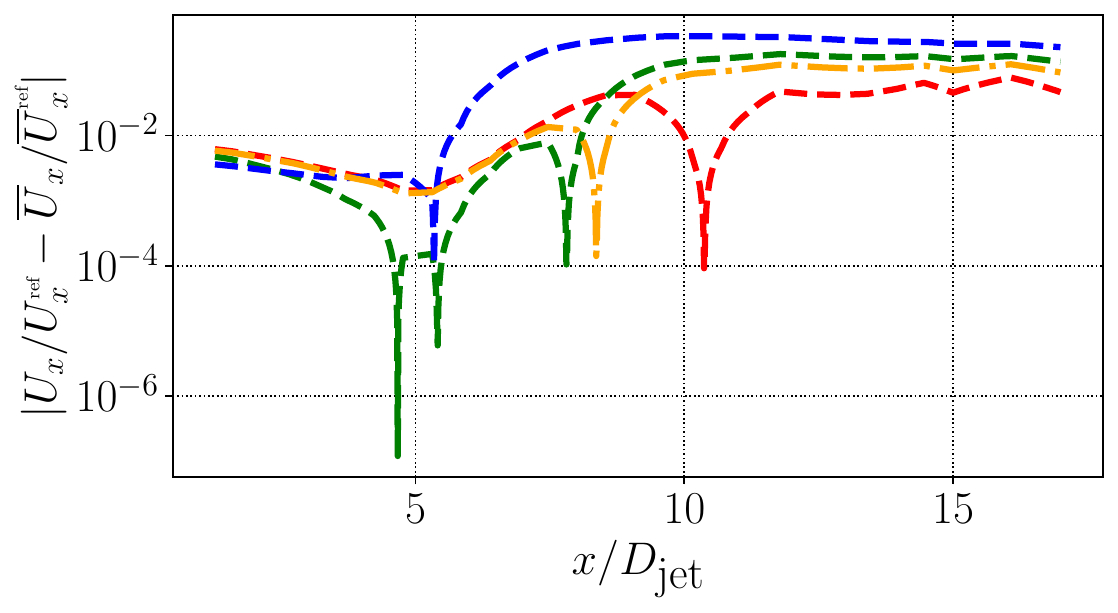}
\caption{Absolute errors along jet-axis}
\label{Fig.ANSJ_VelocityAxisErrors}
\end{subfigure}
\begin{subfigure}{0.5\textwidth}
\includegraphics[scale=0.35]{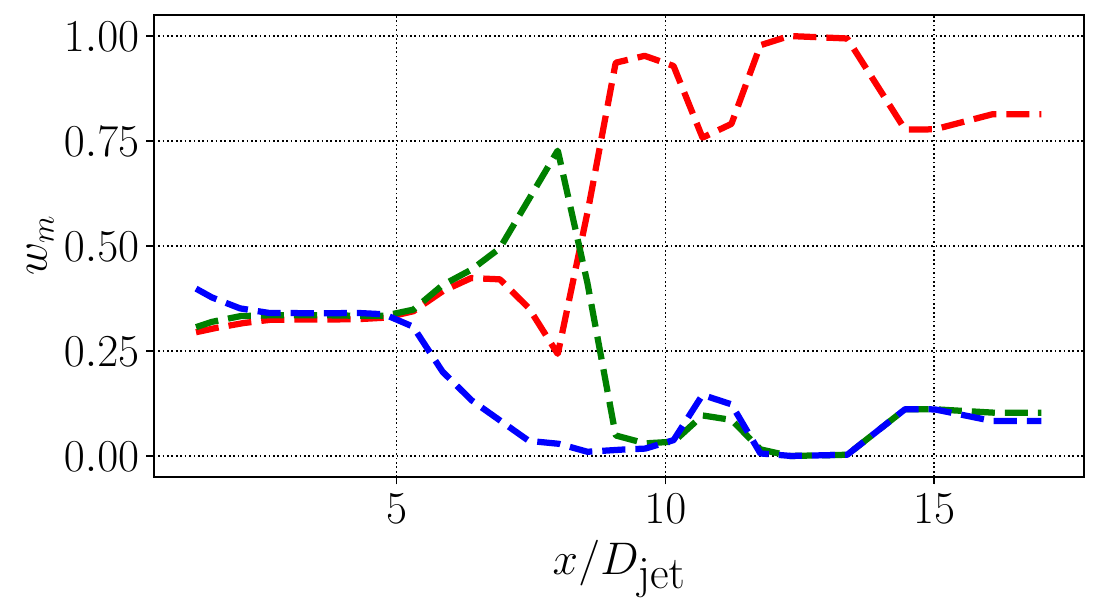}
\caption{Weight distributions along jet axis}
\label{Fig.ANSJ_WeightsAxis}
\end{subfigure}
\caption{\ReviewerOne{Streamwise} velocity, absolute errors and blending weights distribution along the jet axis. $D_{jet}$ is the nozzle radius and the legend denotes the following: (\legendHF) PIV \cite{bridges2010establishing}, (\legendANSJ) \(\MANSJ\), (\legendCHAN) \(\MSST\), (\legendSEP) \(\MSEP\), (\legendintXMA) \(\Mblend\).}
\label{Fig.Velocity_and_weights_along_axis_ANSJ}
\end{figure}

Further analysis of the velocity distribution along the jet axis (Figure  \ref{Fig.ANSJ_VelocityAxis}) shows that the $\MANSJ$ model provides the closest match to the ground truth data, with  $\MSST$  underpredicting the velocity and  $\MSEP$  significantly underperforming. The blended model offers a reasonable compromise between the $\MANSJ$ and $\MSST$ models, capturing the essential flow characteristics with relatively low errors (Figure \ref{Fig.ANSJ_VelocityAxisErrors}). The weight distribution along the axis (Figure \ref{Fig.ANSJ_WeightsAxis}) further confirms the dynamic adjustment, with the $\MANSJ$ model taking over in regions where its performance is critical, while the other models contribute less further downstream.

\begin{table}[h!]
\centering \renewcommand{\arraystretch}{1.2}
\begin{tabular}{lcccc}
\toprule
QoI & $\MANSJ$ & $\MSST$ & $\MSEP$ & $\Mblend$ \\
\midrule
$U/U_{\textrm{ref}}$ & \cellcolor[gray]{0.95} $0.04$ & \cellcolor[gray]{0.78} $0.09$ & \cellcolor[gray]{0.55} $0.16$ & \cellcolor[gray]{0.85} $0.07$ \\
\bottomrule
\end{tabular}
\caption{\textit{mae} on the normalized streamwise velocity in the near-sonic axisymmetric jet flow. Shading represents error magnitude, with darker shades indicating higher error.}
\label{tab:mae_jet}
\end{table}

\subsubsection{Separated Flow Cases}
\ReviewerOne{
Hereafter, we present selected results from three separated-flow training cases, whose key parameters are summarized in Table~\ref{tab:sepFlowCases}, along with the original test case references. The converging-diverging channel case \cite{laval2011direct} is characterized by $Re=12600$, based on inlet conditions, the channel half-height $L=h/2$, the maximum inlet velocity $U_{\max}$, and the inlet kinematic viscosity. The flow is fully developed at the inlet, with a friction Reynolds number $\mathrm{Re}_\tau \approx 600$. The channel geometry contracts and then diverges smoothly, leading to a strong adverse-pressure gradient and resulting separation. The periodic hills case \cite{breuer2009flow} consists in the flow through a channel with periodic restrictions (hills), and exhibits $Re=10595$ (based on the hill height and the bulk velocity at the crest of the first hill) ; here, the flow separates behind each curved hill and reattaches before the next crest. Finally, the curved backward facing step case \cite{bentaleb2012large}, with $Re=13700$ based on the step height $L$ and center-channel inlet velocity $U_{\mathrm{in}}$, features boundary-layer separation over the curved step. In this configuration, the upstream duct is $8.52\,L$ high, and measurements at $x/L=-7.34$ show a boundary layer with $\mathrm{Re}_\theta \approx 1190$ and a thickness of about $0.8\,L$. As shown in Table~\ref{tab:sepFlowCases}, each case uses a characteristic length $L$ and a specific velocity scale ($U_{\max}$, $U$, or $U_{\mathrm{in}}$) for the Reynolds number definition. Grid sizes used in the 2D RANS simulations are also reported.}

\begin{table}[h!]
    \centering
     \renewcommand{\arraystretch}{1.2}
    \begin{tabular}{lccccc}
    \toprule
    \textbf{Case} & \textbf{Re} & \textbf{Re definition} 
                  & \textbf{$L$} & \textbf{Grid Size (2D)} & \textbf{Reference} \\
    \midrule
    \textbf{CD}   
      & $12600$  
      & $Re = {U_{\max}\,L}/{\nu}$ 
      & Channel half-height
      & $140\times 100$  
      & \cite{laval2011direct}\\
    \textbf{PH}   
      & $10595$  
      & $Re = {U_b\,L}/{\nu}$ 
      & Hill height
      & $120\times 130$ 
      & \cite{breuer2009flow}\\
    \textbf{CBFS} 
      & $13700$  
      & $Re = {U_{\text{in}}\,L}/{\nu}$ 
      & Step height
      & $140\times 150$
      &  \cite{bentaleb2012large}\\
    \bottomrule
    \end{tabular}
    \caption{Overview of the separated-flow cases and their Reynolds number definitions based on the  
             characteristic lengths, and the 2D grid resolutions used in RANS simulations.}
    \label{tab:sepFlowCases}
\end{table}

\begin{figure}[hbtp!]
  \centering
  \begin{subfigure}[T]{0.48\textwidth}
    \includegraphics[width=\textwidth]{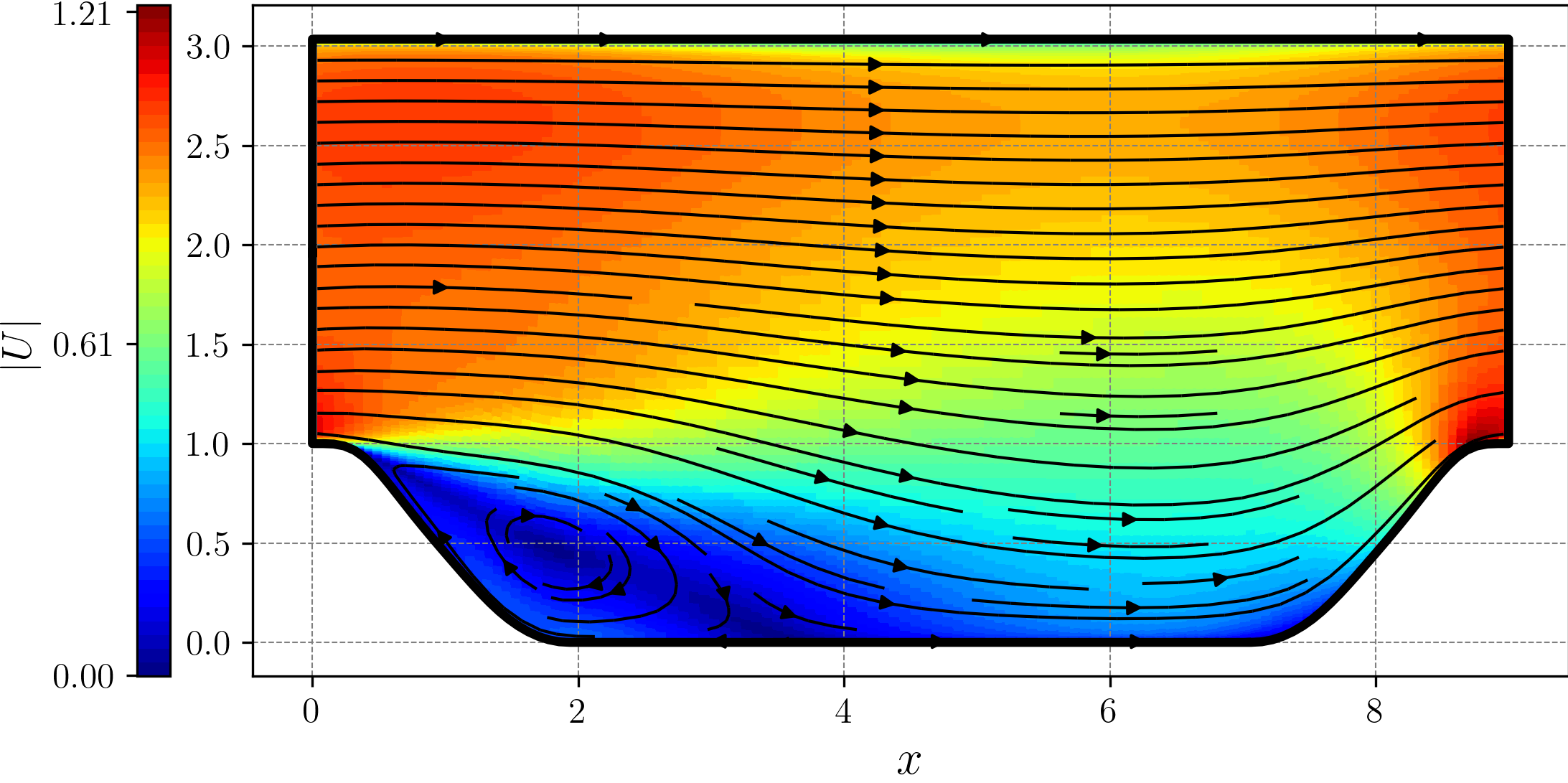}
    \caption{LES long-time-average velocity \cite{temmerman_sgs_models_2001}}
    \label{Fig. PH_VelocityField}
  \end{subfigure}\hfill
  \begin{subfigure}[T]{0.48\textwidth}
    \includegraphics[width=\linewidth, trim=0 0 0 8, clip]{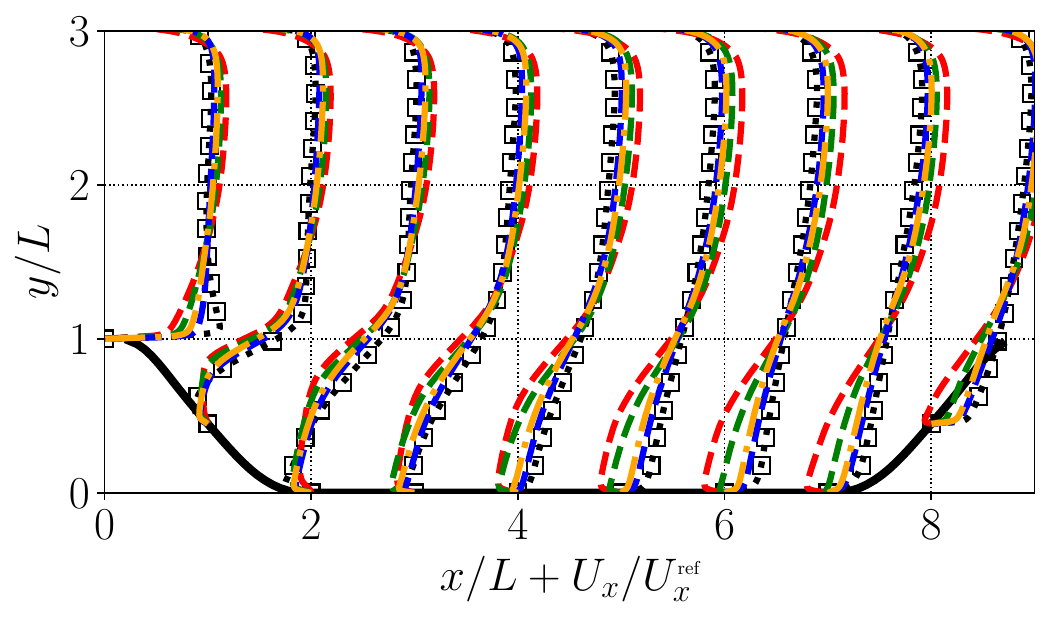}
    \caption{\ReviewerOne{Streamwise} velocity.}
    \label{Fig. PH_VelocityProfiles}
  \end{subfigure} \\
  \begin{subfigure}[T]{0.48\textwidth}
    \includegraphics[width=\linewidth]{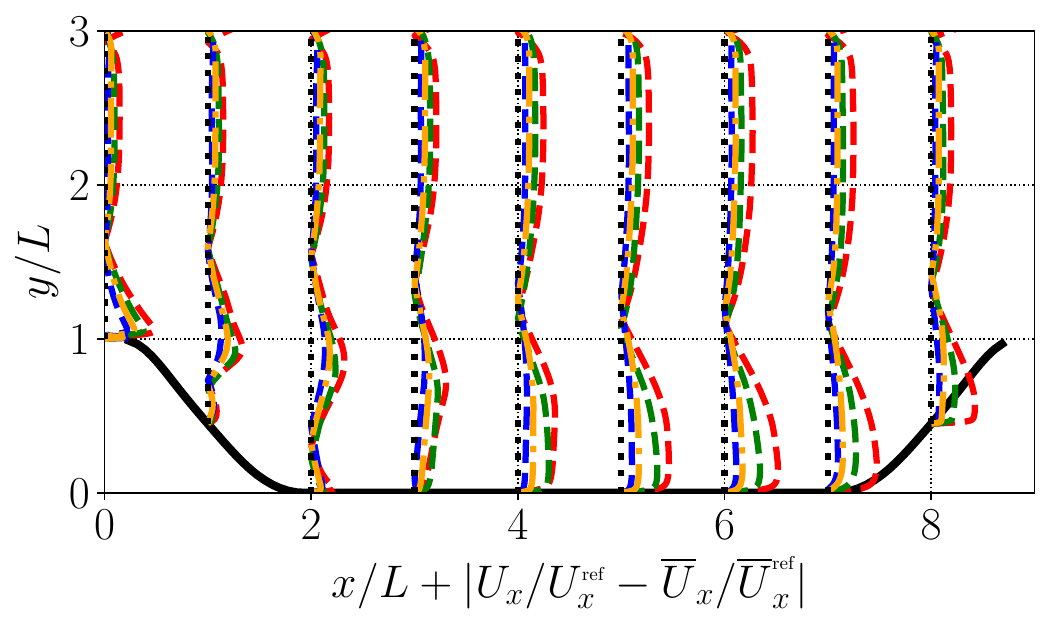}
    \caption{Absolute errors}
    \label{Fig. PH_VelocityErrors}
  \end{subfigure}\hfill
  \begin{subfigure}[T]{0.48\textwidth}
    \includegraphics[width=\linewidth]{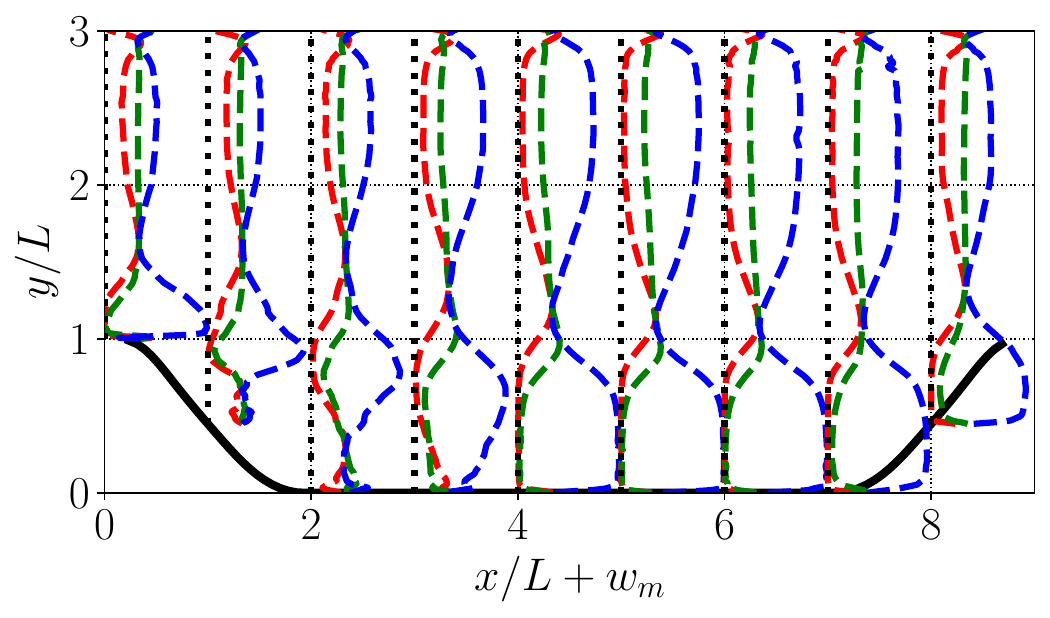}
    \caption{Weight distributions.}
    \label{Fig. PH_WeightDistributions}
  \end{subfigure}
  \caption{\ReviewerOne{Streamwise} velocity, absolute errors and blending weights for the PH case, at various stations $x/L\in\{0, 1, 2, 3 ,4 ,5 ,6 ,7 ,8 ,9\}$ with $L$ the channel half-height (full channel height = $2L$). The legend denotes the following: (\legendHF) LES \cite{temmerman_sgs_models_2001}, (\legendANSJ) \(\MANSJ\), (\legendCHAN) \(\MSST\), (\legendSEP) \(\MSEP\), (\legendintXMA) \(\Mblend\).}
  \label{Fig. Velocity profile and weights PH}
\end{figure}

\begin{figure}[hbtp!]
  \centering
  \begin{subfigure}[T]{0.48\textwidth}
    \includegraphics[width=\textwidth]{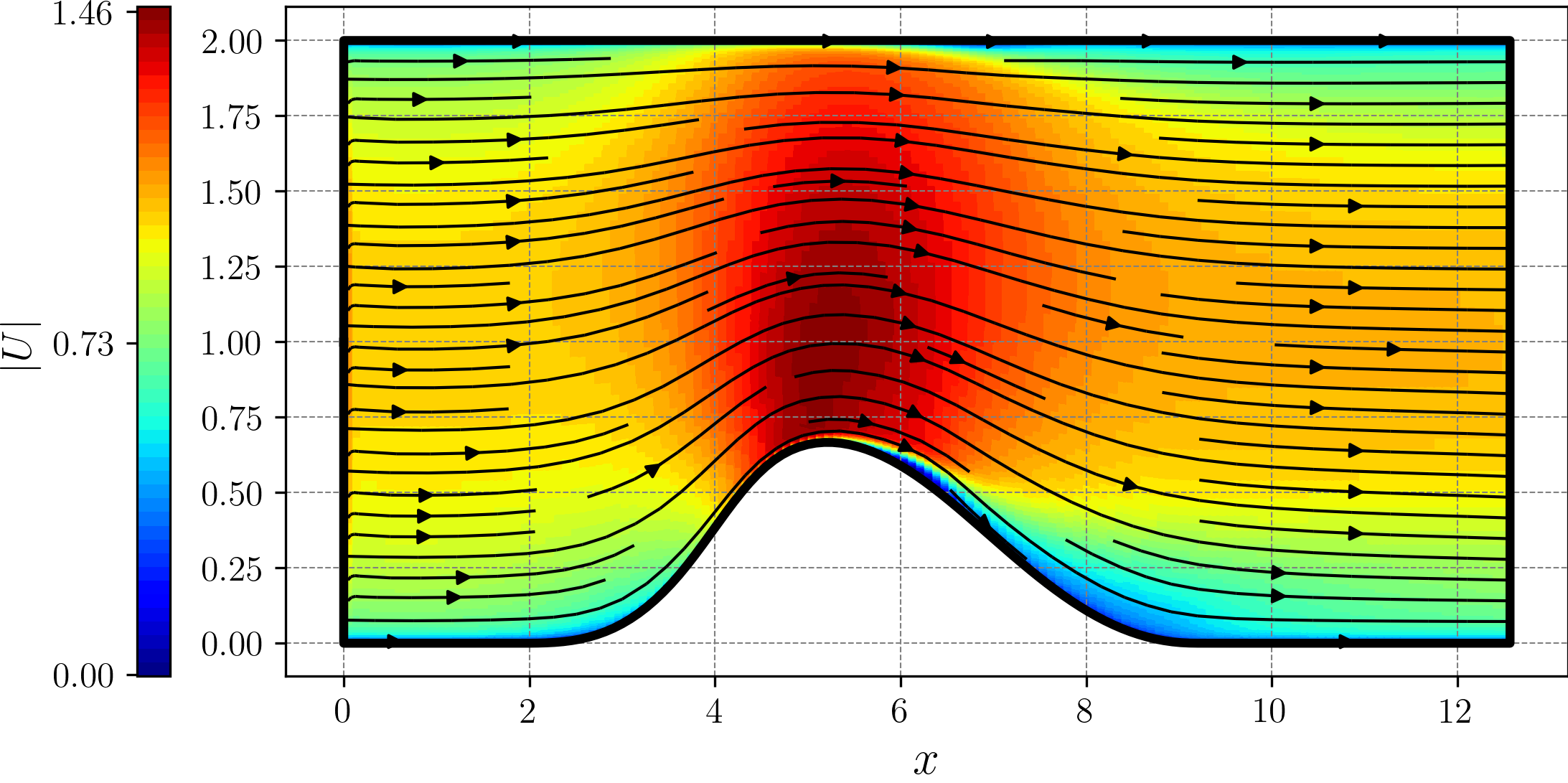}
    \caption{DNS long-time average velocity \cite{laval2011direct}}
    \label{Fig. CD_VelocityField}
  \end{subfigure}\hfill
  \begin{subfigure}[T]{0.48\textwidth}
    \includegraphics[width=\linewidth, trim=0 0 0 8, clip]{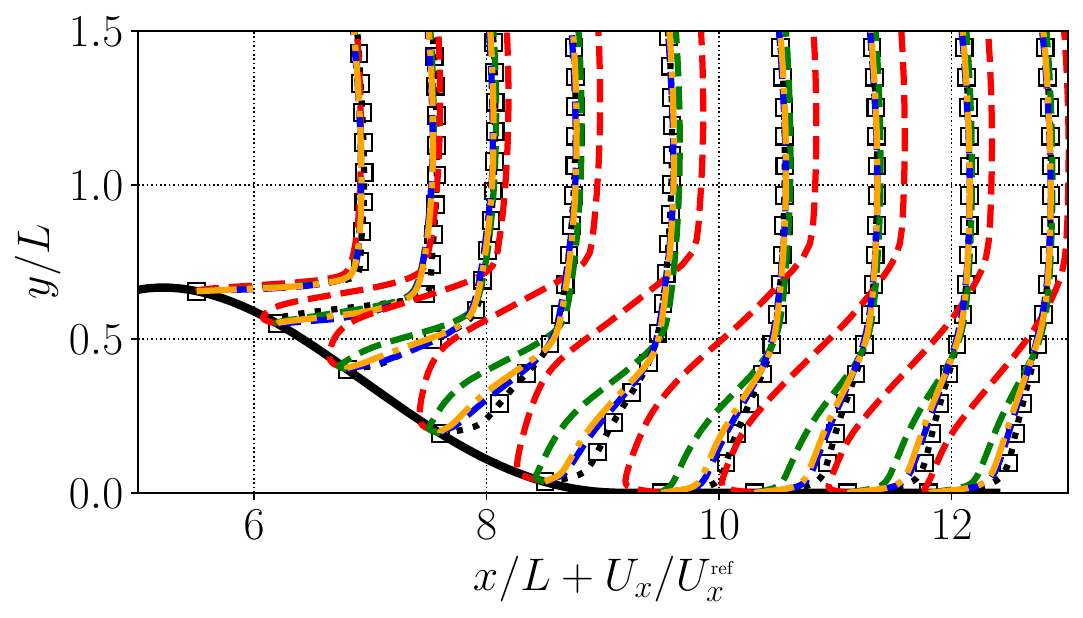}
    \caption{\ReviewerOne{Streamwise} velocity}
    \label{Fig. CD_VelocityProfiles}
  \end{subfigure} \\
  \begin{subfigure}[T]{0.48\textwidth}
    \includegraphics[width=\linewidth]{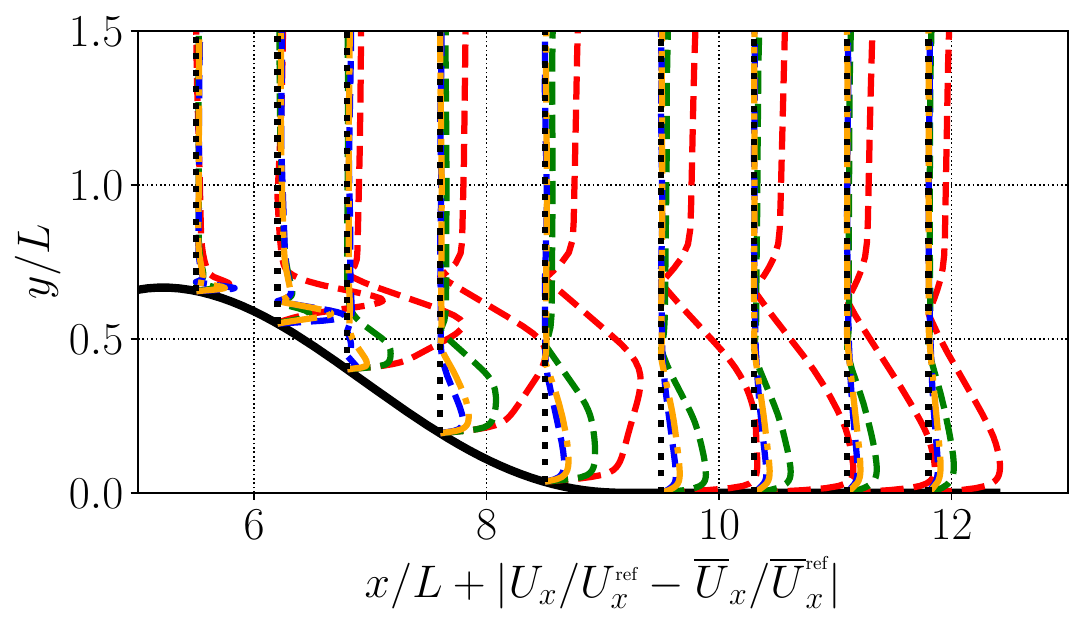}
    \caption{Absolute error}
    \label{Fig. CD_VelocityErrors}
  \end{subfigure}\hfill
  \begin{subfigure}[T]{0.48\textwidth}
    \includegraphics[width=\linewidth]{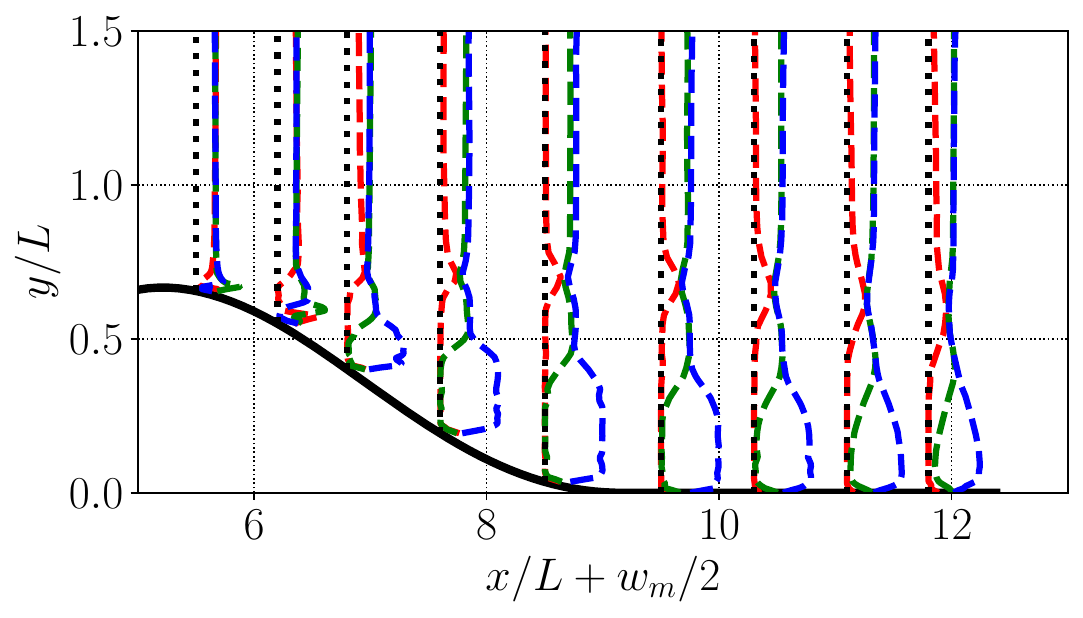}
    \caption{Weight distributions}
    \label{Fig. CD_WeightDistributions}
  \end{subfigure}
  \caption{\ReviewerOne{Streamwise} velocity, absolute errors and blending weights for the CD case, at various stations $x/L\in\{5.5, 6.2, 6.8, 7.6, 8.5, 9.5, 10.3, 11.1, 11.8\}$ with $L$ the channel half-height. The legend denotes the following: (\legendHF) DNS \cite{laval2011direct}, (\legendANSJ) \(\MANSJ\), (\legendCHAN) \(\MSST\), (\legendSEP) \(\MSEP\), (\legendintXMA) \(\Mblend\).}
  \label{Fig. Velocity profile and weights CD}
\end{figure}

As expected, the $\MSEP$ model consistently performs well across all the considered separated flow training cases, particularly in regions of separation and reattachment. The velocity profiles are in good agreement with the high fidelity data for the PH case (Fig. \ref{Fig. PH_VelocityProfiles}) and the CBFS (not shown).  However, the CD case is more challenging, because it is characterized by a small separation bubble in the divergent that cannot be captured by RANS model, even after data-driven correction. Nonetheless, the $\MSEP$ improves the baseline model (see Fig. \ref{Fig. CD_VelocityProfiles}), even if the solution exhibits discrepancies with respect to the reference. 
On the contrary, the $\MANSJ$ model exhibits very poor performance for all the separated cases, for which it incorrectly predicts massive separation.
The relative performance of the three expert models are better reflected in the error profiles, displayed in Figures   \ref{Fig. PH_VelocityErrors} and \ref{Fig. CD_VelocityErrors}.\\

The blended model offers a fair performance across all separated flow cases. It effectively captures the flow dynamics in both attached and separated flow regions. This results in improved accuracy for the predicted velocity profiles with respect to the baseline model, especially in the separated regions. Inspection of the weight profiles (Figures \ref{Fig. PH_WeightDistributions} and \ref{Fig. CD_WeightDistributions}) reveals that the $\MSEP$ model is consistently assigned the highest weight, particularly in regions of strong separation, while the $\MSST$ model is promoted in attached flow regions. The $\MANSJ$ model is always assigned low weights, reflecting its limited applicability in wall-bounded flow scenarios.

%The blended model also improves significantly the predicted skin friction along the bottom wall, reported in Figure \ref{Fig. Cf all} for the three separated flow cases. The predominant contribution of  $\MSEP$  in separated regions, helps the blended model predicting strong separation and reattachment regions more effectively. On the other hand,  $\MSST$ is the main contributor in attached flow regions, thus ensuring better overall accuracy. 

\begin{figure}[hbtp!]
\centering
\begin{subfigure}{0.5\textwidth}
\includegraphics[scale=0.35]{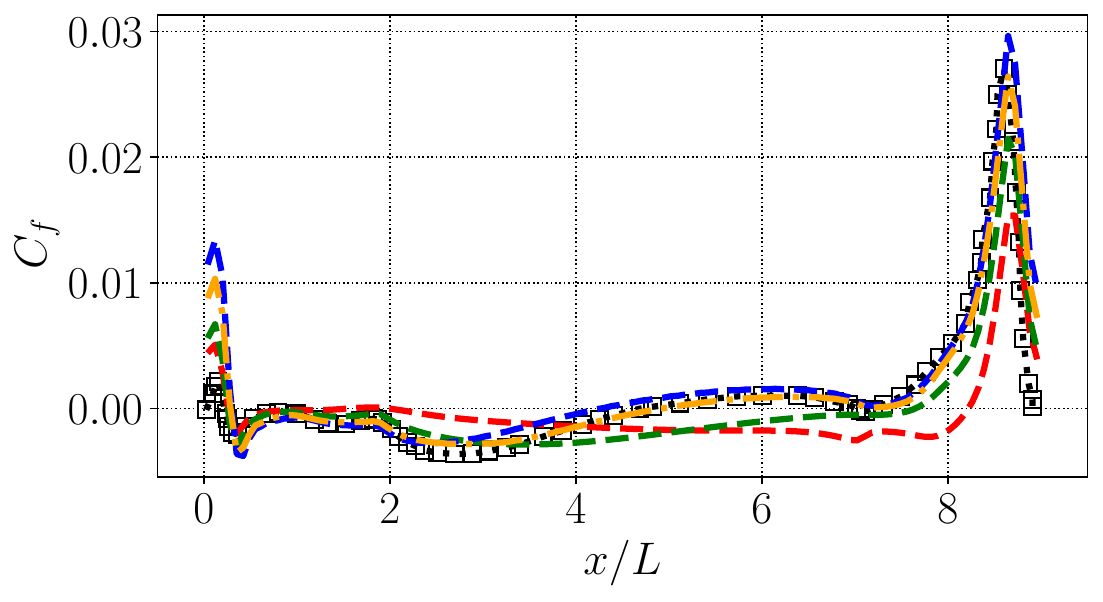}
\caption{PH}
\label{Fig. PH_Cf}
\end{subfigure}\hfill
\begin{subfigure}{0.5\textwidth}
\includegraphics[scale=0.35]{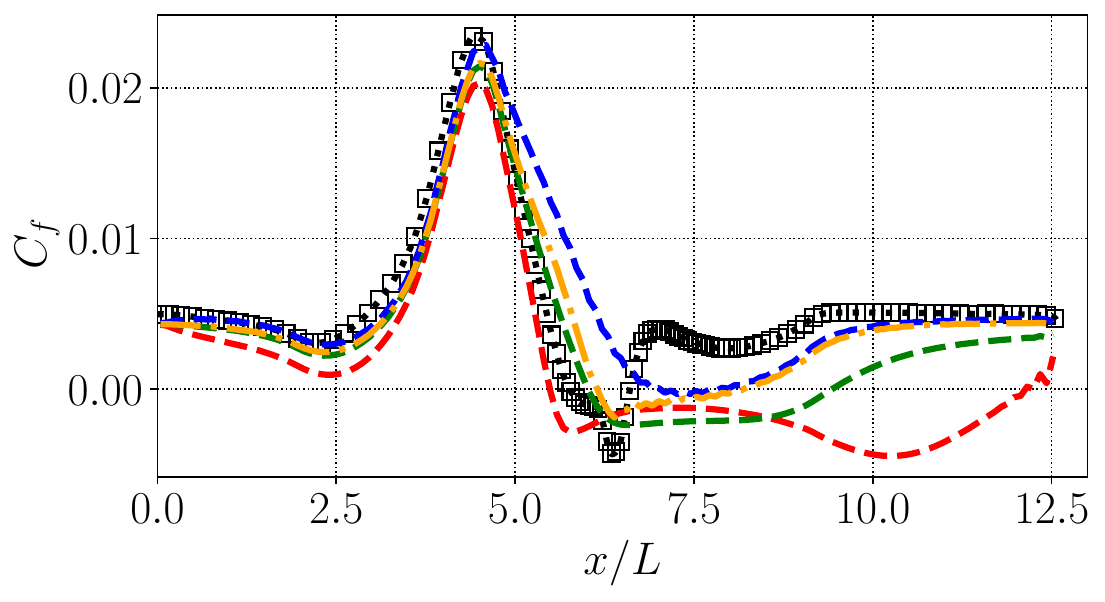}
\caption{CD}
\label{Fig. CD_Cf}
\end{subfigure}

\begin{subfigure}{0.5\textwidth}
\includegraphics[scale=0.35]{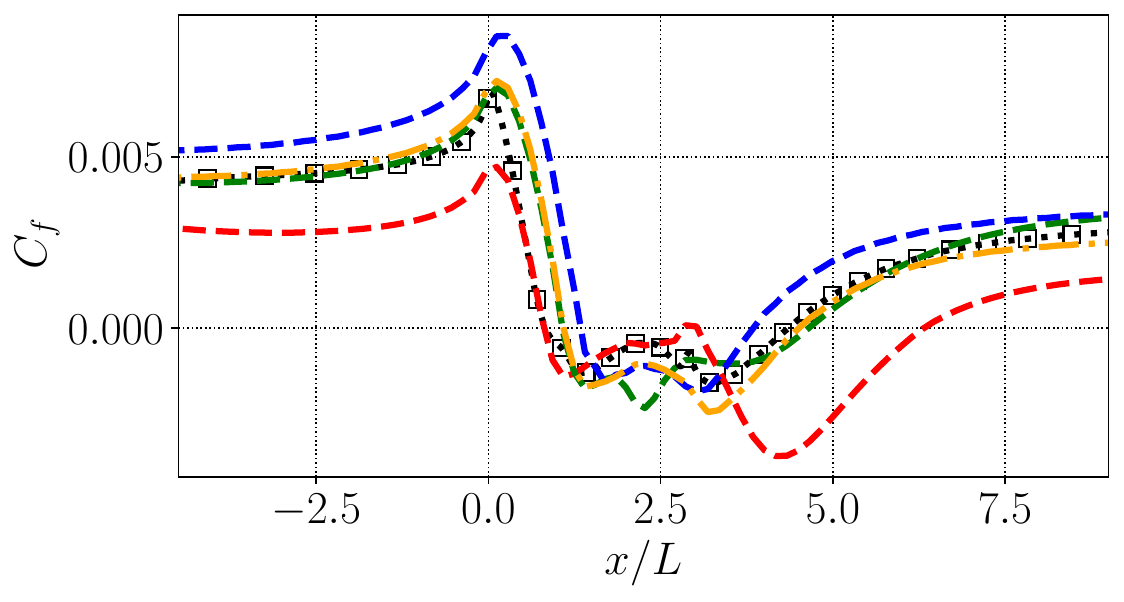}
\caption{CBFS}
\label{Fig. CBFS_Cf}
\end{subfigure}
\caption{Skin friction distributions along the bottom wall \NewModif{ for the PH, CD, and CBFS} cases. $L$ refers to channel half-height for CD, the step height for CBFS and hill crest for PH. The legend denotes the following: (\legendHF) DNS for CD \cite{laval2011direct}, LES for PH \cite{frohlich_LES_2005} and LES for CBFS \cite{bentaleb2012large, lardeau_leschziner_2011}, (\legendANSJ) \(\MANSJ\), (\legendCHAN) \(\MSST\), (\legendSEP) \(\MSEP\), (\legendintXMA) \(\Mblend\).}
\label{Fig. Cf all}
\end{figure}

\ReviewerOne{
The blended model also improves the predicted skin friction along the bottom wall, as shown in Figure~\ref{Fig. Cf all} for the three separated-flow cases. By leveraging the dominant contribution of \(\MSEP\) in the separated regions, the blended model more effectively captures the effects of the strong adverse-pressure gradients and the reattachment zones. In the attached flow portions of these internal flows, \(\MSST\) typically remains the main contributor, ensuring accuracy in regions where the baseline model already performs well. This complementary interplay between \(\MSEP\) and \(\MSST\) is further illustrated in Table~\ref{Tab:bubble_measure_separated}, which compares the predicted separation bubbles against available LES/DNS reference data. The varying degrees of gray shading emphasize the error intensity of the bubble prediction with the various models.

Although each expert model tends to over- or under-predict specific aspects of the separation or reattachment, the blended solution generally reduces the error across all three configurations. In particular, \(\MSEP\) alone often yields excessively shortened or delayed recirculation zones in certain cases, while \(\MSST\) can exaggerate bubble lengths. \(\MANSJ\), which was trained with jet-like free-shear features in mind, generally performs poorly in these wall-driven separations, significantly overestimating the length of the recirculation. In contrast, \(\Mblend\) blends the beneficial traits of each expert, usually aligning more closely with the reference data from LES or DNS. 

}

For a better quantification of the predictive accuracy of the various models, we report in Table \ref{tab:MAE} the \textit{mae} for all models and cases, \NewModif{ computed on velocity profiles and skin friction along the bottom walls}. It is noteworthy that, although the $\MSEP$ model is more accurate than the other experts in separated regions, it can still lead to higher overall errors than the $\MSST$ because of its lower performance in attached regions.  On the other hand, the blended model generally outperforms the specialized models, showing that the features/weight mapping effectively selects the most suitable experts according to the local flow physics.

\begin{table}[h]
    \centering
    \renewcommand{\arraystretch}{1.2}
    \newcommand{\shadecell}[2]{\cellcolor[gray]{#1} #2}
    \begin{tabular}{lccc}
        \toprule
        Model & Start $x/L$ & End $x/L$ & Length \\
        \midrule
        \multicolumn{4}{c}{\textbf{Periodic Hill (PH)}} \\
        LES \cite{frohlich_LES_2005} & \shadecell{0.95}{0.21} & \shadecell{0.95}{4.7} & \shadecell{0.95}{4.49} \\
        $\MANSJ$ & \shadecell{0.80}{0.25} & \shadecell{0.70}{8.23} & \shadecell{0.70}{7.98} \\
        $\MSST$ & \shadecell{0.80}{0.25} & \shadecell{0.75}{7.67} & \shadecell{0.75}{7.42} \\
        $\MSEP$ & \shadecell{0.85}{0.25} & \shadecell{0.90}{4.4} & \shadecell{0.90}{4.15} \\
        $\Mblend$ & \shadecell{0.88}{0.265} & \shadecell{0.92}{4.82} & \shadecell{0.92}{4.555} \\
        \midrule
        \multicolumn{4}{c}{\textbf{Converging-Diverging (CD) Channel}} \\
        DNS \cite{laval2011direct} & \shadecell{0.95}{5.75} & \shadecell{0.95}{6.60} & \shadecell{0.95}{0.85} \\
        $\MANSJ$ & \shadecell{0.85}{5.48} & \shadecell{0.70}{12.13} & \shadecell{0.70}{6.65} \\
        $\MSST$ & \shadecell{0.88}{6.01} & \shadecell{0.80}{9.42} & \shadecell{0.85}{3.41} \\
        $\MSEP$ & \shadecell{0.80}{7.02} & \shadecell{0.85}{7.85} & \shadecell{0.95}{0.83} \\
        $\Mblend$ & \shadecell{0.85}{6.14} & \shadecell{0.82}{8.19} & \shadecell{0.90}{2.05} \\
        \midrule
        \multicolumn{4}{c}{\textbf{Curved Backward-Facing Step (CBFS)}} \\
        LES \cite{bentaleb2012large, lardeau_leschziner_2011} & \shadecell{0.95}{0.81} & \shadecell{0.95}{4.36} & \shadecell{0.95}{3.55} \\
        $\MANSJ$ & \shadecell{0.85}{0.81} & \shadecell{0.70}{6.33} & \shadecell{0.70}{5.52} \\
        $\MSST$ & \shadecell{0.88}{1.06} & \shadecell{0.90}{4.65} & \shadecell{0.95}{3.59} \\
        $\MSEP$ & \shadecell{0.80}{1.33} & \shadecell{0.85}{3.84} & \shadecell{0.80}{2.51} \\
        $\Mblend$ & \shadecell{0.88}{1.08} & \shadecell{0.92}{4.52} & \shadecell{0.90}{3.44} \\
        \bottomrule
    \end{tabular}
    \caption{Separation bubble measure for PH, CBFS, and CD cases. Light gray shading represents deviation from LES/DNS, with darker shades indicating higher error.}\label{Tab:bubble_measure_separated}
\end{table}

\begin{table}[h!]
\centering
\renewcommand{\arraystretch}{1.2}
\hspace*{-1.2cm}
\begin{tabular}{lcccc}
\toprule
QoI & $\MANSJ$ & $\MSST$ & $\MSEP$ & $\Mblend$ \\
\midrule
\multicolumn{5}{c}{\textbf{Periodic Hill (PH)}} \\
$U/U_{\textrm{ref}}$ & \cellcolor[gray]{0.55} $2.01$ & \cellcolor[gray]{0.68} $1.37$ & \cellcolor[gray]{0.95} $0.06$ & \cellcolor[gray]{0.94} $0.09$ \\
$C_f$ & \cellcolor[gray]{0.55} $3.53 \cdot 10^{-3}$ & \cellcolor[gray]{0.87} $2.37 \cdot 10^{-3}$ & \cellcolor[gray]{0.79} $2.65 \cdot 10^{-3}$ & \cellcolor[gray]{0.95} $2.07 \cdot 10^{-3}$ \\
\midrule
\multicolumn{5}{c}{\textbf{Converging-Diverging (CD) Channel}} \\
$U/U_{\textrm{ref}}$ & \cellcolor[gray]{0.55} $1.93$ & \cellcolor[gray]{0.94} $0.07$ & \cellcolor[gray]{0.95} $0.03$ & \cellcolor[gray]{0.95} $0.03$ \\
$C_f$ & \cellcolor[gray]{0.55} $4.31 \cdot 10^{-3}$ & \cellcolor[gray]{0.82} $2.83 \cdot 10^{-3}$ & \cellcolor[gray]{0.86} $2.61 \cdot 10^{-3}$ & \cellcolor[gray]{0.95} $2.15 \cdot 10^{-3}$ \\
\midrule
\multicolumn{5}{c}{\textbf{Curved Backward-Facing Step (CBFS)}} \\
$U/U_{\textrm{ref}}$ & \cellcolor[gray]{0.55} $0.03$ & \cellcolor[gray]{0.95} $0.01$ & \cellcolor[gray]{0.75} $0.02$ & \cellcolor[gray]{0.95} $0.01$ \\
$C_f$ & \cellcolor[gray]{0.55} $1.51 \cdot 10^{-3}$ & \cellcolor[gray]{0.95} $4.83 \cdot 10^{-4}$ & \cellcolor[gray]{0.72} $1.08 \cdot 10^{-3}$ & \cellcolor[gray]{0.94} $5.05 \cdot 10^{-4}$ \\
\midrule
\end{tabular}
\caption{\textit{mae} on the normalized streamwise velocity $U/U_{\textrm{ref}}$ and the skin friction coefficient $C_f$ for the PH, CD, and CBFS cases. Shading represents error magnitude, with darker shades indicating higher error.}
\label{tab:MAE}
\end{table}

\ReviewerOne{
\section{Application to untrained scenarios}
In this section, we assess the predictive capabilities of the blended data-driven model for three test cases, selected from the NASA turbulence modeling testing challenge \cite{rumsey2022nasa}, that were not included in the databases used for training the experts or the weighting functions. The configuration and grid resolutions used for this set of RANS simulations are reported in Table \ref{tab:testCasesConfig}.}

\begin{table}[h!]
     \renewcommand{\arraystretch}{1.2}
    \begin{tabular}{lcccc}
    \toprule
    \textbf{Case} & \textbf{Re} %& \textbf{Re Definition} 
                  & \textbf{ $L$} 
                  & \textbf{Grid Size (2D)} 
                  & \textbf{Reference}\\
    \midrule
    \textbf{ZPGBL} 
      & $5\times 10^6$
      %& %$Re = {U_{\infty}\,L}/{\nu}$
      & 1 unit length
      & $545\times 385$ 
      &  \cite{wieghardt1951turbulent}\\
    \textbf{WMH}   
      & $9.36\times 10^5$  
      %& %$Re = {U_{\infty}\,L}/{\nu}$
      & hump's chord
      & $409\times 109$ 
      & \cite{greenblatt2006experimental1}\\
    \textbf{NACA0012} 
      & $6\times 10^6$  
      %& %$Re = {U_{\infty}\,L}/{\nu}$ 
      & airfoil chord
      & $225\times 65$ 
      & \cite{mccroskey1987critical}\\
    \bottomrule
    \end{tabular}
    \caption{Overview of the test cases and their characteristic Reynolds numbers based on $L$, 
             along with the 2D grid resolutions used in the RANS simulations. 
             For the three cases the Reynolds number is based on the freestream conditions and the reference length $L$ given in the table. 
             The NACA0012 is simulated
             over a range of angles of attack from $0^\circ$ to $18^\circ$.}
    \label{tab:testCasesConfig}
\end{table}

\ReviewerOne{
The first test case is an incompressible 2D zero-pressure-gradient boundary layer (ZPGBL) with unit Reynolds number of $5\times10^6$, originally reported in \cite{wieghardt1951turbulent} and later included in the 1968 AFOSR-IFP Stanford Conference on turbulent flows \cite{coles1968computation}, which serves to check non-regression of the baseline model's accuracy in attached wall-bounded flows . This flow develops under essentially constant pressure conditions along the streamwise direction and represents a canonical validation for turbulence models designed to handle boundary layers without significant separations or compressibility effects.

The second test case is the 2D NASA Wall-Mounted Hump (WMH) flow at $Re=936\,000$ \cite{greenblatt2006experimental1}. This more complex configuration exhibits flow acceleration on the fore portion of the hump, boundary-layer separation, reattachment downstream of the crest, and eventual redevelopment of the boundary layer. These features test the model’s ability to capture strong adverse-pressure gradients, separation, and subsequent reattachment, a well-known challenge for classical eddy viscosity approaches.

The third test case is the flow over a NACA0012 airfoil at a chord-based Reynolds number of $6$ million and at various angles of attack ranging from $0^\circ$ to $18^\circ$. A summary of experimental data can be found in \cite{mccroskey1987critical}. This configuration is frequently used to evaluate turbulence models for external aerodynamic applications, especially under conditions that can range from mildly separated flows at moderate angles of attack to stronger separations. As the angle of attack increases, the flow may encounter significant adverse-pressure gradients at the leading edge or the upper surface, making it a demanding test of the blended model’s capacity to predict separation and complex lift-related phenomena.

}

\subsection{Assessing training features coverage in the latent space}
Prior to applying the blended model to unseen cases, we first analyze the distribution of the flow features used as input to the RFR during training. The \NewModif{ vector \( \bm{\eta} \) contains 11 physical features, each representing different aspects of the flow. To inspect the capability of the RFR model to generalize effectively to new, unseen cases, we reduce this 11-dimensional feature space into a two-dimensional latent space, with the latent variable denoted \( \bm{\zeta} \). }
This dimensionality reduction helps to visually assess whether the flow cases used for training adequately capture the variability of the test cases. 
Before applying the dimensionality reduction techniques, we perform a pre-processing step using $k$-means clustering to reduce the density of data points in feature space. This is done to enhance the clarity of the subsequent analysis and allow for a more interpretable visualization of the latent structures. After clustering, we apply three dimensionality reduction techniques: PCA (Principal Component Analysis) \cite{sirovich1987turbulence}, $t$-SNE ($t$-distributed Stochastic Neighbor Embedding) \cite{maaten2008visualizing}, and autoencoders \cite{rumelhart1986learning,hinton2006reducing}, to examine the distribution of features.

\ReviewerTwo{ In the latent space, for clarity, we represent the spread and variance of the latent features using covariance regions, visualized as ellipses. These regions are constructed by calculating the covariance matrix of the latent variables \(  \bm{\zeta} \) for each dataset, capturing how the data points are distributed in the two-dimensional space. The covariance matrix \( \Sigma \) is computed as:
\[
\Sigma = \frac{1}{N} \sum_{i=1}^{N} (\bm{\zeta}_i - \overline{\bm{\zeta}})(\bm{\zeta}_i - \overline{\bm{\zeta}})^T,
\]
where \( N \) is the number of data points, \( \bm{\zeta}_i \) are the latent variables, and \( \overline{\bm{\zeta}} \) is their mean. The covariance matrix describes the extent to which the latent variables co-vary with each other. 
To visualize the variability, we compute the eigen-decomposition of \( \Sigma \) to determine the axes of the ellipses. 
%The scaling factor is derived from the chi-square distribution, ensuring that the ellipses correspond to a 95\% confidence region. Specifically, for a bivariate Gaussian distribution, the axes are scaled by a factor \( 2.45 \sqrt{\lambda_i} \).
%
These ellipses provide an intuitive representation of the latent space structure, illustrating how well the test cases are covered by the training data.
For PCA, $t$-SNE or autoencoders, the axes of the ellipses correspond to the principal components, which represent the directions of maximum variance. The lengths of the axes are determined using the eigenvalues \( \lambda_i \) of the covariance matrix, scaled by a confidence factor derived from the chi-square distribution.  
Specifically, the ellipses define a ``95\% confidence region'' that encloses approximately 95\% of the probability mass under the assumption of a multivariate Gaussian distribution. For a bivariate Gaussian distribution, the scaling factor is given by \( \sqrt{\chi^2_2(0.95)} \), where \( \chi^2_2(0.95) \approx 5.99 \). This leads to a scaling factor of \( \sqrt{5.99} \approx 2.45 \), meaning that the ellipse axes are given by:
$
\text{width}, \text{height} = 2.45  (\sqrt{\lambda_1}, \sqrt{\lambda_2})
$.

}

\begin{figure}[hbtp!]
\centering
\begin{subfigure}{0.32\textwidth}
\includegraphics[scale=0.3]{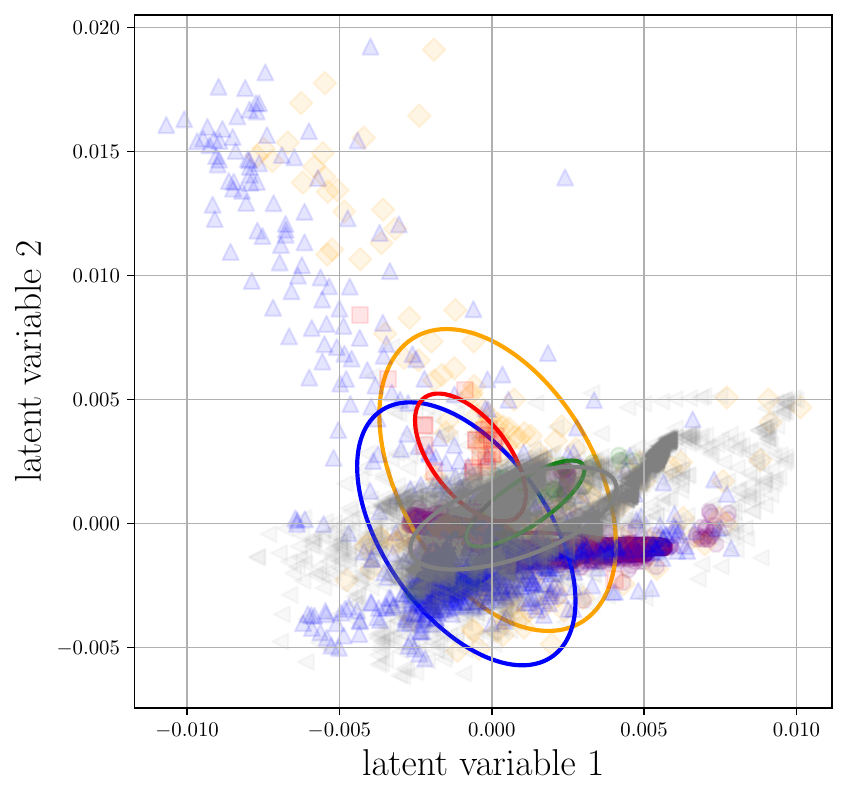}
\caption{PCA}\label{PCA_latent}
\end{subfigure}\hfill
\begin{subfigure}{0.32\textwidth}
\includegraphics[scale=0.3]{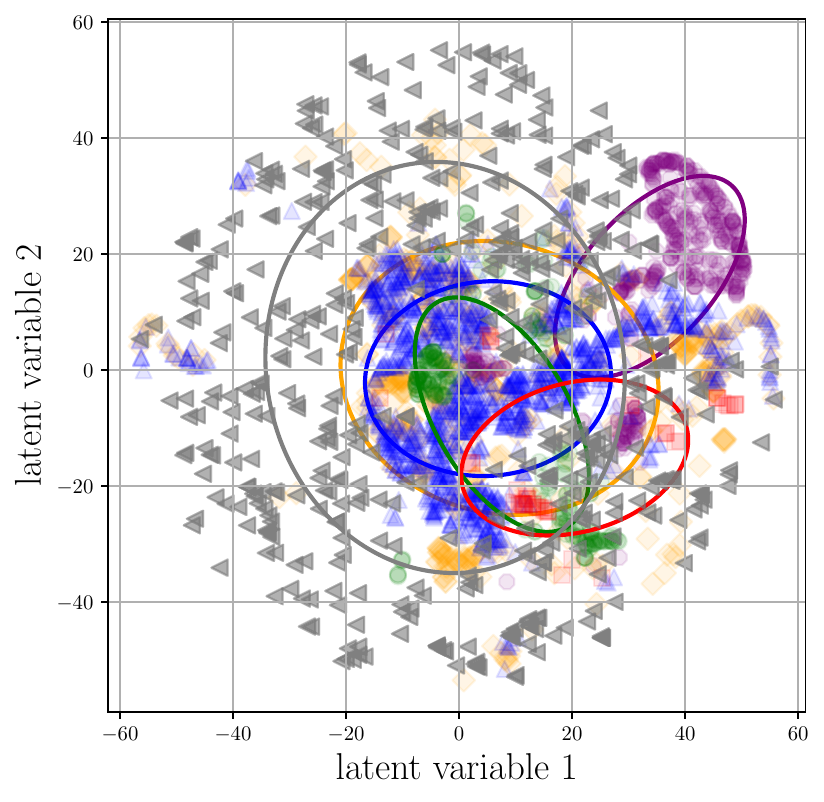}
\caption{$t$-SNE}\label{tSNE_latent}
\end{subfigure}\hfill
\begin{subfigure}{0.32\textwidth}
\includegraphics[scale=0.3]{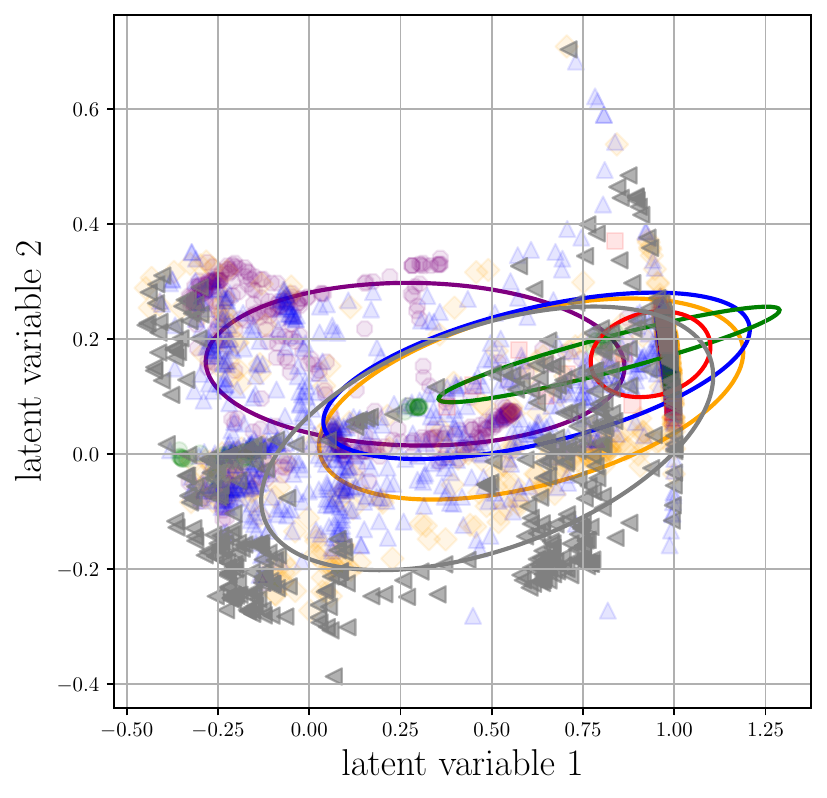}
\caption{Autoencoders}\label{Encoder_latent}
\end{subfigure}
\centering
\includegraphics[scale=0.7]{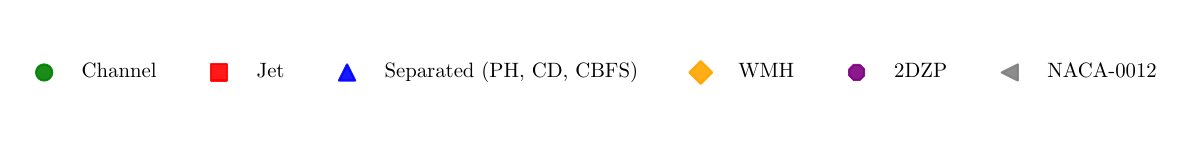}
\caption{Latent space projection of the physical flow features using PCA, $t$-SNE, and autoencoders. The ellipses represent variance regions based on the covariance matrix of the latent variables, with axes proportional to the square roots of the eigenvalues. These ellipses visualize the 95\% confidence region, indicating where the majority of the training data is concentrated.}
\label{PCA features}
\end{figure}

\ReviewerOne{
To analyze features in the latent space, we begin by visualizing the data distribution using principal component analysis (PCA), which projects high-dimensional features onto directions of maximum variance. As shown in Figure~\ref{PCA_latent}, all three test cases (the 2D zero-pressure-gradient boundary layer, the wall-mounted hump, and the NACA0012 at angles of attack ranging from $0^\circ$ to $18^\circ$) lie within the overall variance regions spanned by the training data. This indicates that, in terms of large-scale global structures, the training set provides a reasonable coverage of the feature space. Nonetheless, we do observe a set of outliers, particularly for the NACA0012, suggesting that some aspects of high-lift flow physics are less densely represented in the original training database.

To gain insight into potential non-linear dependencies that PCA might overlook, we apply the $t$-SNE ($t$-distributed stochastic neighbor embedding) algorithm \cite{van2008visualizing}, which prioritizes preserving the local neighborhood of each point in the high-dimensional space. Figure~\ref{tSNE_latent} shows that $t$-SNE clusters the test features more distinctly, while still revealing that the majority of points for all three test cases remain embedded in or near the training data regions. Here, the distribution of outliers associated with the NACA0012 case become more apparent, underscoring that a subset of the airfoil flow conditions, especially at higher angles of attack, deviate from the more moderate conditions captured by the training data. The 2D boundary layer and wall-mounted hump cases, by contrast, show less pronounced outliers, suggesting that their underlying flow features are more in line with the training-set physics.
Autoencoders further illustrate the distribution of test data within the latent space by learning non-linear representations that facilitate near-accurate reconstruction of the original features. As shown in Figure~\ref{Encoder_latent}, the autoencoder effectively maps both the 2D boundary layer and wall-mounted hump test data closely within the general variance regions spanned by the training cases. While the NACA0012 flow still presents apparent outliers. 

Altogether, these dimensionality-reduction techniques confirm that the training data broadly encompasses the feature space relevant to the test scenarios, with a certain degree of extrapolation required, most prominently for the NACA0012. While the 2D projections cannot fully capture every subtlety of the 11-dimensional feature space, the combined evidence from PCA, $t$-SNE, and autoencoders indicates that the training data is sufficiently representative to support robust prediction on these untrained flows. Any residual outliers in the latent space highlight potential avenues for refining the training database to better capture the non-linear flow physics encountered in more complex flows.

}

\subsubsection{2D Zero-pressure-gradient boundary layer flow (ZPGBL)}
\begin{figure}[hbtp!]
\centering

\begin{subfigure}{0.5\textwidth}
\includegraphics[scale=0.35]{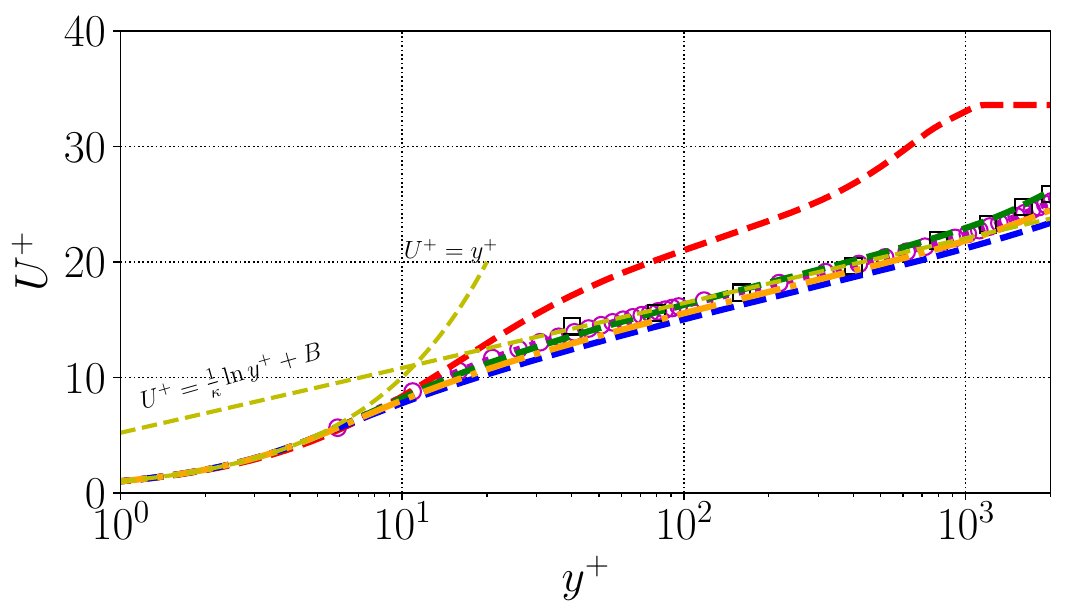}
\caption{Wall-normal profiles of the velocity.}
\label{Fig. FlatPlate_UPlus}
\end{subfigure}\hfill
\begin{subfigure}{0.5\textwidth}
\includegraphics[scale=0.35]{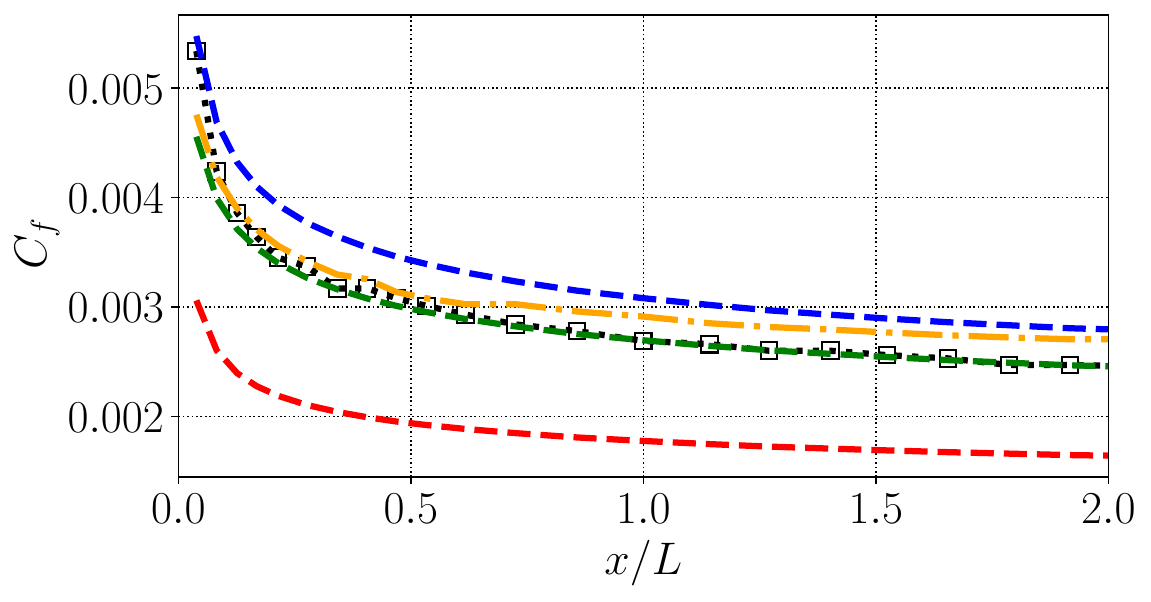}
\caption{Friction coefficient \(C_f\).}
\label{Fig. FlatPlate_Cf}
\end{subfigure}\\
%\begin{subfigure}{0.5\textwidth}
%\includegraphics[scale=0.35]{figures/TurbFlatePlate/figure_absErrors_UPlus_vs_yPlus.pdf}
%\caption{Absolute errors.}
%\label{Fig. FlatPlate_AbsErrorsUPlus}
%\end{subfigure}\\
\begin{subfigure}{0.5\textwidth}
\includegraphics[scale=0.35]{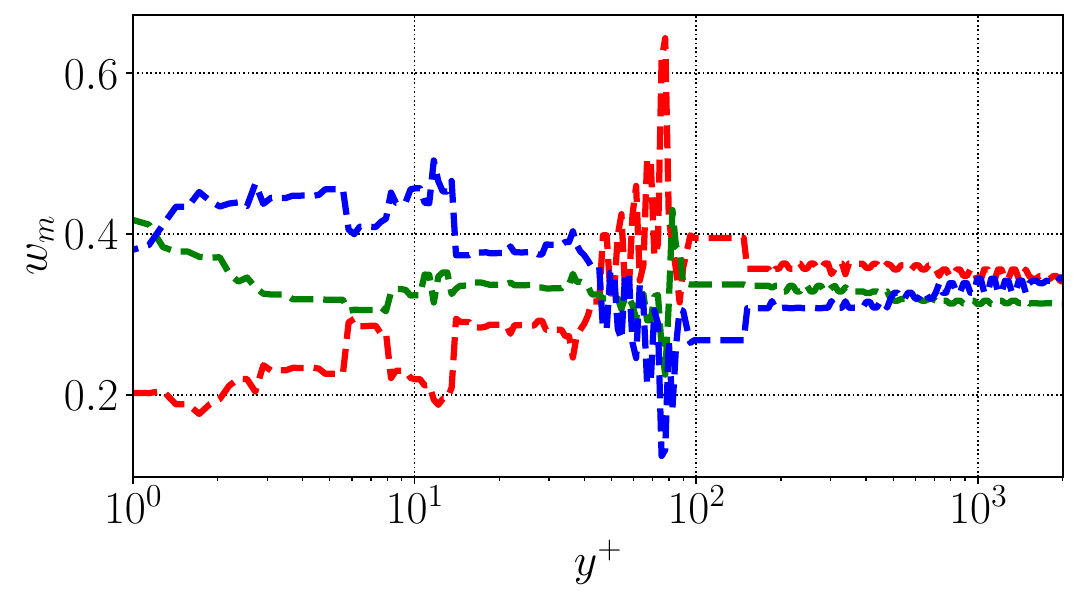}
\caption{Weight distribution along $y+$.}
\label{Fig. FlatPlate_WeightsYPlus}
\end{subfigure}\hfill
\begin{subfigure}{0.5\textwidth}
\includegraphics[scale=0.35]{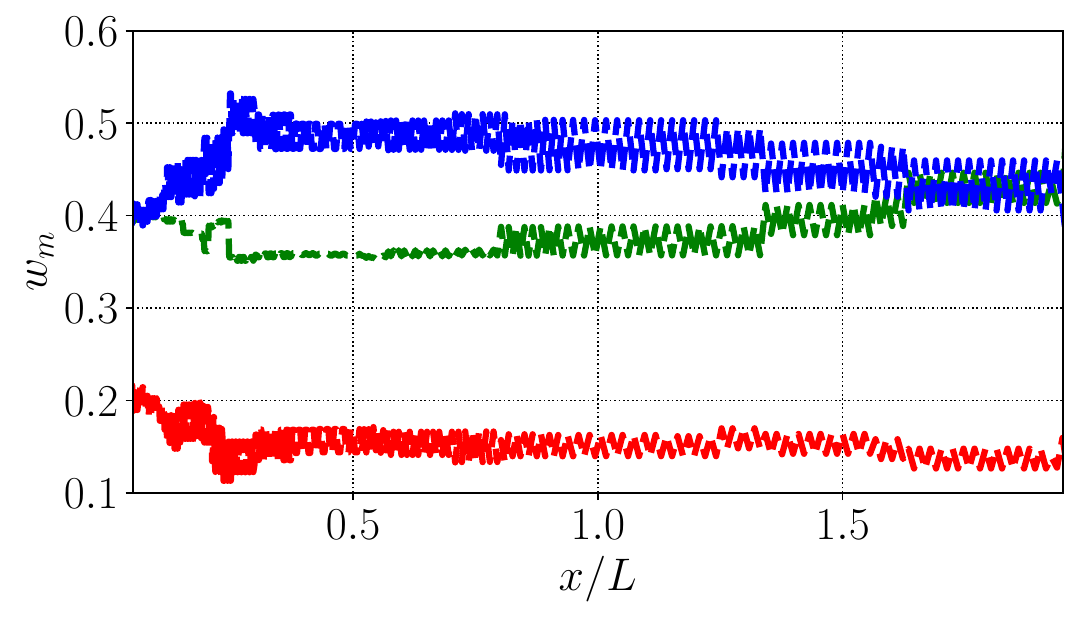}
\caption{Weights along the wall.}
\label{Fig. FlatPlate_WeightsWall}
\end{subfigure}
\caption{2D Zero-pressure-gradient boundary layer flow case. Top row: velocity profile and skin friction distribution. Bottom row: blending weights along the wall normal coordinate and along the wall. The legend denotes the following: (\legendHFm) law-of-the-wall based on Coles' mean velocity profile \cite{coles1956law, Bardina_NASA_TM_110446_1997}, (\legendHF) experimental data of Wieghardt \cite{wieghardt1951turbulent}, (\legendANSJ) \(\MANSJ\), (\legendCHAN) \(\MSST\), (\legendSEP) \(\MSEP\), (\legendintXMA) \(\Mblend\), and (\legendtheory) theoretical prediction with the Kármán constants \(\kappa = 0.41\) and \(B = 5.2\).}
\label{Fig. U+ vs y+}
\end{figure}

We first focus on the 2D ZPGBL flow.
\ReviewerOne{ The velocity profiles normalized with respect to wall units at \( x=0.97 \) are presented in Figure \ref{Fig. FlatPlate_UPlus}. Similar to the training channel flow case, the \( \MSST \) model exhibits excellent agreement with the experimental data from \cite{wieghardt1951turbulent} and theoretical predictions, supported by its low \textit{mae}, as shown in Table \ref{tab:FlatPlate_MAE}. The \( \MSEP \) model, while slightly under-predicts the velocity in the logarithmic region, still manages to capture the overall boundary layer trend reasonably well. In contrast, the \( \MANSJ \) model significantly over-predicts the velocity across the entire boundary layer.
%In addition to Figure \ref{Fig. FlatPlate_UPlus}, Figure \ref{Fig. FlatPlate_AbsErrorsUPlus} provides deeper insight into the behavior of the blended model. It shows that, while 
The blended model on the other hand strikes a good balance, outperforming both the \( \MSEP \) and \( \MANSJ \) models even if it does not quite reach the accuracy of the \( \MSST \) model.}

Regarding the blending weight behavior along $y^+$, Figure \ref{Fig. FlatPlate_WeightsYPlus} shows that, contrary to expectations, the \( \MSST \) model does not dominate the weight distribution along \( y^+ \). Instead, the \( \MANSJ \) model gains significant weight in the logarithmic region. This indicates that the shifts in model contributions in the log-law region have little impact on the overall accuracy.
%, as all models ultimately yield similar corrections. 
Despite these variations, the blended model remains robust due to the balanced contributions from all models, ensuring consistent and reliable predictions across different regions of the boundary layer.
When looking at the blending behavior along the wall (Figure \ref{Fig. FlatPlate_WeightsWall}), we observe that the \( \MSEP \) and \( \MSST \) models dominate, with \( \MSEP \) contributing most strongly, followed by \( \MSST \). Meanwhile, the \( \MANSJ \) model has minimal influence along the wall.
\ReviewerOne{ The weights distributions allows the blended model to achieve a balanced prediction of \( C_f \), resulting in a low \textit{mae} after the best performing model $\MSST$.  
%(Figure \ref{Fig. FlatPlate_Cf} and Table \ref{tab:FlatPlate_MAE}). 

Both the prediction of $ C_f $ and $U^+$ illustrate the idea of automatic model adaptation. In fact, without prior knowledge of the best performing individual model, the blending approach delivers quite close results to this last by automatically weighting different models appropriately. We also notice that, while the weight distribution exhibits small oscillations, probably due to irregularity in the random forest regressor, these do not appear to  translate into oscillations in the numerical solution. In the future, such defect could be improved by replacing random forests with a smoother regressor, e.g. Gaussian Processes.

}%

\begin{table}[h!]
\centering \renewcommand{\arraystretch}{1.2}
\hspace*{-.8cm}
\begin{tabular}{lcccc}
\toprule
QoI & $\MANSJ$ & $\MSST$ & $\MSEP$ & $\Mblend$ \\
\midrule
$C_f$ & \cellcolor[gray]{0.55} $1.1 \times 10^{-3}$ & \cellcolor[gray]{0.91} $8.4 \times 10^{-5}$ & \cellcolor[gray]{0.75} $3.7 \times 10^{-4}$ & \cellcolor[gray]{0.88} $1.7 \times 10^{-4}$ \\
$U^+$ & \cellcolor[gray]{0.55} $6.27$ & \cellcolor[gray]{0.95} $0.42$ & \cellcolor[gray]{0.82} $1.45$ & \cellcolor[gray]{0.90} $0.7$ \\
\bottomrule
\end{tabular}
\caption{\textit{mae} for $C_f$ and $U^+$ for the Turbulent Flat Plate flow. Shading represents error magnitude, with darker shades indicating higher error.}
\label{tab:FlatPlate_MAE}
\end{table}

\subsubsection{Wall-Mounted Hump (WMH)}

\begin{figure}[hbtp!]
  \centering
  \begin{subfigure}[T]{0.48\textwidth}
    \includegraphics[width=\textwidth]{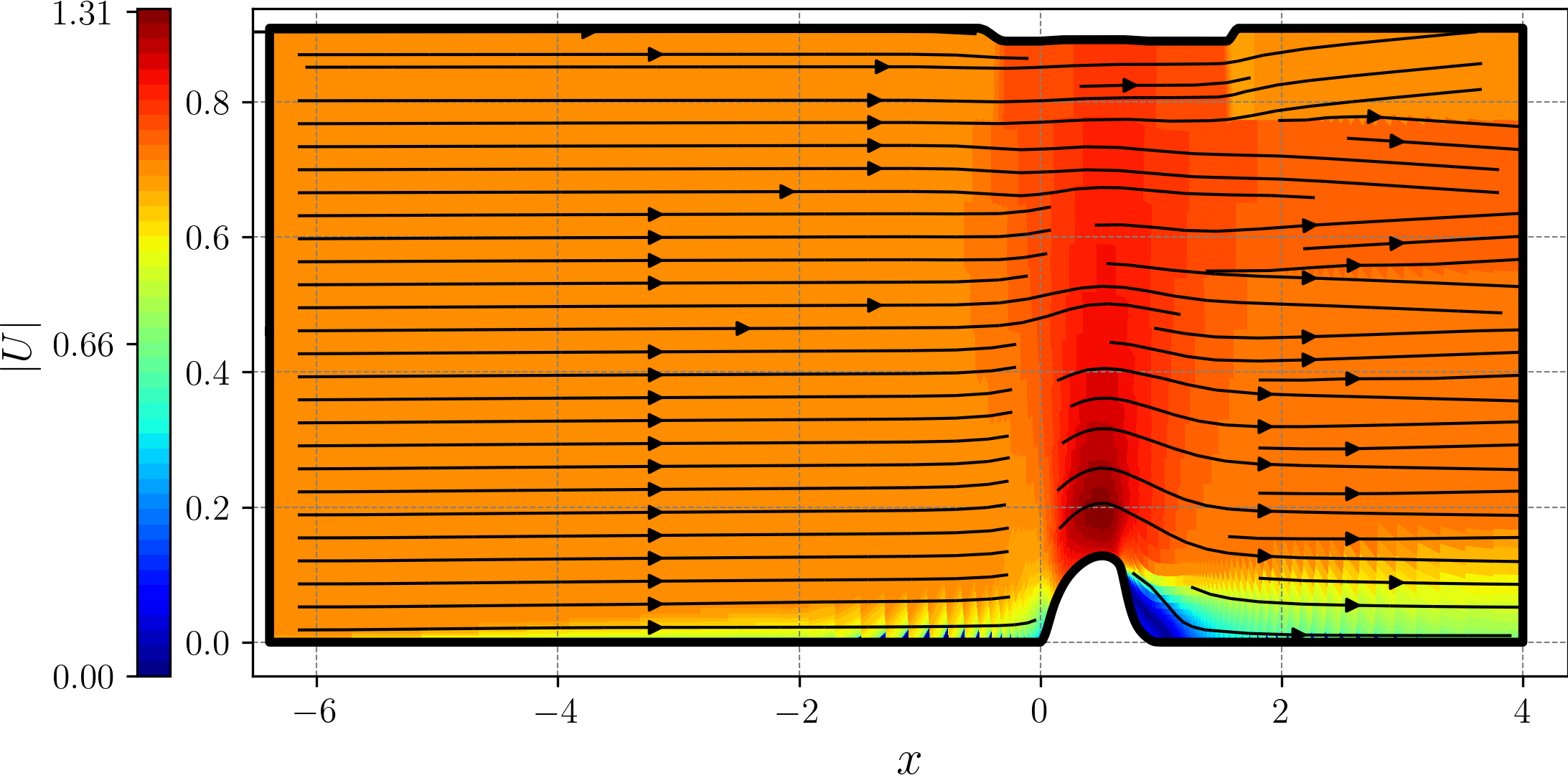}
    \caption{RANS velocity obtained by $\Mblend$}
    \label{Fig. WMH_VelocityField}
  \end{subfigure}\hfill
  \begin{subfigure}[T]{0.48\textwidth}
    \includegraphics[width=\linewidth, trim=0 0 0 8, clip]{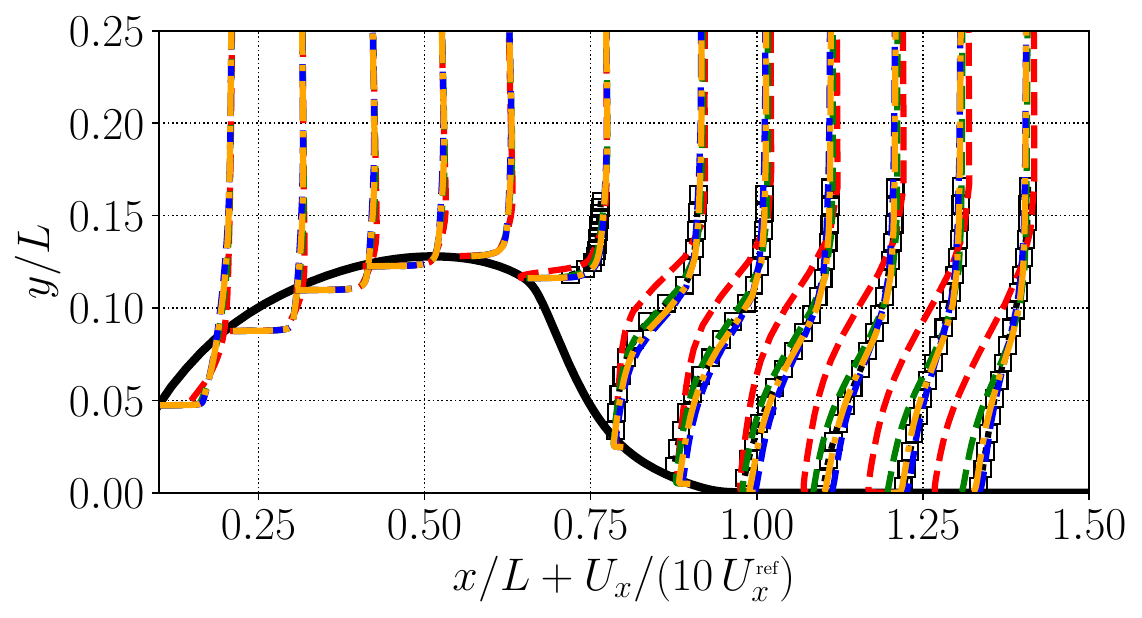}
    \caption{Velocity profiles across $x/L$ locations.}
    \label{Fig. WMH_VelocityProfiles}
  \end{subfigure} \\
  \begin{subfigure}[T]{0.48\textwidth}
    \includegraphics[width=\linewidth]{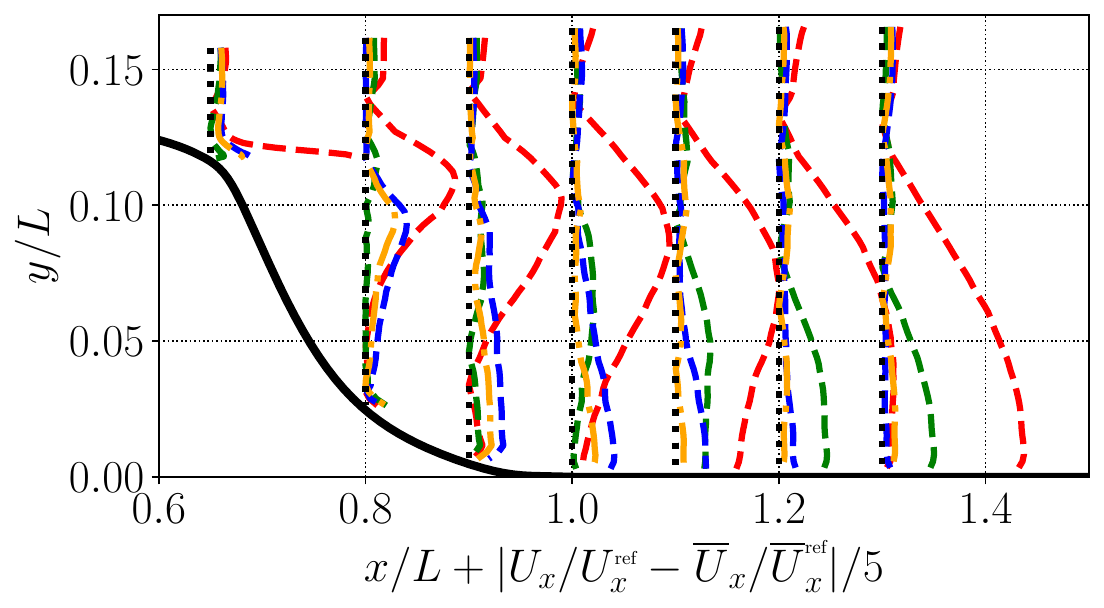}
    \caption{Absolute errors.}
    \label{Fig. WMH_VelocityErrors}
  \end{subfigure}\hfill
  \begin{subfigure}[T]{0.48\textwidth}
    \includegraphics[width=\linewidth]{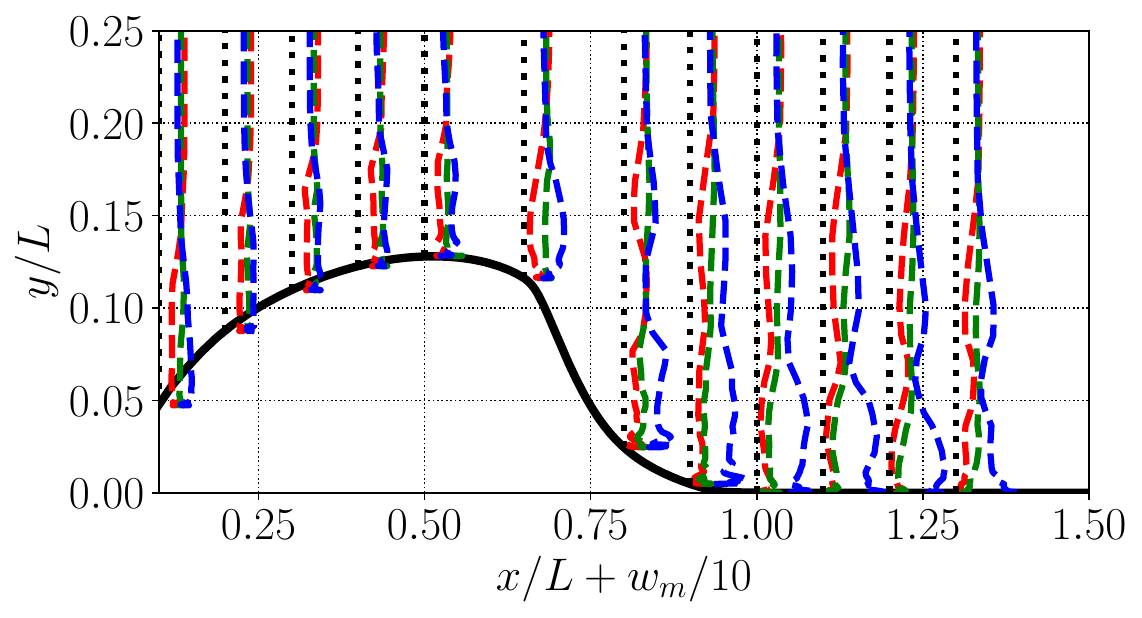}
    \caption{Weight distributions}
    \label{Fig. WMH_WeightDistributions}
  \end{subfigure}
  \caption{Horizontal velocity, \textit{mae} and weighting functions at various streamwise stations of the WMH flow domain. $L$ is the the bump "chord" and the legend denotes the following: (\legendHF) experimental data \cite{greenblatt2006experimental1,greenblatt2006experimental2}, (\legendANSJ) \(\MANSJ\), (\legendCHAN) \(\MSST\), (\legendSEP) \(\MSEP\), (\legendintXMA) \(\Mblend\).}
  \label{Fig. Velocity profile and weights WMH}
\end{figure}

Our second test case is the turbulent flow through a channel with a wall-mounted hump. The velocity profiles and their respective absolute errors at various streamwise stations are presented in Figures \ref{Fig. WMH_VelocityProfiles} and \ref{Fig. WMH_VelocityErrors}. For this case, the \( \MSEP \) model exhibits overall good performance compared to the other expert models. Specifically, it improves the representation of the velocity profiles in the separated region downstream of the hump with respect to the baseline model \( \MSST \), although it predicts earlier reattachment than observed in the reference experimental data \cite{greenblatt2006experimental1,greenblatt2006experimental2, NaughtonWMH2006}. In contrast, the \( \MANSJ \) model performs similarly to the training separated cases and largely overestimates the extent of the separation bubble.

The weight profiles (Figure \ref{Fig. WMH_WeightDistributions}) show that \( \MSST \) and \( \MSEP \) are nearly equally weighted in the attached flow region upstream of the separation, while \( \MSEP \) takes precedence in the recirculation region and beyond. The \( \MANSJ \) model, on the other hand, is assigned lower weights throughout both the attached boundary layer and the separated region. Interestingly, all weights tend to converge towards a value of 1/3 in the core flow, indicating that this region is insensitive to the specific turbulence model used. The blended model’s solution closely mirrors that of \( \MSEP \), with additional contributions from the other expert models helping to improve the prediction of the reattachment point.

\begin{figure}[hbtp!]
\centering
\begin{subfigure}{0.5\textwidth}
\includegraphics[scale=0.35]{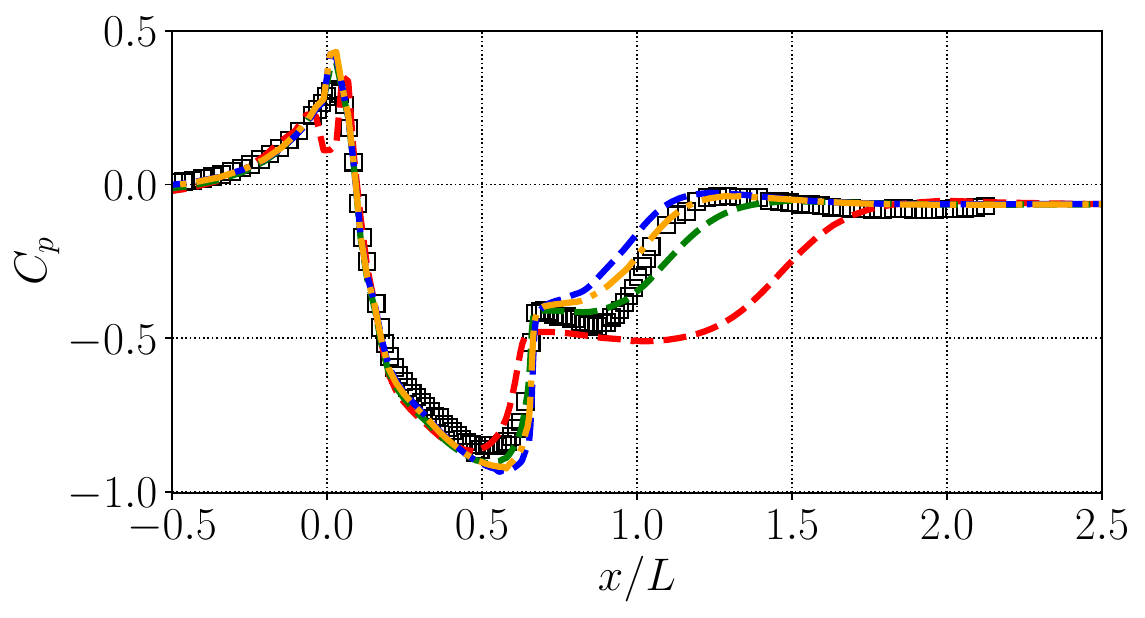}
\caption{Pressure coefficient $C_p$.}
\label{Fig. WMH_Cp}
\end{subfigure}\hspace*{0.2cm}
\begin{subfigure}{0.5\textwidth}
\includegraphics[scale=0.35]{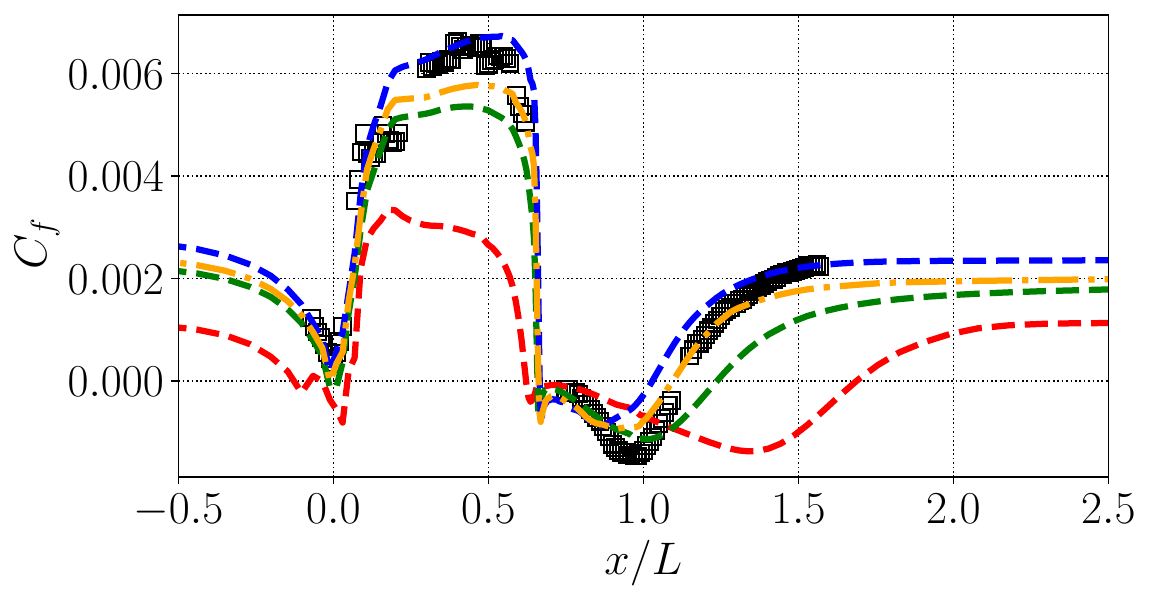}
\caption{Friction coefficient $C_f$.}
\label{Fig. WMH_Cf}
\end{subfigure}

\begin{subfigure}{0.5\textwidth}
\includegraphics[scale=0.35]{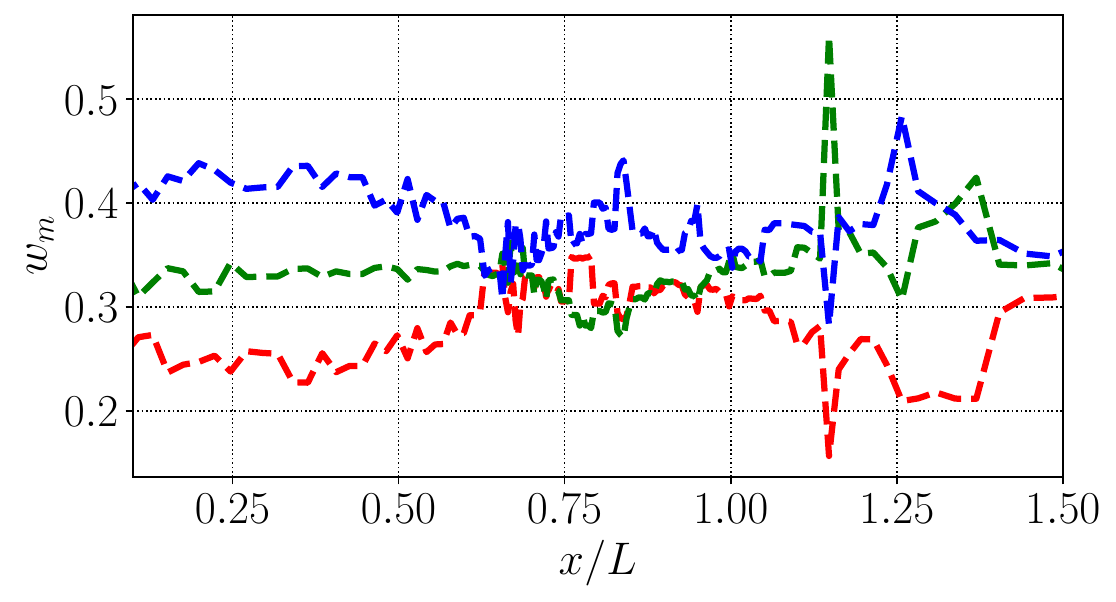}
\caption{Weights distribution along the wall.}
\label{Fig. WMH_weights_wall}
\end{subfigure}
\caption{Pressure coefficient and skin friction coefficient with the weights distributions along the bottom wall for the WMH flow case. $L$ is the the bump "chord" and the legend denotes the following: (\legendHF) experimental data \cite{NaughtonWMH2006}, (\legendANSJ) \(\MANSJ\), (\legendCHAN) \(\MSST\), (\legendSEP) \(\MSEP\), ($\legendintXMA$) \(\Mblend\).}
\label{Fig. Cp and Cf WMH}
\end{figure}

\ReviewerOne{
The pressure and skin friction coefficient distributions along the bottom wall (Figures~\ref{Fig. WMH_Cp} and \ref{Fig. WMH_Cf}) further highlight that the blended model strikes a compromise between the baseline \(\MSST\), which overpredicts the extent of the separated region, and the \(\MSEP\) model, which underpredicts it, while minimizing the influence of the outlier \(\MANSJ\) model, as evidenced by Figure~\ref{Fig. WMH_weights_wall}. The blended model’s performance is further reflected in the distribution of weights along the bottom wall, where the \( \MSEP \) model dominates, followed by \( \MSST \), contributing to a balanced and accurate prediction of the skin friction coefficient \( C_f \). The \( \MANSJ \) model has minimal influence in this region, reinforcing the reliability of the blended model's output.
The predicted separation bubble characteristics are summarized in Table~\ref{fig:bubble_measure_WMH}, where the start and end of the separation region, as well as its overall length, are compared against experimental data \cite{NaughtonWMH2006}. The baseline \(\MSST\) model and \(\MANSJ\) both overestimate the bubble length, with \(\MANSJ\) displaying the most pronounced discrepancy, whereas \(\MSEP\) slightly under-predicts the reattachment point. Notably, the blended model \(\Mblend\) closely matches the experimental data, with an accurate prediction of the bubble length, underscoring how the dynamic weighting scheme effectively capitalizes on the strengths of each expert model and mitigates their weaknesses.
%}
%
}

In summary, the blended model performs well across the different quantities evaluated in this test case. It benefits from the strengths of each expert model, particularly \( \MSEP \) and \( \MSST \), while effectively minimizing the errors contributed by \( \MANSJ \). This balanced behavior is evident in both the velocity profiles and skin friction distributions. The blended model achieves the lowest \textit{mae} for two out of the three metrics considered, as highlighted in Table \ref{tab:WMH_MAE}, confirming its superior overall performance.

\begin{table}[h!]
\centering \renewcommand{\arraystretch}{1.2}
\hspace*{-1.25cm}
\begin{tabular}{lcccc}
\toprule
QoI & $\MANSJ$ & $\MSST$ & $\MSEP$ & $\Mblend$ \\
\midrule
$U/U_{\textrm{ref}}$ & \cellcolor[gray]{0.55} $0.24$ & \cellcolor[gray]{0.91} $0.06$ & \cellcolor[gray]{0.91} $0.06$ & \cellcolor[gray]{0.95} $0.04$ \\
$C_f$ & \cellcolor[gray]{0.55} $2.17 \cdot 10^{-3}$ & \cellcolor[gray]{0.87} $7.53 \cdot 10^{-4}$ & \cellcolor[gray]{0.94} $4.21 \cdot 10^{-4}$ & \cellcolor[gray]{0.95} $3.86 \cdot 10^{-4}$ \\
$C_p$ & \cellcolor[gray]{0.55} $0.09$ & \cellcolor[gray]{0.95} $0.03$ & \cellcolor[gray]{0.82} $0.05$ & \cellcolor[gray]{0.88} $0.04$ \\
\bottomrule
\end{tabular}
\caption{\textit{mae} on the normalized streamwise velocity $U/U_{\textrm{ref}}$, $C_f$, and $C_p$ for the WMH flow. Shading represents error magnitude, with darker shades indicating higher error.}
\label{tab:WMH_MAE}
\end{table}

\begin{table}[h]
    \centering
     \renewcommand{\arraystretch}{1.2}
    \begin{tabular}{lccc}
        \toprule
        Model & Start $x/L$ & End $x/L$ & Length \\
        \midrule
        Experiment \cite{NaughtonWMH2006} & \cellcolor[gray]{0.95} 0.62 & \cellcolor[gray]{0.95} 1.15 & \cellcolor[gray]{0.95} 0.53 \\
        $\MANSJ$ & \cellcolor[gray]{0.95} 0.62 & \cellcolor[gray]{0.55} 1.7 & \cellcolor[gray]{0.55} 1.08 \\
        $\MSST$ & \cellcolor[gray]{0.95} 0.62 & \cellcolor[gray]{0.88} 1.24 & \cellcolor[gray]{0.88} 0.62 \\
        $\MSEP$ & \cellcolor[gray]{0.95} 0.62 & \cellcolor[gray]{0.86} 1.03 & \cellcolor[gray]{0.86} 0.41 \\
        $\Mblend$ & \cellcolor[gray]{0.95} 0.62 & \cellcolor[gray]{0.94} 1.14 & \cellcolor[gray]{0.94} 0.52 \\
        \bottomrule
    \end{tabular}
    \caption{Separation Bubble measure for Wall-Mounted Hump (WMH) Case. Shading represents error magnitude with respect to experiment, with darker shades indicating higher error.}\label{fig:bubble_measure_WMH}
\end{table}

\ReviewerOne{
\subsubsection{NACA0012}
As a final test case, we consider the flow over a NACA0012 airfoil with a chord length \(L = 1\,\si{m}\), at a Reynolds number of \(Re = 6\times 10^6 \). Geometry and grid files are provided by the NASA Turbulence Modeling Resource website \cite{nasa_turbulence_data}. As this configuration involves an external aerodynamic flow at a higher Reynolds number than the training data, it presents an \emph{extrapolative} scenario to test and push to the limit the generalizability of the proposed blending turbulence-modeling approach.

\noindent
Figure~\ref{fig:NACA12_streamlines} presents isocontours of the velocity magnitude and selected streamlines of the baseline model solution at three angles of attack: \(0^\circ\), \(10^\circ\), and \(15^\circ\). Across this range, the boundary layer on the suction side thickens, and incipient or fully separated flow regions may form as the incidence increases. To benchmark model predictions, the NACA0012 airfoil was tested at angles of attack ranging from \(0^\circ\) up to \(18^\circ\). This wide sweep allows us to evaluate each turbulence model’s performance under both lightly loaded and more heavily separated conditions.

\noindent
Figure~\ref{fig:NACA12_combined} compares the drag and lift coefficients obtained from the baseline \(\MSST\) model, the separated‐flow–oriented \(\MSEP\) model, the jet‐flow–inspired \(\MANSJ\) model, and the blended strategy \(\Mblend\) against experimental data from Ladson~\cite{Ladson1988}, Gregory~\&~O’Reilly~\cite{Gregory1970}, and Abbott~\&~von~Doenhoff~\cite{Abbott1959}. From these curves, \(\MANSJ\) exhibits the largest discrepancies, especially at moderate to high angles of attack, whereas both \(\MSST\) and \(\MSEP\) align more closely with the measured trends.
Of note, the \(\MSST\) performs better than \(\MSEP\) for most of the present flows, the latter model being too dissipative. This indicates that \(\MSEP\) should be replaced or supplemented with another expert model better suited for the flow at stake.

\noindent
Despite the underperfomance of \(\MSEP\), the \(\Mblend\) model adaptively weights contributions from \(\MSST\), \(\MSEP\), and \(\MANSJ\), suppressing poorly suited corrections in regions of the flow where they degrade accuracy. Figure~\ref{fig:NACA12_CpCf_all} show examples of surface pressure \(C_p\) and skin friction \(C_f\) at three angles of attack, together with local blending weights along the upper and lower airfoil surfaces. As expected, \(\MANSJ\) receives minimal weights across most of the chord, while \(\MSST\) and \(\MSEP\) dominate depending on the local flow regime. In moderately separated regions on the suction side, the \(\MSEP\) corrections become more influential, reflecting its specialization for adverse-pressure-gradient and separated flows. Meanwhile, \(\MSST\) retains its presence in more attached zones.

\noindent
Overall, by combining multiple expert models, \(\Mblend\) avoids the dominant errors of any single approach. Despite the higher Reynolds number and distinct flow geometry, the blended framework succeeds in replicating key aerodynamic characteristics, including lift-curve slope, drag behavior, and surface-pressure profiles. This indicates a promising ability to generalize beyond the original training data. Such robustness is indispensable in industrial and research applications where designers must reliably predict off-design conditions or new configurations. By adaptively filtering out unhelpful contributions from less-relevant models (here, \(\MANSJ\) is rarely favored), \(\Mblend\) offers a pragmatic, automatic alternative to relying on a single turbulence model across diverse aerodynamic flow regimes.

\begin{figure}[h!]
    \centering
    \begin{subfigure}[t]{0.32\textwidth}
        \centering
        \includegraphics[width=\textwidth]{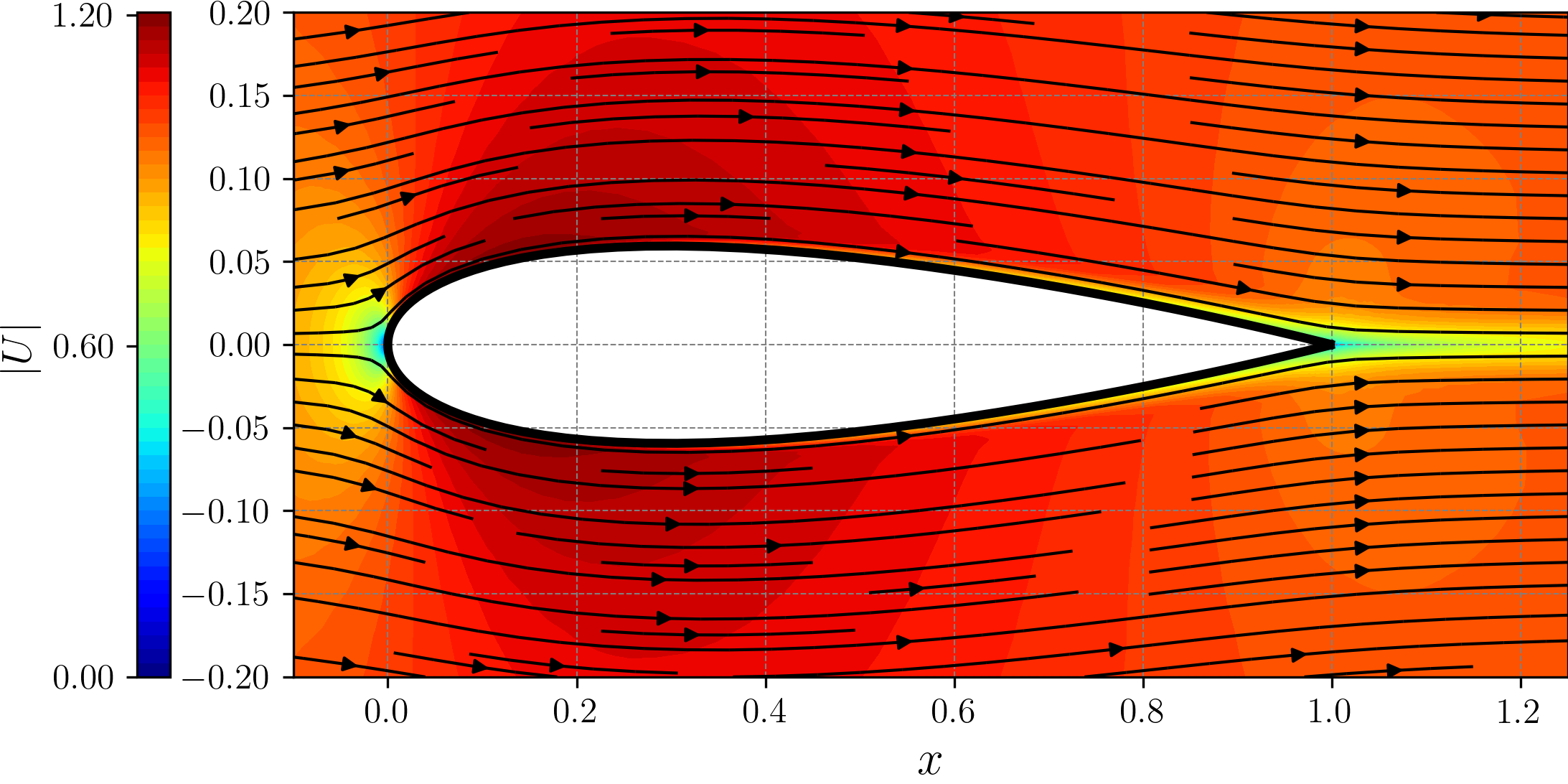} 
        \caption{Angle $0^\circ$}
    \end{subfigure}
    \hfill
    \begin{subfigure}[t]{0.32\textwidth}
        \centering
        \includegraphics[width=\textwidth]{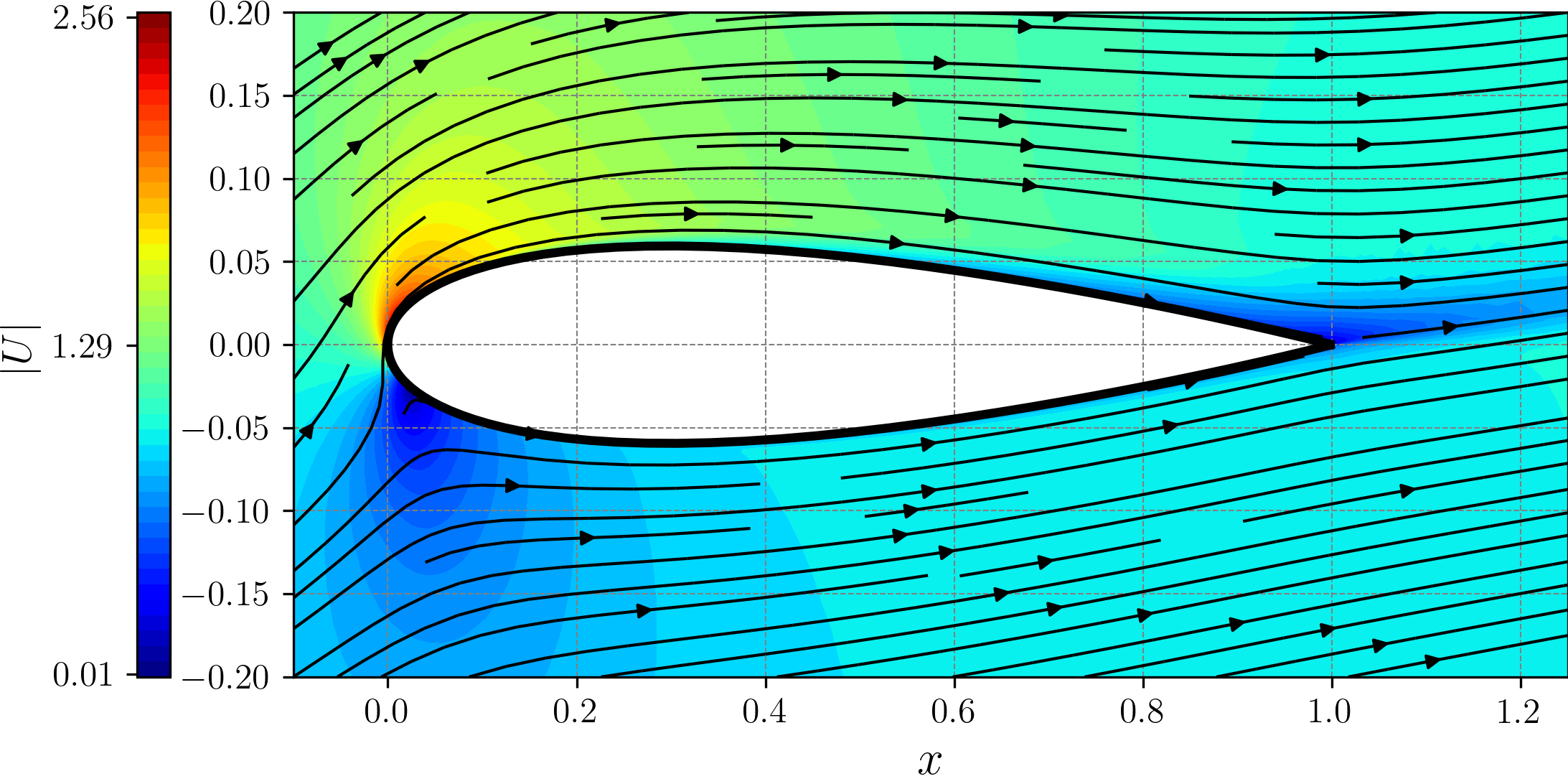} 
        \caption{Angle $10^\circ$}
    \end{subfigure}
    \hfill
    \begin{subfigure}[t]{0.32\textwidth}
        \centering
        \includegraphics[width=\textwidth]{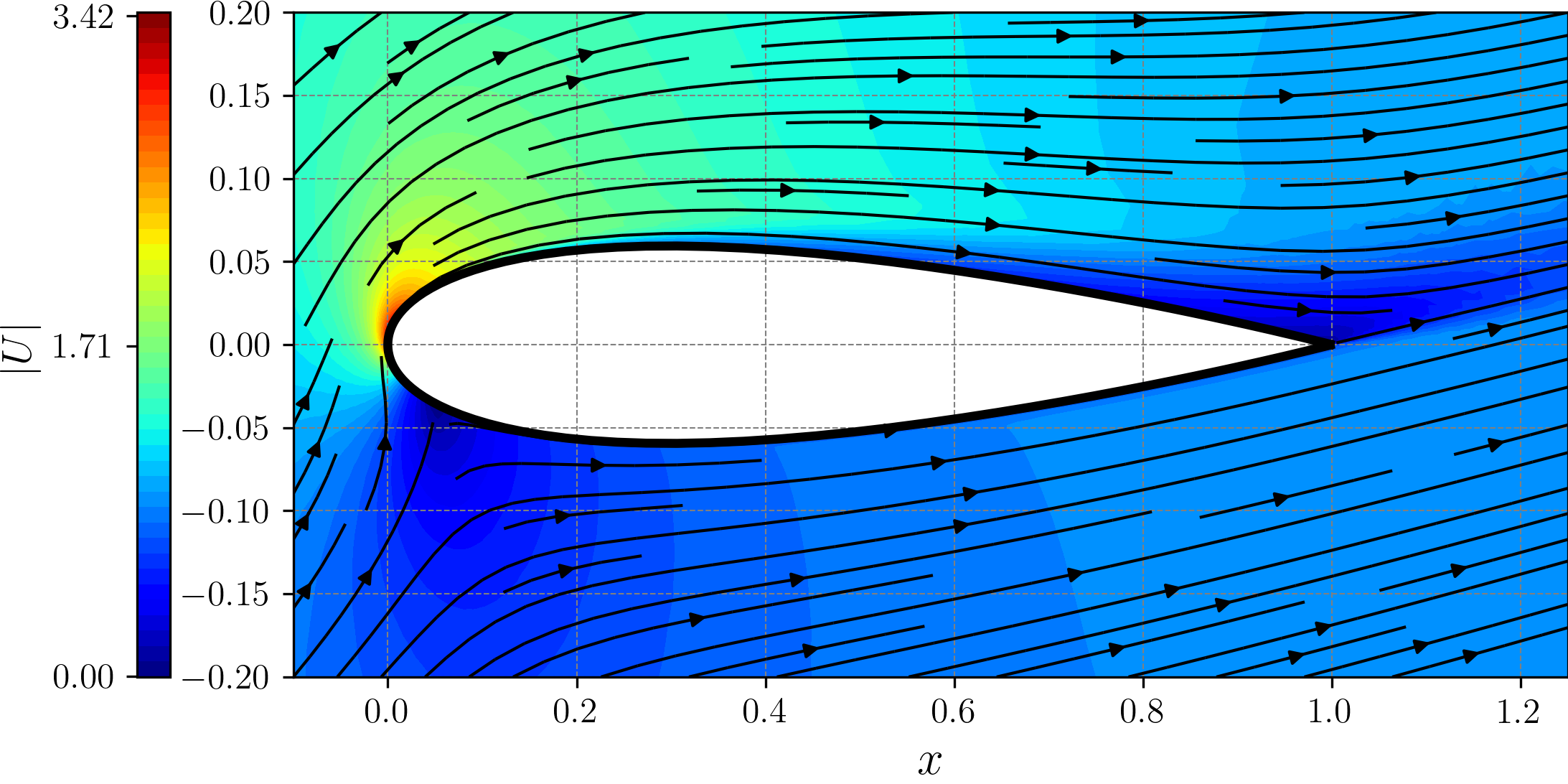} 
        \caption{Angle $15^\circ$}
    \end{subfigure}
    \caption{Streamlines of the 2D NACA0012 flow for the blended model \(\Mblend\) at different angles of attack.}
    \label{fig:NACA12_streamlines}
\end{figure}

\begin{figure}[h!]
    \centering
    
    % --- First Row: All Models ---
    \begin{subfigure}[t]{\textwidth}
        \centering
        \begin{subfigure}[t]{0.32\textwidth}
            \centering
            \includegraphics[width=\textwidth]{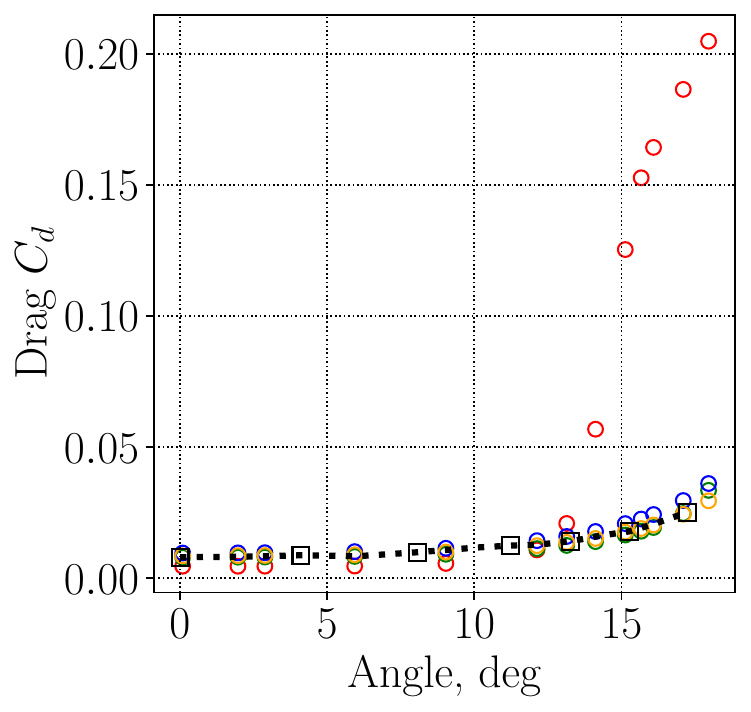}
            
        \end{subfigure}
        \hfill
        \begin{subfigure}[t]{0.32\textwidth}
            \centering
            \includegraphics[width=\textwidth]{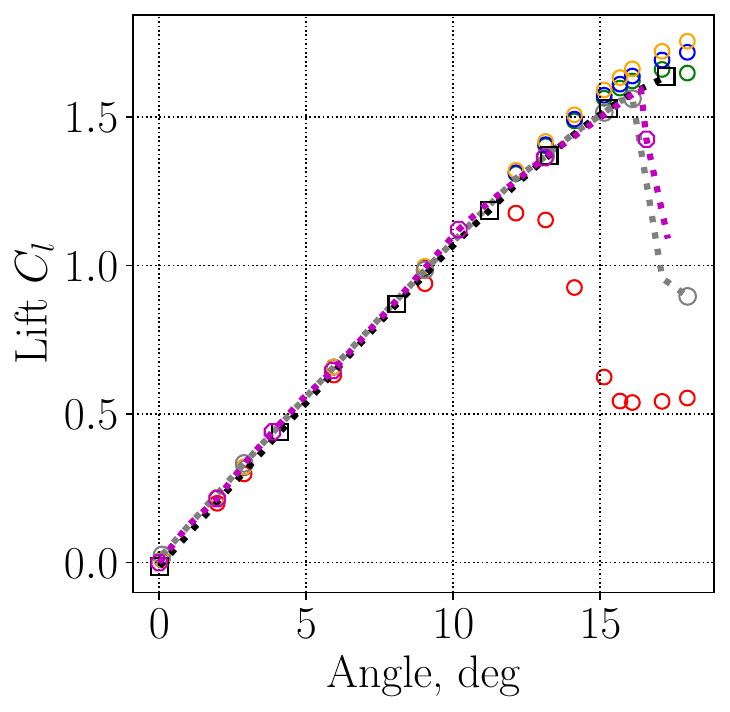}
            
        \end{subfigure}
        \hfill
        \begin{subfigure}[t]{0.32\textwidth}
            \centering
            \includegraphics[width=\textwidth]{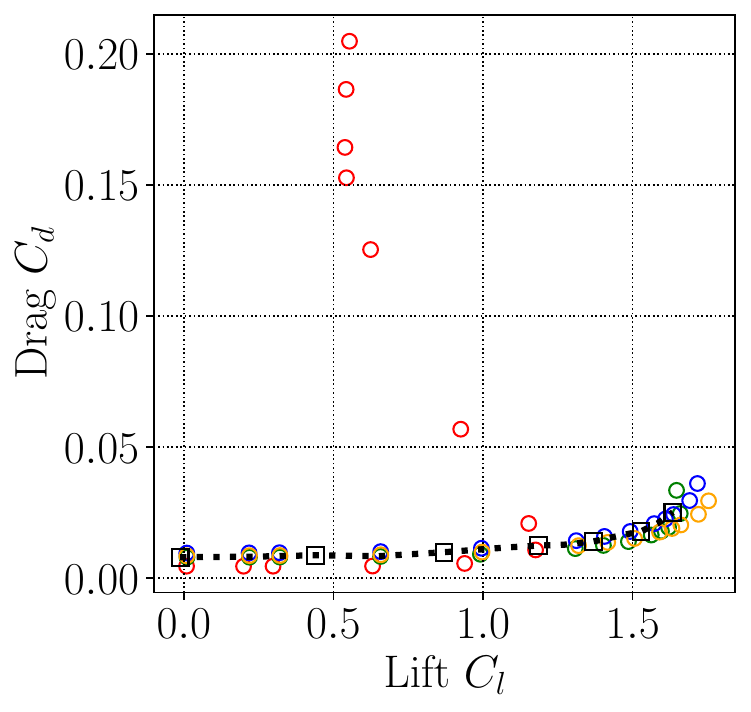}
            
        \end{subfigure}
        \caption{Drag and lift coefficients for the NACA~0012. }
        \label{fig:NACA12_draglift}
    \end{subfigure}
    
    \vspace{1em}
    
    % --- Second Row: ANSJ Omitted ---
    \begin{subfigure}[t]{\textwidth}
        \centering
        \begin{subfigure}[t]{0.32\textwidth}
            \centering
            \includegraphics[width=\textwidth]{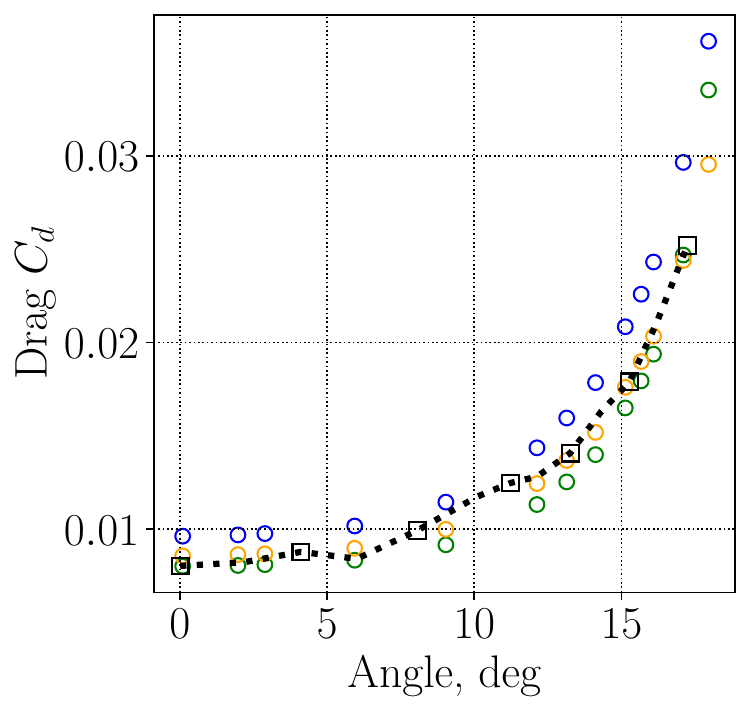}
            
        \end{subfigure}
        \hfill
        \begin{subfigure}[t]{0.32\textwidth}
            \centering
            \includegraphics[width=\textwidth]{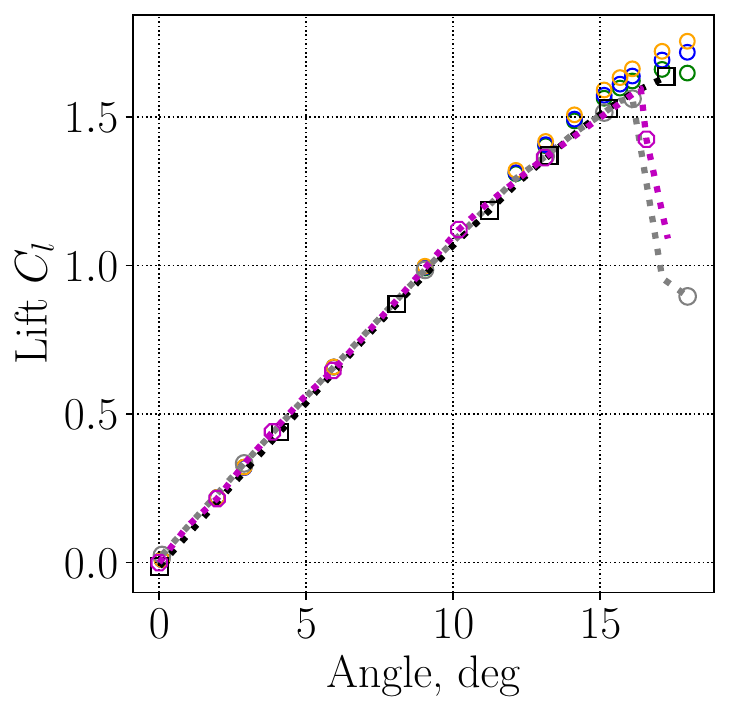}
            
        \end{subfigure}
        \hfill
        \begin{subfigure}[t]{0.32\textwidth}
            \centering
            \includegraphics[width=\textwidth]{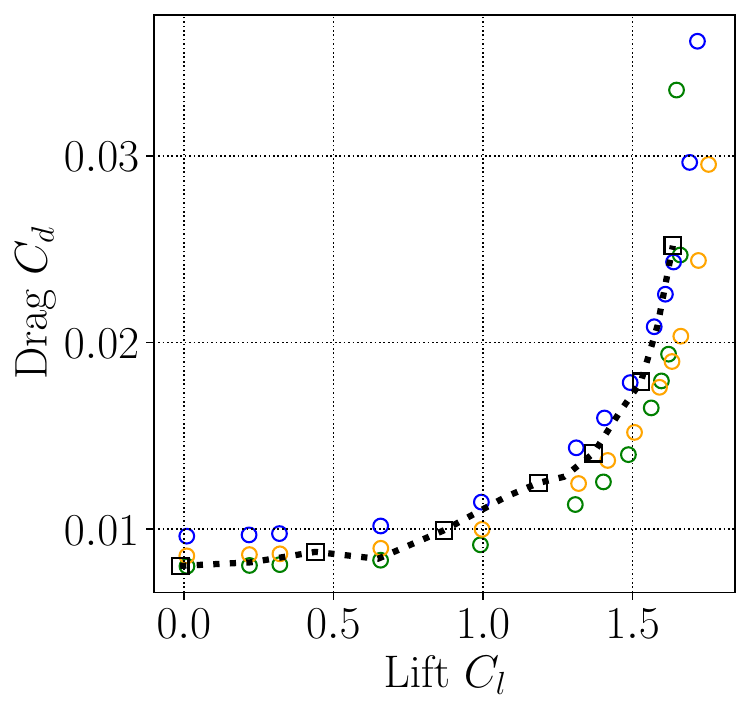}
            
        \end{subfigure}
        \caption{Drag and lift coefficients for the NACA~0012 with \(\MANSJ\) omitted for clarity. 
                 }
        \label{fig:NACA12_dragliftClean}
    \end{subfigure}
    
    \caption{Comparison of turbulence model predictions for the NACA0012 airfoil. 
    In (a), all models (\(\MANSJ\), \(\MSST\), \(\MSEP\), \(\Mblend\)) are shown, 
    while (b) omits the \(\MANSJ\) model for better clarity. Symbols for reference data: (\legendHF)  Ladson~\cite{Ladson1988}, (\legendHFgray) Gregory \& O'Reilly~\cite{Gregory1970}, and (\legendHFm) Abbott \& von Doenhoff~\cite{Abbott1959}. 
                 Symbols for turbulence models: (\legendCHANMarker) \(\MSST\), 
                 (\legendSEPMarker) \(\MSEP\), 
                 and (\legendintXMAMarker) \(\Mblend\).}
    \label{fig:NACA12_combined}
\end{figure}

\begin{figure}[h!]
    \centering
    % -- Row for alpha=0 --
    \begin{subfigure}[t]{\textwidth}
        \centering
        % First row (alpha=0) four plots side by side
        \begin{subfigure}[t]{0.24\textwidth}
            \centering
            \includegraphics[width=\textwidth]{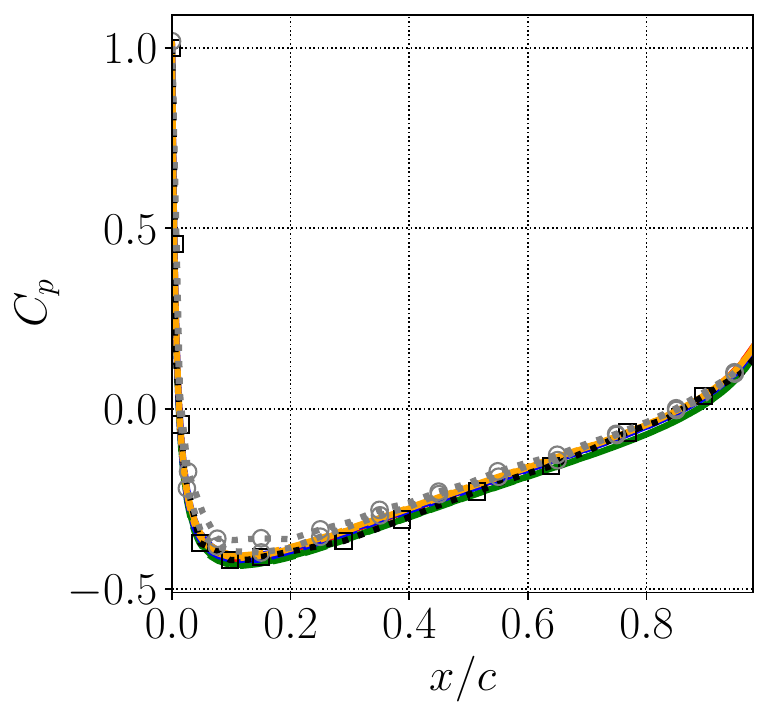}
            \label{fig:NACA12_Cp0}
        \end{subfigure}
        \hfill
        \begin{subfigure}[t]{0.24\textwidth}
            \centering
            \includegraphics[width=\textwidth]{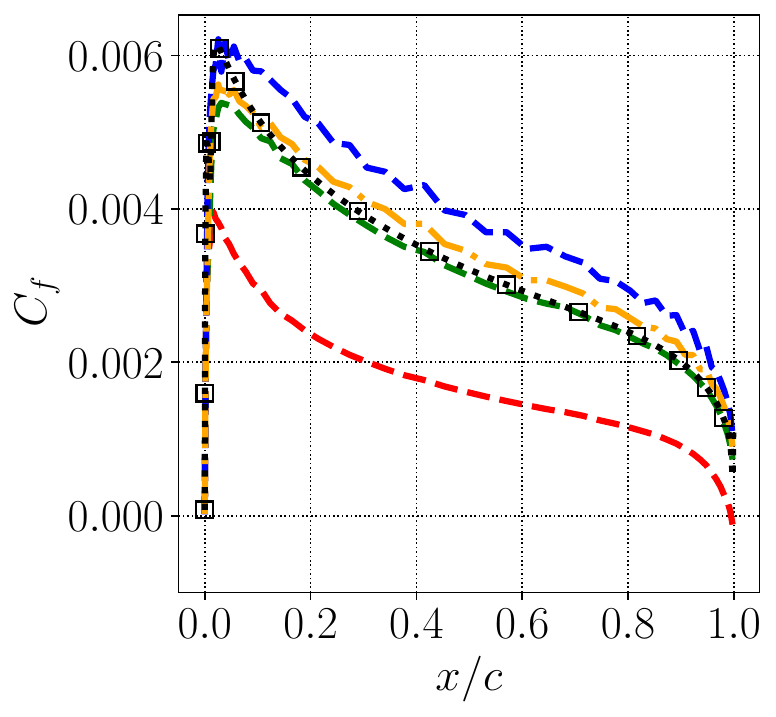}
            \label{fig:NACA12_Cf0}
        \end{subfigure}
        \hfill
        \begin{subfigure}[t]{0.24\textwidth}
            \centering
            \includegraphics[width=\textwidth]{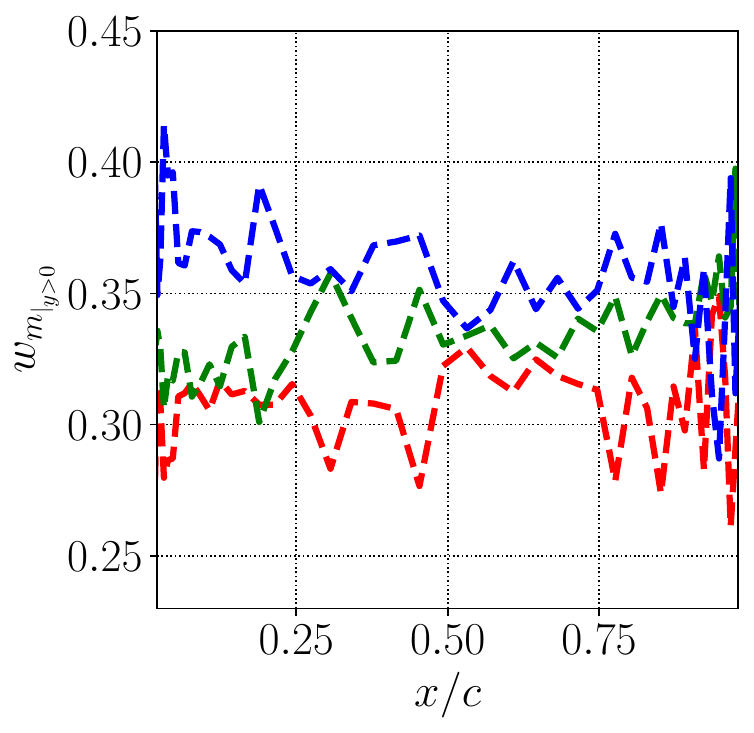}
            \label{fig:NACA12_weightsUpper0}
        \end{subfigure}
        \hfill
        \begin{subfigure}[t]{0.24\textwidth}
            \centering
            \includegraphics[width=\textwidth]{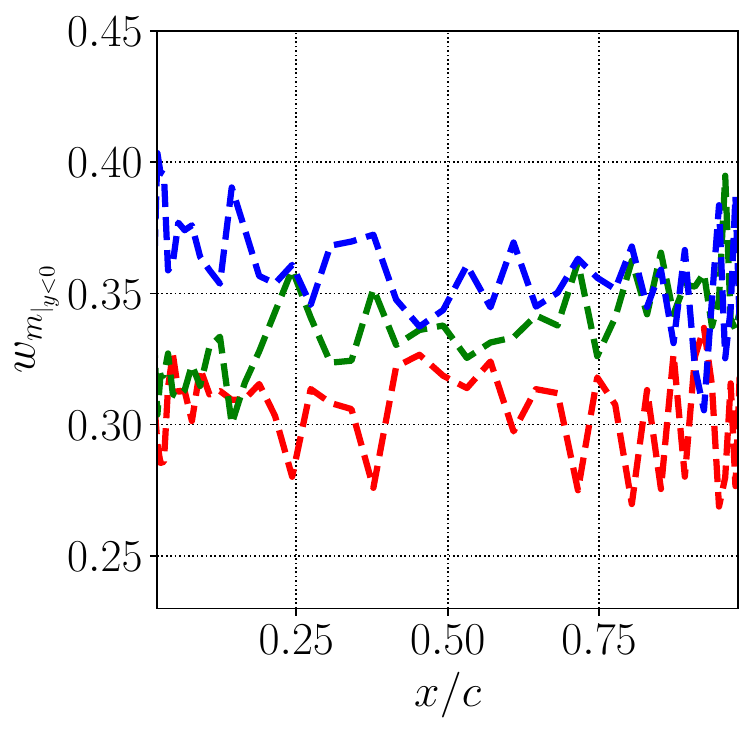}
            \label{fig:NACA12_weightsLower0}
        \end{subfigure}
        \caption{Angle of attack $0^\circ$}
        \label{fig:NACA12_alpha0}
    \end{subfigure}
    
    \vspace{1em} % Some vertical space between rows
    
    % -- Row for alpha=10 --
    \begin{subfigure}[t]{\textwidth}
        \centering
        % Second row (alpha=10) four plots side by side
        \begin{subfigure}[t]{0.24\textwidth}
            \centering
            \includegraphics[width=\textwidth]{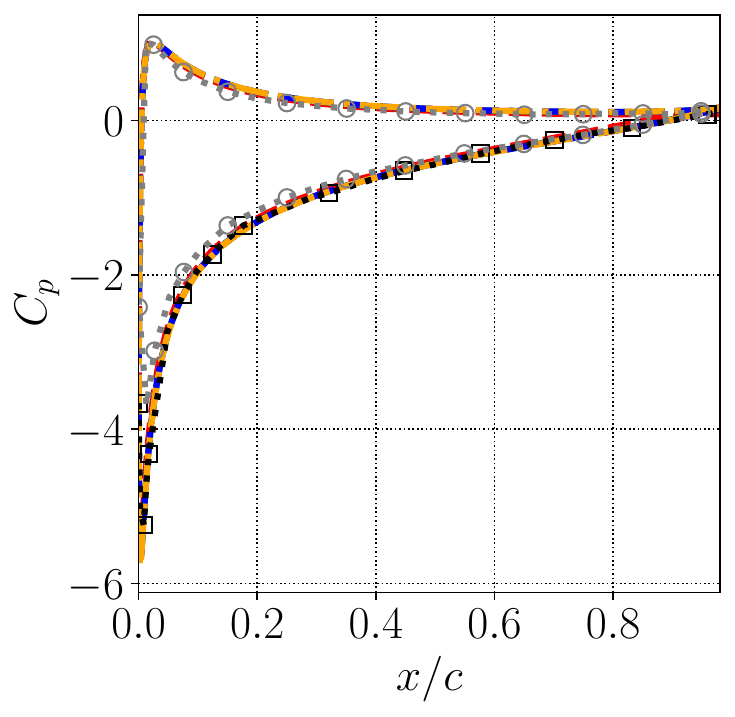}
            \label{fig:NACA12_Cp10}
        \end{subfigure}
        \hfill
        \begin{subfigure}[t]{0.24\textwidth}
            \centering
            \includegraphics[width=\textwidth]{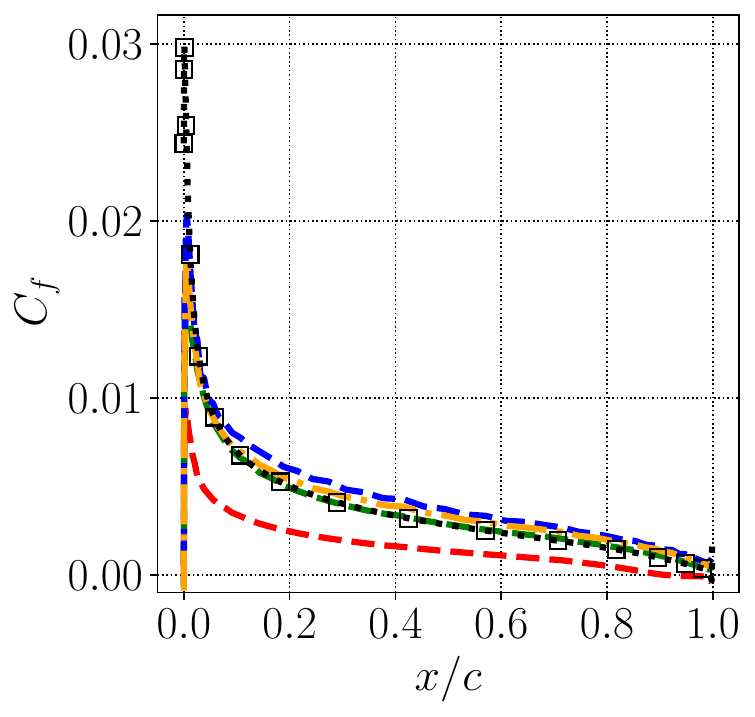}
            \label{fig:NACA12_Cf10}
        \end{subfigure}
        \hfill
        \begin{subfigure}[t]{0.24\textwidth}
            \centering
            \includegraphics[width=\textwidth]{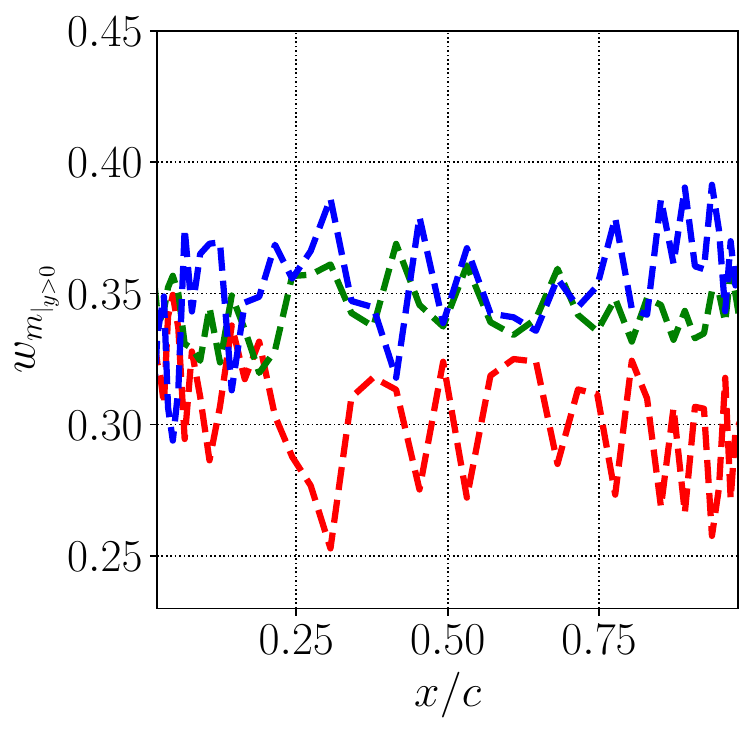}
            \label{fig:NACA12_weightsUpper10}
        \end{subfigure}
        \hfill
        \begin{subfigure}[t]{0.24\textwidth}
            \centering
            \includegraphics[width=\textwidth]{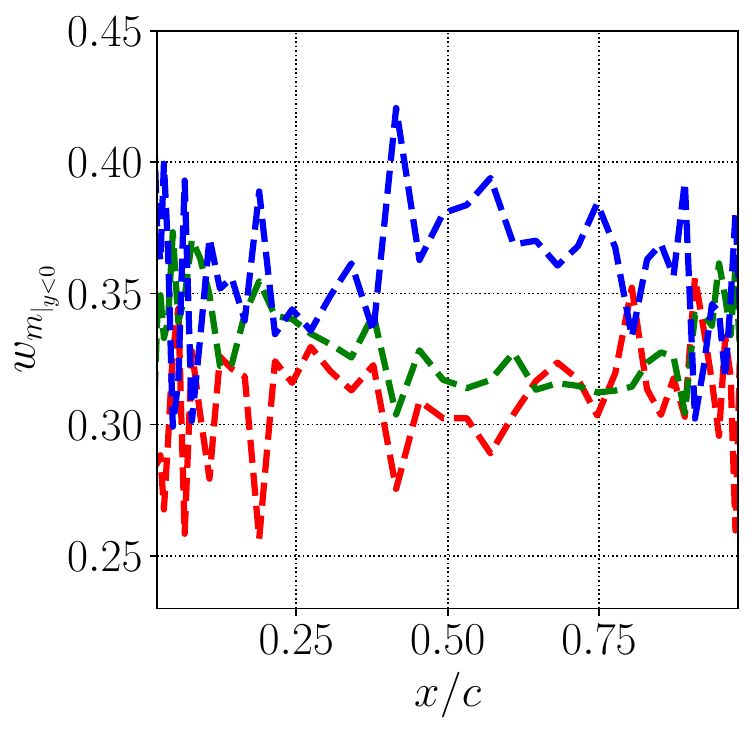}
            \label{fig:NACA12_weightsLower10}
        \end{subfigure}
        \caption{Angle of attack $10^\circ$}
        \label{fig:NACA12_alpha10}
    \end{subfigure}
    
    \vspace{1em} % Some vertical space between rows
    
    % -- Row for alpha=15 --
    \begin{subfigure}[t]{\textwidth}
        \centering
        % Third row (alpha=15) four plots side by side
        \begin{subfigure}[t]{0.24\textwidth}
            \centering
            \includegraphics[width=\textwidth]{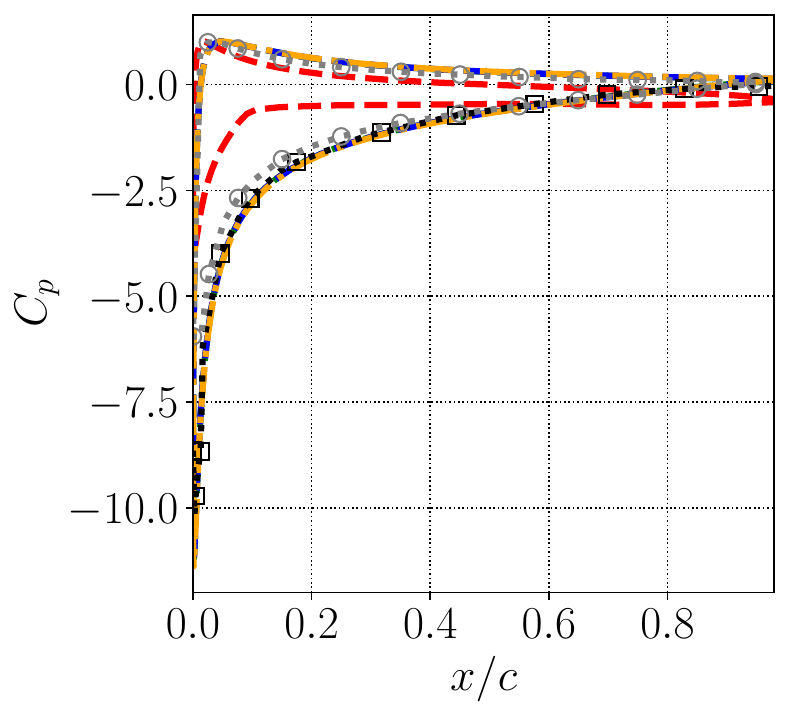}
            \label{fig:NACA12_Cp15}
        \end{subfigure}
        \hfill
        \begin{subfigure}[t]{0.24\textwidth}
            \centering
            \includegraphics[width=\textwidth]{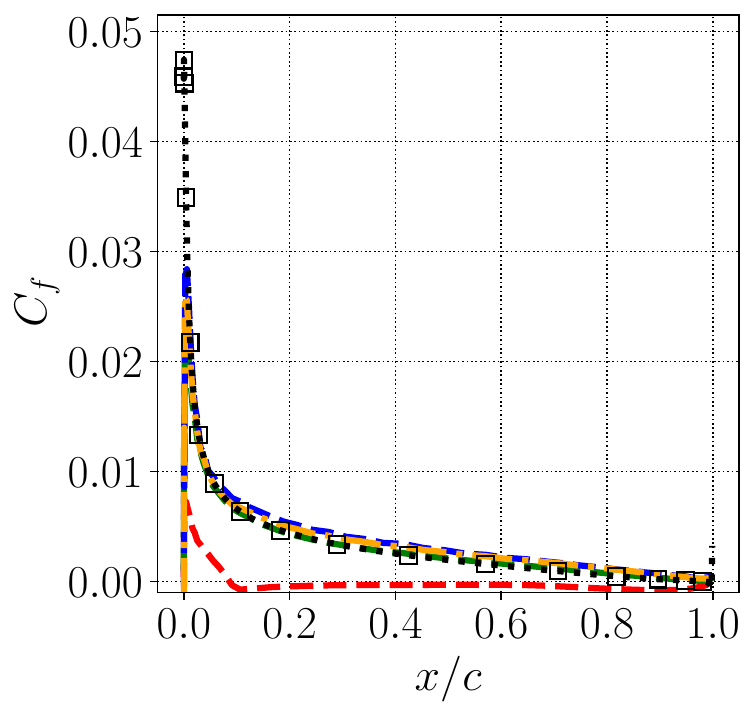}
            \label{fig:NACA12_Cf15}
        \end{subfigure}
        \hfill
        \begin{subfigure}[t]{0.24\textwidth}
            \centering
            \includegraphics[width=\textwidth]{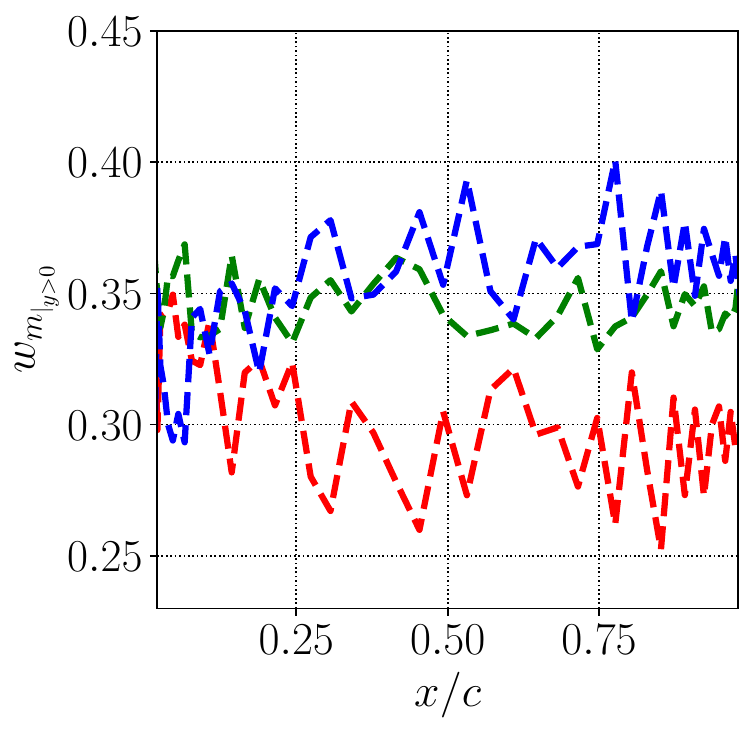}
            \label{fig:NACA12_weightsUpper15}
        \end{subfigure}
        \hfill
        \begin{subfigure}[t]{0.24\textwidth}
            \centering
            \includegraphics[width=\textwidth]{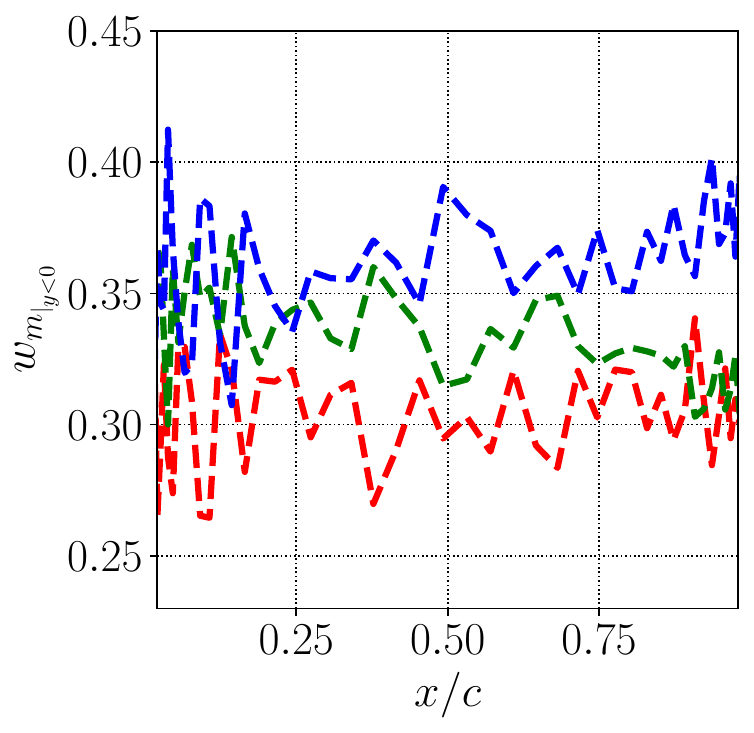}
            \label{fig:NACA12_weightsLower15}
        \end{subfigure}
        \caption{Angle of attack $15^\circ$}
        \label{fig:NACA12_alpha15}
    \end{subfigure}
    
    % Global caption and legend
    \caption{Surface pressure (first column), skin friction (second column), and blending weights on the upper (third column) and lower (fourth column) airfoil surfaces for the NACA0012 at three angles of attack ($0^\circ$, $10^\circ$, and $15^\circ$). Symbols for reference data: (\legendHF) Ladson~\cite{Ladson1988}, (\legendHFgray) Gregory \& O'Reilly~\cite{Gregory1970}, and (\legendHFm) Abbott \& von Doenhoff~\cite{Abbott1959}. Symbols for turbulence models:(\legendANSJ) \(\MANSJ\), (\legendCHAN) \(\MSST\), (\legendSEP) \(\MSEP\), (\legendintXMA) \(\Mblend\).}
    \label{fig:NACA12_CpCf_all}
\end{figure}

}

\section{Conclusions}
This work introduced a novel machine learning-based methodology for blending expert data-driven models\NewModif{, trained for specific flow classes,} to improve the performance of the \( k \)-\( \omega \) SST model across diverse flow configurations. The key feature of this approach is the \textit{intrusive internal} blending mechanism within the RANS equations, which differs from existing \textit{non-intrusive external} blending methods \cite{cherroud2025space}. This internal blending provides a unified solution by applying model corrections directly via weighting functions predicted by RFR, which are trained using the solutions of the expert models.

The proposed blending approach operates in two distinct phases: an offline \textit{learning} phase and an online \textit{prediction} phase. During the offline phase, a mixture of available experts is selected, where each expert model is tailored to specialize in a specific flow regime. RFR are employed to dynamically adjust the contribution of each expert model based on local flow features. The available expert models used in this paper were trained across various flow regimes, including attached boundary layers, separated flows, and free-shear flows (such as an axisymmetric jet). The flexibility of the weighting functions ensures that the most relevant expert models dominate in regions where they perform best, while less suitable models are downweighted. In the online phase, the computational process is streamlined into two steps. First, a baseline simulation is run to extract key flow features. Then, the blended correction formed by the RFR  is propagated within the RANS solver.

To evaluate the effectiveness of the proposed methodology, the model was tested on several training and test flow cases from the NASA turbulence modeling testing challenge \cite{rumsey2022nasa}. \ReviewerOne{ The test cases include the 2D zero-pressure-gradient flat plate flow, the separated flow over a wall-mounted hump, and a NACA0012 airfoil flow at high Reynolds number. The results demonstrate that the blended model outperforms the baseline \( \MSST \) model in flow configurations where the baseline model struggles, and that are sufficiently close to the training flows. Notably, in certain cases, the blended model even surpasses the best-performing individual expert models. Moreover, its accurate predictions of lift, drag, and surface-pressure distributions on the untrained NACA0012 flows, which differ significantly from the training configurations, highlight the model’s ability to generalize beyond the flow configurations present in the training data.}

Looking ahead, future work should focus on \ReviewerTwo{ developing a more optimal set of input features characterizing flow regions dominated by different physical dynamics and hence allowing to better allocate weights to the expert models. For instance, automatic clustering techniques could be employed to identify homogeneous data sets, allowing for more targeted expert model training \cite{cecileETMM}. Feature normalization is another critical aspect worth of further investigation, since this could contribute improving generalizability by limiting extrapolation.} \ReviewerOne{ Extending the methodology to more complex three-dimensional flows by incorporating additional expert models will be essential for validating the approach in more challenging configurations. This will require training expert model corrections accounting for 3D effects}. Another important avenue for future work is exploring online training of the weighting functions and optimizing feature selection to better capture the flow characteristics. \ReviewerTwo{Finally, nonlinear model combinations could also be considered, such as transformer models \cite{fang2024exploiting}), hierarchical mixtures of experts \cite{JordanJacobs_1994expertregions}, and Kolmogorov-Arnold Neural Networks \cite{liu2024kan}.}

Addressing these challenges will enhance the accuracy and adaptability of the blended model, ensuring its robustness across a broader range of flow configurations.

\section*{Funding information}
The authors gratefully acknowledge the support of the Institute for Scientific Computing and Data Science of Sorbonne University for providing the necessary resources and environment for this research under internal project ``LearnFLUIDS'' (Machine-LEARNing for FLUID Simulations).

\appendix

\section{Input features}
\label{ext:methodo}
%\section{Physical flow features used for regression}
To incorporate information about the local flow physics, we selected a set of physical flow features among those initially introduced in \cite{ling2015evaluation}, and summarized in Table \ref{Table_features}. 
The latter are supplemented by an additional feature proposed in Ref. \cite{girimaji2022NASA}: 
$$\eta_{11} = \frac{P_k}{P_k + \epsilon}$$
which introduces information about turbulent flow regimes for which the baseline model $k-\omega$ SST delivers reliable predictions. Specifically, $\eta_{11}$ is valuable in representing scenarios such as freely decaying turbulence, characterized by $P_k \rightarrow 0$ (hence, $\eta_{11} \rightarrow 0$), and equilibrium turbulence, where $P_k \approx \epsilon$, leading to $\eta_{11} \rightarrow \frac{1}{2}$. 

\ReviewerTwo{ These heuristic features were chosen based on expert judgement for their relevance to turbulence modeling. For instance, the $Q$-criterion is used to detect vortical regions, the turbulent kinetic energy, and its ratios to local strain or dissipation rate are used to detect the local state of turbulence, etc.

Of note, the present choice is mostly based on practical implementation reasons and previous work by the present and other authors, but it is most likely not optimal. In addition, while the present features are, by construction, bounded within -1 and 1 to improve generalizability, other nondimensionalization choices could be imagined.
Finally, unsupervised machine learning and latent space analysis could be use to automatically extract a best set of features from the training data, but such features would hardly be physically interpretable. 

A systematic framework called LIFE (``Learning and
Inference assisted by Feature-space Engineering'') for selecting suitable features for turbulence and transition model augmentation has been recently proposed in \cite{srivastava2021generalizable}. 
More research is warranted in the future for finding a set of more optimal, non-dimensional and interpretable features.
}

\begin{table*}%[H]%
% \setstretch{1.55}
\centering
\centerline{
\resizebox{1.1\textwidth}{!}{%
\begin{tabular}{ c|c|c||c|c|c }
%\hline
Feature & Description & Formula & Feature & Description & Formula \\ \hline
$\expcomp_1$ & Normalized $Q$ criterion & $\dfrac{||\Ot||^2 - ||\St||^2 }{||\Ot||^2 + ||\St||^2}$ & 
$\expcomp_6$ & Viscosity ratio & $ \dfrac{\nu_T}{100\nu + \nu_T}$ \\ \hline
$\expcomp_2$&Turbulence intensity&$\dfrac{k}{0.5\vel_i\vel_i+k}$&$\expcomp_7$
& \begin{tabular}[c]{@{}l@{}}Ratio of pressure\\ normal stresses to\\ normal shear
stresses\end{tabular} &$\dfrac{\sqrt{\DPS\derp{\pres}{x_i} \derp{\pres}{x_i}}}{\DPS\sqrt{\derp{\pres}{x_j}
\derp{\pres}{x_j}} + 0.5\rho\derp{\vel_k^2 }{x_k}}$ \\ \hline
$\expcomp_3$&\begin{tabular}[c]{@{}l@{}}Turbulent Reynolds\\number\end{tabular}&
$\min{\left(\dfrac{\sqrt{k}\lambda}{50 \nu}, 2\right)}$ & $\expcomp_8$
&\begin{tabular}[c]{@{}l@{}}Non-orthogonality\\ marker between velocity\\ and its gradient
\cite{gorle2013framework}\end{tabular} &$\dfrac{\DPS\left|\vel_k\vel_l\derp{\vel_k}{x_l} \right|}{\DPS\sqrt{
\vel_n \vel_n\vel_i\derp{\vel_i}{x_j} \vel_m\derp{\vel_m}{x_j}} +\left| \vel_i\vel_j
\derp{\vel_i}{x_j} \right|}$ \\ \hline
$\expcomp_4$&\begin{tabular}[c]{@{}l@{}}Pressure gradient\\along streamline\end{tabular}&
$\dfrac{ \DPS\vel_k\derp{\pres}{x_k} }{\DPS \sqrt{\derp{\pres}{x_j}\derp{\pres}{x_j} \vel_i \vel_i} + \left|
\vel_l\derp{\pres}{x_l }\right|}$ & $\expcomp_{9}$ & \begin{tabular}[c]{@{}l@{}}Ratio of convection
to\\production \ReviewerOne{ of $k$}\end{tabular} &$\dfrac{\DPS\vel_i\derp{k}{x_i}}{\DPS|
\overline{u_j' u_l'}S_{jl} | + \vel_l \derp{k}{x_l} }$ \\ \hline
$\expcomp_5$& \begin{tabular}[c]{@{}l@{}}Ratio of turbulent\\time scale to mean\\strain time
scale\end{tabular} & $\dfrac{ ||\St|| k}{ ||\St|| k + \varepsilon }$ & $\expcomp_{10}$ 
& \begin{tabular}[c]{@{}l@{}}Ratio of total Reynolds\\ stresses to normal\\ Reynolds
stresses\end{tabular} & $ \dfrac{||\overline{u_i' u_j'}||}{ k + ||\overline{u_i' u_j'}|| } $ \\
%\hline
\end{tabular}%
}}
\caption{ List of input features used to train the weighting functions for the blended model.}\label{Table_features}
\end{table*}

\section{Constants and auxiliary relations for the  \komegasst{}}\label{Appendix_kwSST_csts}
The constants of \komegasst{} model are given bellow
\begin{align}	
	 &\hspace*{-0.4cm}F_1 = \text{tanh}\left[\left(\min\left\{\min \left[\max \left(\frac{\sqrt{k}}{\beta^* \omega y}, \frac{500\nu}{y^2\omega} \right), \frac{4\alpha_{\omega 2}k}{CD^+_{k\omega}y^2} \right] ;10\right\}\right)^4\right],\nonumber\\
	 &CD^+_{k\omega} = \max \left( CD_{k\omega}, 10^{-10} \right), \nonumber\\
	 &F_2 = \tanh \left[ \left(\min\left\{ \max \left( \frac{2\sqrt{k}}{\beta^* \omega y}, \frac{500\nu}{y^2 \omega} \right) ;100\right\}\right)^2  \right], \nonumber\\
  &F_3 = 1 - \text{tanh}\left[\left(\min\left\{ 150 \frac{\nu}{\omega y^2} ; 10\right\}\right)^4\right], \nonumber\\
  & F_{23} = F_2 \,F_3, \text{ if }F_3 \text{ is enabled, or } F_2 \text{ otherwise}, \nonumber\\
	 %&\Phi = F_1\Phi_1 + (1-F_1)\Phi_2,
  &D_{k}^{\text{Eff}} = \nu + \alpha_k(F1) \nu_t, \nonumber\\
  &D_{\omega}^{\text{Eff}} = \nu + \alpha_{\omega}(F1) \nu_t \nonumber
	 \end{align}
\noindent where $y$ is blending function distance and $\alpha_k(F1)$, $\alpha_{\omega}(F1)$ are blending functions given by
$$\alpha_k(F_1) = F_1 \alpha_{k_1} + (1-F_1) \alpha_{k_2}$$
$$\alpha_{\omega}(F_1) = F_1 \alpha_{{\omega}_1} + (1-F_1) \alpha_{{\omega}_2}$$
and 
\begin{align*}
	&\alpha_k = (0.85,1), \alpha_\omega = (0.5,0.856),\\
	&\gamma = (5/9,0.44), \beta = (0.075,0.0828).
\end{align*}

$$\beta^*=0.09, a_1 = 0.31, b_1=1, c_1=10$$

\bibliography{cas-refs}

\end{document}